%% file: WageLevelsMinWage_resubmission.tex
\documentclass[12pt,english]{article}
\input{preamble}

\title{\vspace*{2cm}Does regional variation in wage levels identify the effects of a national minimum wage?}

\author{Daniel Haanwinckel	\thanks{Department of Economics, UCLA and NBER. Contact information: haanwinckel@econ.ucla.edu \newline This paper benefited from conversations with Martha Bailey, Patrick Kline, Adriana Lleras-Muney, Maurizio Mazzocco, Yotam Shem-tov, and many others, as well as from comments from the editor, anonymous referees, and participants of the 2023 Columbia Junior Micro/Macro Labor Conference, the 2023 NBER Labor Studies Fall Meeting, the 2024 SOLE Meeting, and the 2024 Meeting of the Brazilian Econometric Society. Filipe Fiedler and Luca Lorenzini provided excellent research assistance.}}
\date{May 2026}

\begin{document}

	\renewcommand{\floatpagefraction}{.7}
	
	\singlespacing
	
	\maketitle
	
	\begin{abstract}
		This paper asks whether regional wage differences can identify the effects of a national minimum wage.
		I study two common exposure-based approaches: effective-minimum-wage designs, which compare the minimum wage to contemporaneous local wages, and fraction-affected/gap designs, which measure pre-reform exposure to the new minimum.
		Using theory, simulations, and evidence from Brazil, I show when these approaches can mislead and how their performance depends on specification choices.
		The results lead to practical recommendations for applied researchers, including when to avoid these designs, how to test their assumptions, which specifications are more reliable, and how similar concerns may apply to other settings.	
	\end{abstract}
	
	\vspace{2cm}
	

	\thispagestyle{empty} \clearpage
	\setcounter{page}{1}
	\onehalfspacing
	
	\section{Introduction}
	
	How can we identify the labor market effects of minimum wage policies? A common approach is to use state- or province-level minimum wage laws as the ``treatment variable'' in panel regressions where the outcomes are employment or wage metrics aggregated at the region-time level. This strategy is used in classic work such as \cite{Card1994}, and newer estimators exploiting this idea continue to be developed \citep[e.g.][]{Cengiz2018,Ghanem2023}. 	
	However, in many contexts, this identifying variation is unavailable because minimum wages are set at the country level. In these cases, a feasible alternative is to exploit the idea that country-wide minimum wage laws are probably more binding in low-wage regions. Thus, following an increase in the national minimum wage, one can infer minimum wage effects by comparing the differential evolution of regions with different wage levels.
	
	There are two main ways to implement this idea, both using panel regressions at the regional level. The first, which I refer to as the ``effective minimum wage design,'' defines treatment intensity as the current gap between the minimum wage and a measure of regional wage levels (typically the median wage). The second---the ``fraction-affected/gap designs''---employ difference-in-differences-style regressions with treatment variables constructed based on wage levels in a pre-reform period.
	Recent applications of these designs include Mexico \citep{Bosch2010}, South Africa \citep{Dinkelman2012}, Germany \citep{Dustmann2021}, the US in the 1960s and 1970s \citep{Bailey2021}, Brazil \citep{Engbom2022}, and the UK \citep{Giupponi2024}.
	
	This paper examines the validity of these estimators when the minimum wage is set at the national level. This examination is necessary because the original papers proposing these methods---\cite{Lee1999} and \citet{Card1992}, respectively---studied the United States in the 1980s and 1990s, a setting with substantial identifying variation from state-level minimum wage laws. When the data do not include such variation, the source of identification is less transparent: unspecified economic factors that shift regional wage levels, combined with functional form assumptions mapping regional wage levels to treatment intensity.
	
	The paper's first contribution is to document potential sources of biases not discussed in previous work. I use simple economic theory to list plausible margins through which minimum wages may affect labor markets, and then evaluate whether the estimators can capture those effects. I show that these estimators' assumptions and functional forms are sometimes incompatible with how minimum wages may affect the economy, leading to misspecification biases.
	
	The second contribution is to quantify biases in typical applications. In addition to the new sources of bias discussed above, I also assess the relevance of minor violations of the identification assumptions discussed in previous work. I include those issues because they are often not tested in empirical applications, and because---as I argue in the paper---they may be common.
	
	I employ two approaches to quantifying biases. The first is a battery of simulation exercises based on hundreds of economic models calibrated to generate realistic data sets. The simulations span a broad range of plausible minimum wage effects: disemployment in the lower tail of the wage distribution, large wage increases with no employment effects, positive effects on both wages and employment, and changes in the returns to skill in a ``canonical'' model in the style of \citet{Katz1992}.	
	The second approach is a detailed study of the rise of the Brazilian federal minimum wage starting in 1995. Unlike in simulations, we do not observe the underlying data-generating process in a real application, so biases cannot be measured directly. However, if equally plausible econometric strategies exploiting similar variation lead to large differences in estimated causal effects, that pattern would still be consistent with misspecification in some (or all) of those strategies. To assess this possibility, I collect many empirical estimates of minimum wage effects in Brazil---some from existing literature and others from original data analysis. 
	
	The third contribution is to distill the paper's results into practical recommendations for applied researchers. I argue that effective minimum wage designs should be avoided when the data do not include region-specific minimum wages. Fraction-affected/gap designs can perform well when the parallel pre-trends assumption holds. However, that assumption can be violated due to several plausible economic factors in minimum wage settings, so it should not be assumed to hold without supporting evidence. I provide specific guidance on testing, functional forms, and other specification choices. 
	
	The paper is organized as follows. The remainder of this introduction discusses how the paper relates to other papers on the econometrics of minimum wages. Section~\ref{sec:definitions} defines terms used throughout the paper, starting from the treatment effect to be identified from the data. Section~\ref{sec:eff-min-w} presents and studies the effective minimum wage design, while Section~\ref{sec:fagap} does the same for the fraction-affected/gap designs. Section~\ref{sec:500-dgps} features the main simulation exercises. The analysis of the Brazilian case follows in Section~\ref{sec:brazil}. Section~\ref{sec:recommendations} provides practical recommendations for applied researchers. Finally, the last section concludes with a discussion of how similar econometric issues may appear in two other contexts: the local labor market effects of recessions and the relationship between employer concentration and labor market outcomes.
	
	\paragraph{Related literature.} Many papers discuss econometric challenges when minimum wage effects take time to materialize.\footnote{Using Canadian data, \citet{Baker1999} document that employment responds differently to low- or high-frequency minimum wage variation. \citet{Sorkin2015} discusses practical challenges in estimating long-run effects using reduced-form approaches. \cite{Meer2016} discuss econometric challenges that arise if minimum wages change employment growth rates instead of employment levels. \citet{Vogel2023} documents that minimum wage effects ``trickle up'' the wage distribution over a few years.} I focus on a two-period model without dynamics, thus ruling out those issues by construction in my setting. That makes my findings distinct from, and complementary to, that strand of literature. 
	
	The structural models I study also do not include measurement error, so the issues in this paper differ from the mechanical bias discussed in \citet{Autor2016}.	My model also imposes complete independence between regions, ruling out spillover effects coming from, e.g., migration responses \citep{Cadena2014,Huang2019}.
	
	The issues I discuss complement recent work on the econometrics of difference-in-differences models. Because the analysis is restricted to a two-period setting, concerns related to staggered treatment timing do not apply \citep{Chaisemartin2020,Roth2023}. My analysis is closer to papers studying continuous treatment intensity \citep{Callaway2024} and biases arising from heterogeneous treatment effects \citep{deChaisemartin2018}.
	
	A particularly important reference point is the handbook chapter by \citet{Dube2024}, which provides a comprehensive and practically useful synthesis of empirical methods in the modern minimum wage literature. Their discussion of national minimum wage settings describes exposure-based estimators built from pre-reform measures of minimum wage ``bite,'' including fraction-affected and gap-style variables, and emphasizes several identification concerns: parallel trends, spillovers and mobility across groups, and treatment-effect heterogeneity correlated with exposure. This paper complements that discussion by connecting those econometric concerns to the economics of minimum wage incidence. I show that even in otherwise favorable regional designs, scalar exposure measures impose strong restrictions on the economic relationship between the minimum wage and labor market outcomes. Standard models imply that this relationship may be nonlinear, especially for employment outcomes, and the simulations quantify how much bias can result from imposing overly simple functional forms.
	
	Finally, my paper focuses on designs where regions are the unit of analysis. For this reason, I do not evaluate ``fraction-affected/gap'' designs where the unit of analysis is workers \citep[as some specifications in][]{Dustmann2021} or firms \citep{Harasztosi2018}. These estimators are not designed to capture labor market-level treatment effects but rather how outcomes for different groups of firms or workers evolve in response to the minimum wage. A detailed econometric analysis of these methods does not fit within the framework I propose in this paper and is thus left to future work.

	\section{Setup \label{sec:definitions}}
	
	\subsection{Data-generating processes and the outcomes of interest}
	
	We start with a ``potential outcomes'' function $f$ that, given the level of the log minimum wage $mw$ and a vector of regional characteristics $\bm{\theta}$, outputs an outcome $y$, such as employment or wage inequality. That is: $y=f(mw, \bm{\theta})$. The causal effects of the minimum wage in a region are determined by how $f$ varies with $mw$, holding $\bm{\theta}$ constant. The vector $\bm{\theta}$ should be interpreted as including all other determinants of the outcome of interest at the regional level, along with determinants of treatment effect heterogeneity across regions. In simulations, I will impose specific functional forms for $f$ corresponding to different economic models.
	
	We study two-period ($t\in\{0,1\}$) data-generating processes (DGPs) of the following form:
	\begin{align}
		y_{r,t} & =f\left(mw_{t}, \bm{\theta}_{r,t}\right) \label{eq:general-dgp}
		\\
		[\bm{\theta}'_{r,0}, \bm{\theta}'_{r,1}]' & \sim G \notag
	\end{align}
	where $r\in\{1,\dots,R\}$ indexes regions, and $G$ is the distribution of region-specific characteristics. Note that the vector of regional characteristics may change over time, along with the minimum wage. Regional characteristics may correlate over time but are independent across regions.
	
	I assume that $mw_1>mw_0$. There is no loss of generality in the DGP from Equation~\eqref{eq:general-dgp}, as outcomes depend only on current values of $mw_{t}$ and $\bm{\theta}_{r,t}$. There would be loss of generality in models with nominal rigidities or other dynamic concerns, but I abstract from such issues in this paper.
	
	There are two natural ways to define the ceteris paribus causal effects of the rise in the national minimum wage:
	\begin{align*}
		ATE_0 & = \mathbb{E}\left[ f\left(mw_1, \theta_{r,0}\right) - f\left(mw_0, \theta_{r,0}\right) \right] & \qquad \qquad ATE_1 & = \mathbb{E}\left[ f\left(mw_1, \theta_{r,1}\right) - f\left(mw_0, \theta_{r,1}\right)\right]
		\\
		& = \mathbb{E}\left[ f\left(mw_1, \theta_{r,0}\right) \right] - \mathbb{E}\left[\bm{y}_{r,0}\right]
		& & = \mathbb{E}\left[ \bm{y}_{r,1} \right] - \mathbb{E}\left[f\left(mw_0, \theta_{r,1}\right)\right],
		\end{align*}
	where the expectation is taken with respect to the distribution of the $\bm{\theta}_{r,t}$ variables. The first formulation, ${ATE}_0$, requires evaluating a counterfactual where the minimum wage rises from $mw_0$ to $mw_1$ but other characteristics remain at their $t=0$ levels. The second formulation compares the outcomes as of $t=1$ to a counterfactual scenario where the minimum wage remained at the $t=0$ level. The two definitions are identical if the $\theta_{r,t}$ distribution is time-invariant. I use the average of these two definitions as the object of interest to be recovered by the econometric designs:
	\begin{equation}
		ATE = \frac{ATE_0 + ATE_1}{2} \label{eq:ate}
	\end{equation}
	
	Does this definition really correspond to the central object of interest in papers using the effective minimum wage and fraction-affected/gap designs? I claim that the answer is \textit{yes}. Appendix~\ref{appendix:ate-discussion} substantiates this point by showing that all of the recently published papers cited early in the introduction seek to identify an average (or aggregate) effect of this type.
	
	I impose an additional restriction on the data-generating process: there are no trends in overall wage levels. Suppose the minimum wage change is simultaneous with an unobserved shock to total factor productivity (TFP) affecting all regions. In that case, it is only possible to separately identify the average effects of the minimum wage by imposing further assumptions. To abstract from this ``missing intercept'' issue, I rule out common TFP shocks.
	
	\subsection{The Normal-markdown simulation model \label{subsec:normal-markdown}}
	
	Most simulation exercises in the paper are based on the model described below. It has two objectives. First, it must be simple, transparent, and useful for illustrating identification concerns. Second, despite its simplicity, it must realistically approximate the main theoretical mechanisms emphasized in minimum wage literature.
	
	In the simulation model, each region corresponds to an isolated local labor market with a continuum of worker productivity levels. These workers supply their labor inelastically. In the absence of a minimum wage, the log wage distribution would be Normal with mean $\mu_{r,t}$ and standard deviation $\sigma_{r,t}$. I refer below to that distribution as the \textit{latent} log wage distribution, and $w^*$ symbolizes latent log wage levels for different workers.\footnote{Wage distributions often exhibit a thicker upper tail than that implied by the log-normal distribution. Because this paper focuses on minimum wage effects, I abstract from this feature to keep the simulation model as simple as possible.}
	
	\textbf{Truncation and censoring effects.} In the basic model, minimum wages have two effects that are modulated by a \textit{markdown} parameter $m\in(0,1)$. First, workers who would earn a log wage below $mw+\ln m$ lose their jobs. Second, workers with latent wages $w^*\in [mw+\ln m, mw)$ earn the minimum wage instead of their latent log wage. The latter effect creates a ``spike'' at the minimum wage in simulated log wage distributions, a phenomenon commonly observed in settings where the minimum wage binds. 
	
	In Appendix~\ref{appendix:simulation-details}, I present a microfoundation for this simulation model based on idiosyncratic preferences for individual firms, as in \citet{Card2018}. Workers with $w^*>mw$ have a high productivity, such that the minimum wage does not bind for them. In the intermediate range $w^*\in [mw+\ln m, mw)$, the minimum wage binds, but workers are productive enough that firms still hire them even if they have to be paid the minimum wage. That's because unconstrained wages are a constant fraction $m$ of the marginal products of labor. Finally, workers with $w^*<mw+\ln m$ lose their jobs. The perfectly competitive model is approximated as $m\to 1$. With smaller $m$, firms have more monopsony power, allowing minimum wage policies to increase wages with little employment loss.
	
	\begin{figure}
		\centering
		
		\includegraphics[width=0.7\linewidth]{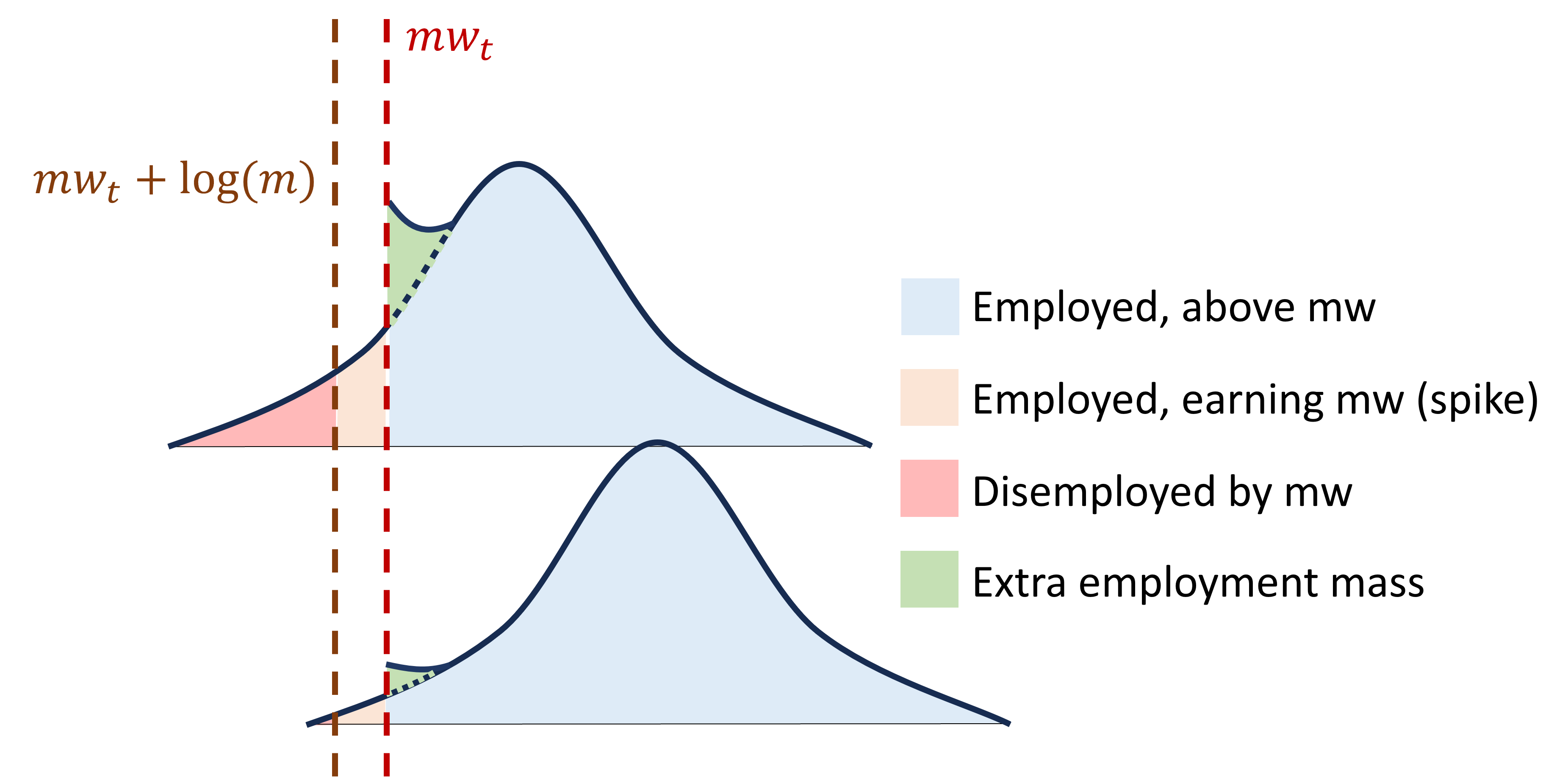}
		\caption{Minimum wage effects in the Normal-markdown simulation model}
		\label{fig:simulation-model}
	\end{figure}
	
	\paragraph{Positive employment effects.} In the full model, the minimum wage may also increase employment at wage levels slightly above the minimum wage. This effect is modeled by adding a triangle-shaped employment mass to the log wage density plot, according to parameters $P_{base}$ and $P_{height}$. The base of the triangle extends from the log minimum wage $mw$ on the left to $mw +P_{base}$ on the right. The height of the triangle at the point where the minimum wage binds is $P_{height}$ multiplied by the latent log-wage density at that point. The added mass is heterogeneous across regions; it may be negligible in high-wage regions where the minimum wage does not bind, but significant in lower-wage regions.
	
	Figure~\ref{fig:simulation-model} illustrates minimum wage effects for two regions with different latent wage levels (corresponding to different values of $\mu_{r,t}$). Truncation corresponds to the red areas, censoring corresponds to the orange areas, and the extra mass corresponds to the green areas.
	
	The full simulation model can approximate a broad range of minimum wage effects discussed in the labor economics literature. In many contexts, minimum wages are believed to have near-zero employment effects but significant wage increases in the lower tail \citep{Dube2024}. That can correspond to either a low level of $m$ (such that there is much more censoring than truncation) or a high $m$ combined with positive employment effects (such that these effects roughly offset truncation). The difference between those two possibilities is that the second also features wage ``spillovers'' that extend beyond the minimum wage itself \citep{Fortin2021}. These spillovers can have several theoretical origins: worker reallocation from low-wage to high-wage firms \citep{Dustmann2021}, increased work effort in response to the threat of disemployment \citep{Leamer1999}, wage bumps for occupations above the minimum wage arising from fairness concerns within the firm \citep{Giuponni2026}, or changes in the marginal product of labor within firms in response to reallocation \citep{Haanwinckel2025}. If added mass effects are large enough, the net employment effect is positive, consistent with minimum wages increasing employment due to a combination of monopsony power and an upward-sloping aggregate labor supply \citep{Robinson1933}. Net positive employment effects may also arise from increased search effort by workers \citep{Adams2022,Piqueras2024}.

	\paragraph{Calibration.} All simulation exercises use samples with 200 regions. Unless otherwise noted, the national log minimum wage increases by 20 log points. Each DGP specifies a joint distribution for the vector $[\mu_{r,0},\sigma_{r,0},\mu_{r,1},\sigma_{r,1}]$ (i.e., how latent log wages differ across regions and correlate over time); the markdown parameter $m$; and the positive employment parameters $P_{base},P_{height}$. The baseline calibration used in simulations from Sections~\ref{sec:eff-min-w} and~\ref{sec:fagap} is based on data from the US Current Population Survey for the years 1989 (corresponding to $t=0$) and 2004 (corresponding to $t=1$), aggregated to the state level. It uses $m=0.7$, corresponding to a significant degree of market power, while abstracting from added employment mass effects. These effects feature prominently in the main simulation exercises in Section~\ref{sec:500-dgps}. See Appendix~\ref{appendix:simulation-details} for details.
	
	\subsection{Exposure-based estimators}
	
	Throughout the paper, we will analyze panel regressions of the form
		\begin{equation}
		y_{r,t} = \alpha_{r} + \delta_{t} + h_t\left(E_{r,t}\right) + \epsilon_{r,t}, \label{eq:general-exposure-formulation}
	\end{equation}
	where $\alpha_r$ and $\delta_t$ are region and time fixed effects, $E_{r,t}$ is an \textit{observable} metric of regional exposure to the national minimum wage, $h_t$ is a simple function capturing minimum wage effects, and $\epsilon_{r,t}$ is a residual term assumed to be uncorrelated with the other three additive components. Specific estimators differ in how they specify $E_{r,t}$ and $h_t$.
	
	A potential motivation for employing those regressions, instead of structural models of worker and firm behavior, is that the applied researcher may prefer models that are agnostic regarding behavioral assumptions, transparent, and easily implementable. Indeed, those regressions are sometimes referred to as ``reduced-form'' econometric approaches. However, in many ways, they are closer to structural economic models than to reduced-form methods. Most importantly, they are not based on a comparison of treated and control regions, given a readily observed treatment status. Rather, these regressions rely on strong functional-form assumptions to define who is treated, and by how much. If these assumptions are misspecified, there is no guarantee that estimated treatment effects will be consistent.
	
	The problems identified in this paper correspond to different ways in which the empirical model~\eqref{eq:general-exposure-formulation} can be misspecified. First, even if the function $h_t$ is approximately correct, the exposure measure $E_{r,t}$ may not be observable. The researcher may then have to rely on an observable proxy, introducing the possibility of biases from measurement error. Second, exposure may be jointly determined by several latent factors, instead of being well-approximated by a scalar exposure $E_{r,t}$ plus random noise. That can generate problems analogous to omitted variable biases. Third, the functions $h_t$ may not provide a good approximation for the underlying economic model. For example, some estimators may impose linearity despite the fact that minimum wage effects are non-monotonic in theoretical models of the minimum wage. 
	
	We now describe those issues in detail in the context of each specific estimator.
	
	\section{The effective minimum wage design \label{sec:eff-min-w}}
	
	\subsection{Definition \label{subsec:eff-min-w-definition}}
	
	Let $w_{q,r,t}$ denote quantile $q$ of the log wage distribution in region $r$ at time $t$. The ``effective minimum wage'' exposure measure is defined as $E_{r,t}=mw_t - w_{0.5,r,t}$, that is, the difference between the log minimum wage and the local median log wage. The function $h_t(E_{r,t})$ mapping exposure to expected outcomes is a quadratic polynomial, assumed to be constant over time. The resulting regression is
	\begin{equation}
		y_{r,t} = \alpha_{r} + \delta_{t} + \beta \left[mw_t - w_{0.5,r,t}\right] 
		\\
		+ \gamma \left[mw_t - w_{0.5,r,t}\right]^2 + \epsilon_{r,t}. \label{eq:effective-min-wage-design}
	\end{equation}
	
	To calculate the predicted treatment effects of a rising national minimum wage in each region, the researcher multiplies the changes in the effective minimum wage (and its square) by the estimated $\hat{\beta}$ and $\hat{\gamma}$ parameters. Those products can then be added up and averaged across regions, yielding an estimate of average treatment effects as defined in Equation~\ref{eq:ate}:
	$$\widehat{ATE} = \frac{1}{R}\sum_r \left\{\hat{\beta} \left[\left(mw_1 - w_{0.5,r,1}\right)-\left(mw_0 - w_{0.5,r,0}\right) \right] + \hat{\gamma} \left[\left(mw_1 - w_{0.5,r,1}\right)^2-\left(mw_0 - w_{0.5,r,0}\right)^2 \right]\right\} $$	
	This regression model was first introduced by \citet{Lee1999}, who focused on quantile gaps as the outcomes of interest. The design was later used to estimate employment effects as well; \citet{Engbom2022} is one example.\footnote{Such regressions follow in the tradition of earlier papers such as \citet{Neumark1992}, which used variation in wage levels to measure employment effects of minimum wages. I focus on the quantile-based effective minimum wage because it is more common in recent work.}
	
	To discuss identifying assumptions, it is helpful to introduce the semiparametric conceptual framework of \citet{Lee1999}. Each region has a latent distribution of log wages in each period---that is, the distribution of log wages that would prevail with no minimum wage regulation. The cumulative distribution function for those latent log wages is:
	$$F_t\left(\frac{w-\mu_{r,t}}{\sigma_{r,t}}\right)$$
	where $\mu_{r,t}$ and $\sigma_{r,t}$ are the \textit{centrality} (or location) and \textit{dispersion} parameters, respectively. This conceptual model nests the Normal-markdown simulation model introduced in Subsection~\ref{subsec:normal-markdown}.
	
	Using this notation, \citet{Lee1999} emphasizes two identification assumptions. First, the deflator used to construct the effective minimum wage---that is, the median wage $w_{0.5,r,t}$ in Equation~\eqref{eq:effective-min-wage-design}---should provide a good approximation for the centrality parameter $\mu_{r,t}$. Second, the location and dispersion parameters should be uncorrelated across regions conditional on $t$.	
	
	Lee's work focused on wage inequality measures as outcomes. If one is interested in employment outcomes instead, one intuitive extrapolation of Lee's discussion would be to impose a third assumption: shocks to $\mu_{r,t}$ are uncorrelated with all other determinants of employment other than the minimum wage. That assumption is necessary because shocks to $\mu_{r,t}$ are the fundamental source of identifying variation in this design, as we discuss below.
	
	Below, I show that these assumptions are not sufficient for consistency of predicted minimum wage effects, due to imperfect measurement of the exposure variable. I also show that the estimator is not robust: minor deviations from the second assumption introduce large biases.

	\begin{figure}
		\centering
		
		\textbf{Panel A:} Good variation arising from a shock to location, $\Delta \mu_{B}$
		
		\includegraphics[width=0.8\linewidth]{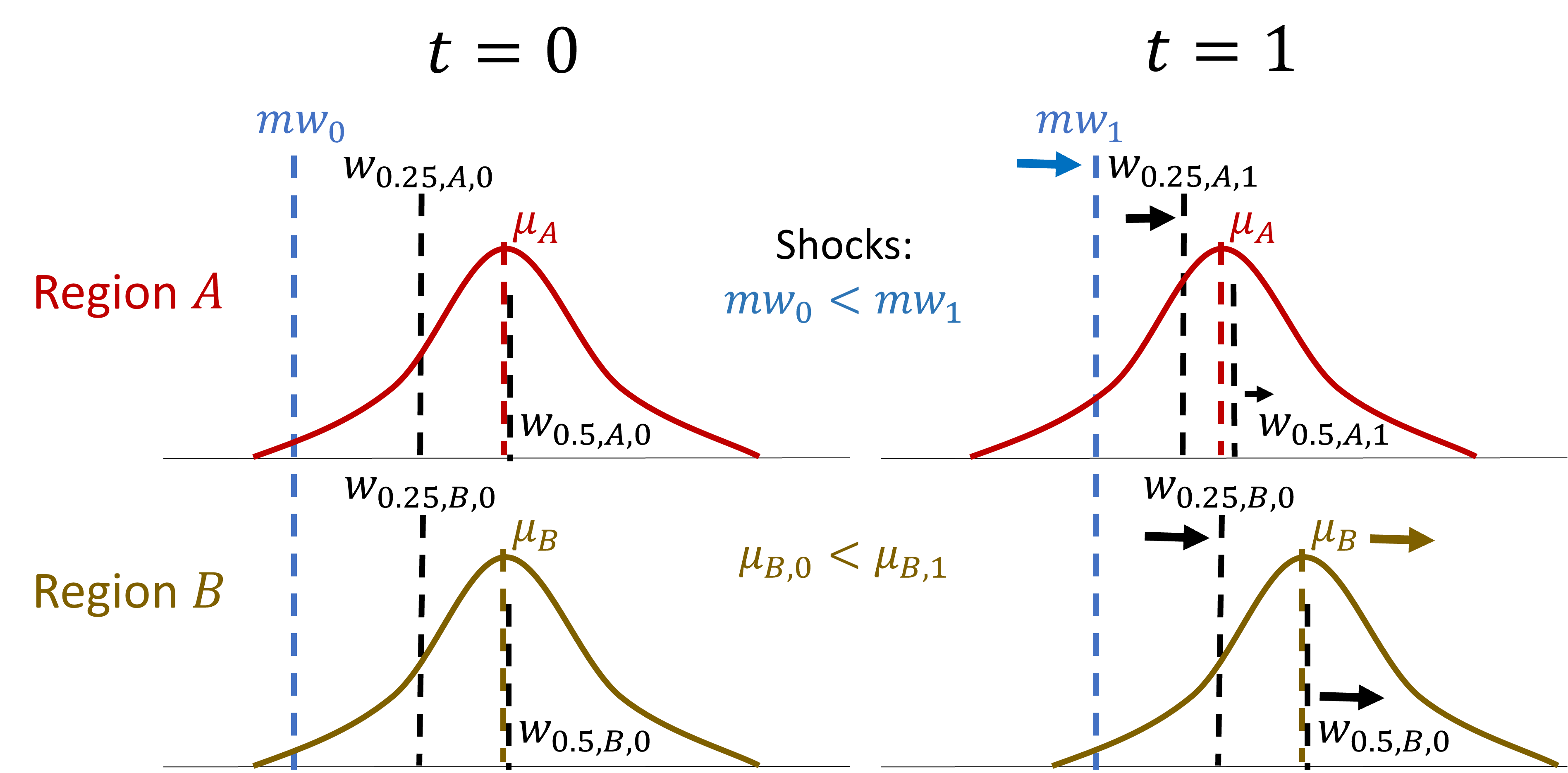}
		
		\bigskip
		
		\textbf{Panel B:} Bad variation arising from differences in dispersion, $\sigma_A>\sigma_B$
		
		\includegraphics[width=0.8\linewidth]{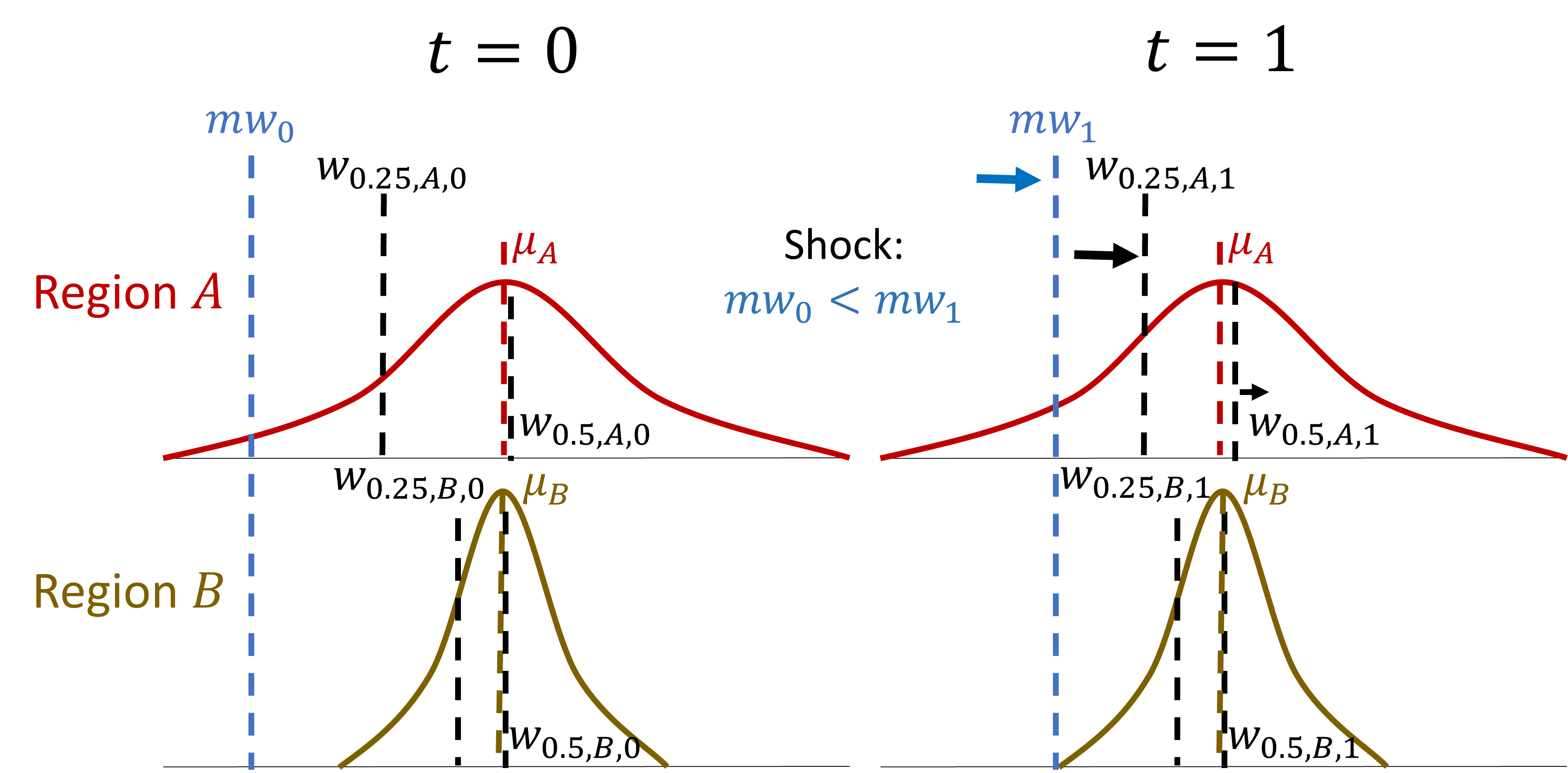}
		
		\bigskip
		
		\textbf{Panel C:} Bad variation arising from a shock to dispersion, $\Delta\sigma_{A}$
		
		\includegraphics[width=0.8\linewidth]{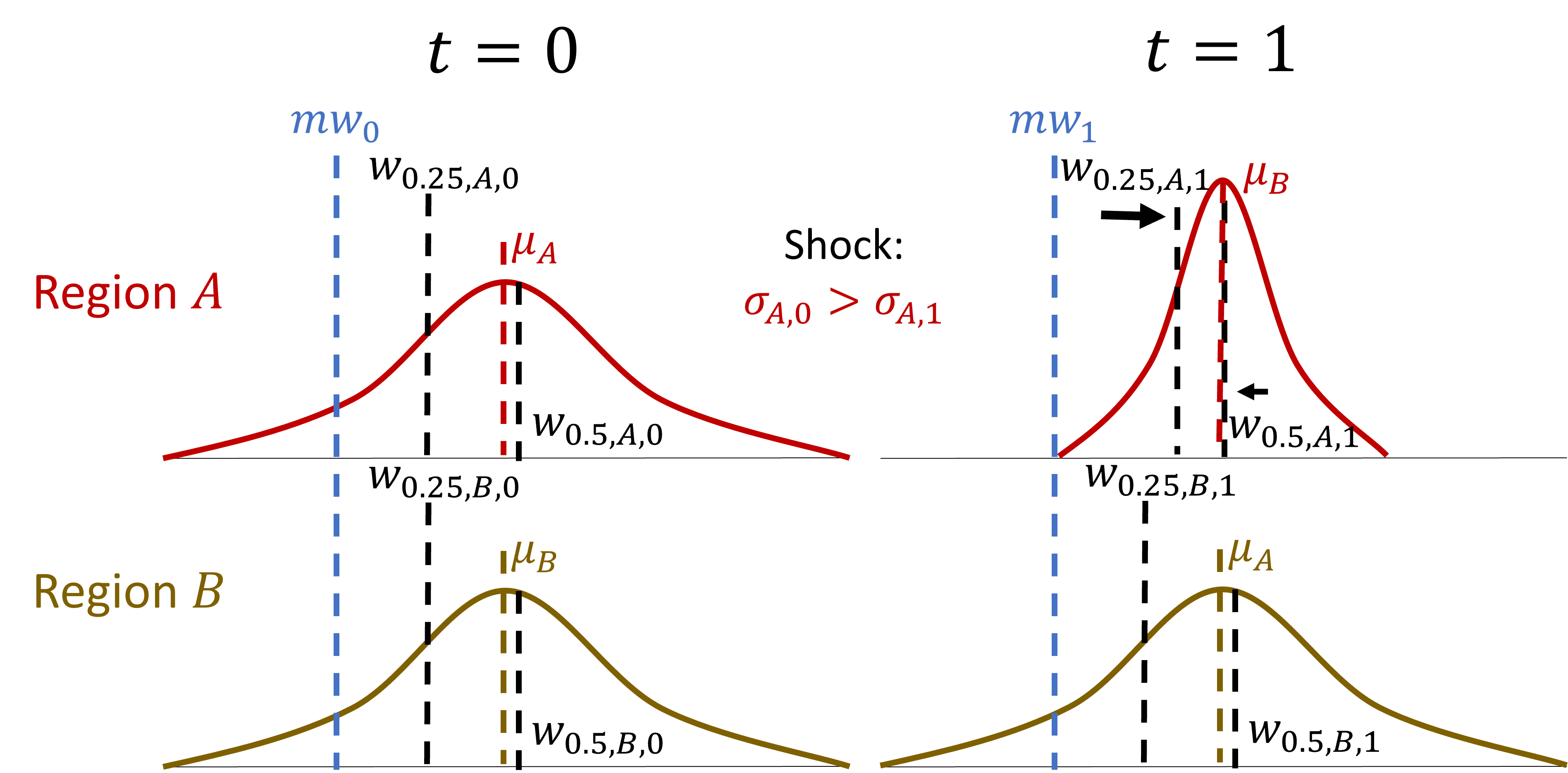}
		\caption{Good and Bad Variation in the Effective Minimum Wage Design}
		\label{fig:good-bad-var}
	\end{figure}

	\subsection{Correlated measurement errors in the exposure variable \label{subsec:good-and-bad}}
	
	The effective minimum wage design is predicated on minimum wage effects being stronger where it bites more into the \textit{latent} wage distribution. In that sense, $E^*_{r,t}=mw_t-\mu_{r,t}$ is the correct but latent exposure measure. The researcher uses $E_{r,t}=mw_t-w_{0.5,r,t}$ instead, assuming that $w_{0.5,r,t}$ is a good proxy for $\mu_{r,t}$. In this subsection, I show that the ``measurement errors'' $w_{0.5,r,t}-\mu_{r,t}$ can introduce large biases even when $w_{0.5,r,t}$ is an excellent proxy for $\mu_{r,t}$ (e.g., when the correlation is above 0.99). The reason is that these measurement errors are correlated with the effects of the minimum wage itself, such that the measurement error is not classical.
	
	\paragraph{Good and bad variation.}	
	The source of identifying variation in the effective minimum wage design is not the national minimum wage increase, but random shocks to the centrality parameters $\mu_{r,t}$. To see why, consider a scenario with two regions, as illustrated in Panel~A of Figure~\ref{fig:good-bad-var}. In period $t=0$, both regions have identical distributions of latent log wages and the national minimum wage is small. In period $t=1$, two things happen. First, the national minimum wage rises, as denoted by the blue arrow. Second, the location parameter in Region $B$ increases (the beige arrow). For simplicity, assume that both shocks have the same magnitude.
	
	In Region $A$, the rising national minimum wage bites more into the latent wage distribution, leading to compression of the realized log wage distribution (note the black arrows indicating movements of log wage quantiles). In $B$, however, the minimum wage binds as much in period $t=1$ as it did in $t=0$ due to the positive shock to $\mu_{r=B,t}$. That shock makes $B$ a perfect ``control group'' for $A$.

	Now, I introduce the idea that heterogeneity in dispersion parameters $\sigma_{r,t}$ generates ``bad variation:'' a spurious link between the observed exposure measure and the outcomes of interest. Again, consider a scenario where Region $A$ is affected by the increase in the national minimum wage, while Region $B$ is not. But now assume that neither location nor dispersion parameters change over time. Region $B$ is unaffected because its time-invariant dispersion $\sigma_{B}$ is minimal, as illustrated in Panel~B of Figure~\ref{fig:good-bad-var}. Latent wages therefore concentrate around the median, so the minimum wage has no bite in that region.
	
	To see what the regression estimates in this scenario, first consider what happens to the exposure measures. The increase in the national minimum wage $mw_t$ mechanically increases $E_{r,t}=mw_t-w_{0.5,r,t}$ in both regions. If exposure rose by the same amount everywhere, region and time fixed effects would absorb that common change. The regression therefore identifies minimum wage effects from \textit{differential changes} in exposure across regions, which in this example can only come from differential changes in median wages $w_{0.5,r,t}$.
	
	The relevant object is therefore the minimum wage's effect on the median. Such effects may be small, but they are unlikely to be exactly zero. They necessarily arise whenever minimum wages have employment effects in the lower tail, as adding or removing workers changes the identity of the median worker. Causal effects on the median may also arise in the absence of employment effects, e.g., in wage-posting models \citep{Burdett1998} or in models of endogenous worker effort \citep{Leamer1999}. I assume in this example that the minimum wage has a small, but \textit{positive}, effect on Region $A$'s median wage, while $B$ is unaffected.
	
	That means the effective minimum wage \textit{falls} in $A$ relative to $B$. Hence, for the regression design, $B$ is more treated than $A$. The predicted treatment effects arising from the regression would thus have the opposite sign relative to the actual causal effects of the minimum wage. The magnitude of the effect would not be informative; indeed, the smaller the causal effects on the median, the larger the measured minimum wage effects, because the estimator infers that small changes in exposure cause large changes in outcomes.
	
	Shocks to dispersion parameters $\sigma_{r,t}$ also introduce bad variation. Panel~C in Figure~\ref{fig:good-bad-var} illustrates a scenario where the only change over time is a fall in the dispersion parameter in Region~$A$. That shock reduces inequality, moving percentile 25 of the log wage distribution closer to the median. Moreover, because the minimum wage becomes less binding, the median log wage falls, so the effective minimum wage rises. Thus, comparing changes between Region~$A$ and Region~$B$, the estimator will estimate a positive relationship between the effective minimum wage and the log wage gap $w_{0.25,r,t}-w_{0.5,r,t}$. However, this link is likely to be significantly overstated relative to the causal minimum wage effects. As in the previous example, the regression attributes the inequality-reducing effects of the $\sigma_{r,t}$ change to a small change in exposure generated by minimum wage effects at the median.
	
	\begin{table}
		\small
		\caption{Effective minimum wage design: good vs. bad variation}
		\centering
		
		\label{tab:iss1}
		\input{tab_new_eff_iss1.tex}	
		\begin{minipage}{1.0\textwidth}
			\footnotesize
			\justify
			\textbf{Notes:}      
			This table summarizes simulation results with 200 regions and two periods. The top row in each panel reports the average of 1,000 simulations of the true $ATE_i$ for different outcomes $i$, corresponding to different columns. The second row shows estimated average treatment effects for each outcome based on the effective minimum wage regressions, averaged over the same 1,000 simulations. The third row shows the average over simulations of the corresponding standard errors, which are clustered at the region level in each simulation. Each panel corresponds to different assumptions on the data-generating process. In \textbf{Panel~A}, regions differ only in the location parameter $\mu_{r,t}$. \textbf{Panel~B} includes differences in the dispersion parameter $\sigma_{r,t}$. \textbf{Panel~C} is like Panel~B but with an increase of the log minimum wage of 0.4 instead of 0.2. \textbf{Panel~D} increases the between-region standard deviation of the $\sigma_{r,t}$ parameters by 50\%. See Appendix~\ref{appendix:simulation-details} for details.
		\end{minipage}		
	\end{table}
	
	\paragraph{Simulations.} To gauge the practical importance of this measurement problem, I conduct a simulation exercise using the baseline calibration of the Normal-markdown model described in Subsection~\ref{subsec:normal-markdown}. The first panel in Table~\ref{tab:iss1} corresponds to a simplified DGP where all regions have the same dispersion parameters $\sigma_{r,t}=\sigma_t$, but differ in $\mu_{r,t}$. This is the ideal scenario with only ``good'' variation. As expected, the estimator performs very well. The true average causal effects from the model are nearly identical to the predicted effects from the regressions, averaged over 1,000 simulations (shown in the second row of each panel of the table). The confidence intervals are also tight, as implied by the standard errors reported in the third row (also averaged over simulations).
	
	Panel B introduces differences in dispersion parameters. The data-generating process still satisfies the structural assumptions emphasized by \citet{Lee1999}: distributions only differ in location and dispersion parameters, not shape, and the location and dispersion parameters have zero correlation. The median wage is also a nearly-perfect proxy for the latent $\mu_{r,t}$ location parameter: Table~\ref{tab:simparams-nm-eff} in Appendix~\ref{appendix:simulation-details} shows that their correlation is 0.999. Still, the estimator displays nontrivial biases. They arise because, though the differences between $w_{0.5,r,t}$ and $\mu_{r,t}$ constitute a tiny share of the variation, they are systematically correlated with the outcomes of interest. 
	
	Panel~C shows that biases are larger when the simulated increase in the federal minimum wage is 40 log points instead of 20 log points, even though the correlation between $\mu_{r,t}$ and $w_{r,t}$ remains above 0.99 (see Table~\ref{tab:simparams-nm-eff}). This exercise reinforces the idea that the ``good'' identifying variation comes not from the change in the minimum wage itself but from idiosyncratic shocks to $\mu_{r,t}$.
	
	Panel D highlights how differences in the dispersion of latent log wages between regions constitute the source of the bias. Increasing those differences by 50\% is enough to double the average bias in the regressions.
	
	\textbf{Are there biases if average employment effects are zero or positive?} The simulations in Table~\ref{tab:iss1} assume minimum wages must cause negative employment effects. It is natural to ask whether biases can be smaller when the minimum wage can cause positive employment effects. The conceptual discussion around Figure~\ref{fig:good-bad-var} suggests that the biases will still be problematic in these cases. If employment effects are generally positive, they will still generate problematic effects on median wages (though in the opposite direction). And even if employment effects are zero \textit{on average}, it is reasonable to expect that these effects are positive in some regions and negative in others, such that the correlated measurement error issue persists. This problem would only disappear under the very strong assumption that employment effects are exactly zero in every local labor market. 
	
	Section~\ref{sec:500-dgps} returns to this issue by quantifying biases using alternative DGPs, many of which have zero or positive employment effects of the minimum wage.

	\subsection{Sensitivity to small correlations between location and dispersion}
	
	The second assumption emphasized by \citet{Lee1999} is independence between the location and dispersion parameters, $\mu_{r,t}$ and $\sigma_{r,t}$, conditional on $t$. The analysis below evaluates the design's robustness to small, but realistic violations of this assumption.\footnote{For an intuition of why this assumption is essential for measuring spillover effects, consider again the ``good variation'' example from the previous subsection (Panel~A in Figure~\ref{fig:good-bad-var}). In that example, Region $A$ was ``treated'' by the minimum wage because its location parameters $\mu_{A,t}$ are constant over time, while Region $B$ is the ``control'' because $\mu_{B,1}-\mu_{B,0}=mw_1-mw_0$. Now, suppose that, along with the increase in location, the dispersion parameter also changes for Region $B$. That would have independent effects on wage inequality, such that $B$ would not be a valid control for $A$ anymore. Conceptually, this issue can be seen as a form of omitted variable bias introduced by failing to account for regional heterogeneity in dispersion. A correlation between location and dispersion parameters is also problematic if the outcome is employment. The reason is that changes in dispersion parameters can make the minimum wage bind more or less in some regions, causing independent effects on the median wage in the presence of a minimum wage.}

	\begin{table}
		\small
		\caption{Correlation between location and dispersion parameters}
		\centering
		
		\label{tab:iss2}
		\input{tab_new_eff_iss2.tex}	
		\begin{minipage}{1.0\textwidth}
			\footnotesize
			\justify
			\textbf{Notes:}      
			See the notes below Table~\ref{tab:iss1} for an explanation of the table's structure. \textbf{Panel~A} uses the same DGP as Panel~B in Table~\ref{tab:iss1}: regions differ in location ($\mu_{r,t}$) and dispersion ($\sigma_{r,t}$) parameters, but they are orthogonal to each other. \textbf{Panel~B} introduces a correlation of 0.076 between location and dispersion parameters.
		\end{minipage}		
	\end{table}
	
	Panel~A in Table~\ref{tab:iss2} shows a baseline scenario where regions differ in dispersion parameters, but location and dispersion parameters are uncorrelated. Panel~B introduces a within-period correlation of 0.076, corresponding to the correlation between state-level median log wages and the standard deviation of log wages using US-CPS data from 1989. That is enough to bring the estimated employment effects to almost zero and make estimated spillover effects much larger than the true ones. Adding that correlation does not significantly affect estimated standard errors; if anything, the estimates become more precise.
	
	These results suggest that even a mild correlation of 0.076 is sufficient to invalidate the effective minimum wage design. But are such correlations common in practice? Below, I argue that applied researchers' priors should be that these correlations are nontrivial.
	
	Empirically, the value of 0.076 was calculated based on US-CPS data from 1989. If one used data from 2004 instead, the value would be 0.264. Using Brazilian data from 1998, I find an even higher value of 0.339 (see Section~\ref{sec:brazil} for details). These examples suggest that 0.076 may be towards the lower end of reasonable values for that parameter. 
	
	A counterpoint to this argument is that correlations between median log wages and standard deviations of log wages do not reveal the corresponding correlations in the \textit{latent} log wage distributions, which are the correct model inputs. That's because observed log wage distributions also incorporate minimum wage effects. To evaluate this concern, I calculate the correlation in the Brazilian data using only microregions where the share of workers earning at most the minimum wage plus 30 log points is 10\% or less (a criterion that selects 88 out of 151 microregions in the full data set). Rather than decreasing, the correlation increases to 0.528.
	
	Theoretical arguments also suggest positive correlations between the location and dispersion of latent log wages. First, if workers are split into education-age groups, higher-wage groups tend to display more within-group inequality. This fact is discussed in detail by \citet{Lemieux2006}, who argues that much of the increase in inequality observed in the US from 1973 to 2003 is a compositional effect deriving from increased educational achievement. The same result has been found in other contexts, such as Brazil \citep{Ferreira2017}. If regions differ in workforce composition, the correlation we discussed above may follow. A second economic factor that could generate this correlation is regional differences in endowments that affect industrial composition, which in turn would lead to non-random sorting of workers and firms to different regions. For example, larger metropolitan areas may feature both high wages (partly due to high local cost of living) and a wider variety of job types, leading to more dispersion in log wages \citep{Papageorgiou2022}. 
	
	\subsection{Alternative specifications and tests as possible solutions}
	
	Existing literature sometimes uses alternative formulations of the effective minimum wage design that are argued to work better. For example, in contexts where the minimum wage is expected to cause spillovers on the median wage, researchers often use higher quantiles of the log wage distribution as the ``deflator'' when constructing the exposure measure. Some papers include additional controls, including region-specific time trends. Appendix~\ref{appendix:eff-min-w-alternatives} discusses whether these alternative specifications can solve the issues highlighted in the previous subsections. I find that these alternative specifications are more likely to amplify biases than solve the problems.
	
	\citet{Lee1999} suggests measuring minimum wage effects on the \textit{upper} tail of the wage distribution as a placebo specification test. In Appendix~\ref{appendix:eff-min-w-alternatives}, I show that this test cannot effectively screen the issues I document due to high rates of both false positives and false negatives. Section~\ref{sec:500-dgps} provides further evidence on this topic. 
	
	This paper focuses on contexts where the minimum wage is set at the national level. One may wonder whether the issues I highlight would apply even in contexts with regional-level changes in minimum wage policies. They do, but in that case, they can be effectively solved using an instrumental variables design that heavily borrows from \citet{Autor2016}. I describe and evaluate that design in Appendix~\ref{appendix:ams2016}.

	\subsection{Taking stock}
	
	The main takeaway is that the effective minimum wage design relies on the existence of latent shocks $\Delta \mu_{r,t}$ with very particular properties: they must shift the location of latent log wage distributions but have no independent effects on their shape or on employment. It is difficult to imagine what economic shock could satisfy these assumptions. Minor violations of those requirements lead to significant biases. Even if these strong assumptions are satisfied, the estimator still suffers from misspecification biases arising from imperfect measurement of the exposure variable. 
	
	If one must use this regression design, the extended discussion in Appendix~\ref{appendix:eff-min-w-alternatives} suggests that the baseline design of \citet{Lee1999} should be preferred. Specifically, the effective minimum wage should be constructed using the median rather than other quantiles of the wage distribution; time fixed effects should always be used; region-specific trends should be avoided; and specifications that remove region fixed effects may in some cases work better. That said, the evidence collected here and in Section~\ref{sec:500-dgps} suggests that the effective minimum wage is unlikely to work well when the data do not include region-specific minimum wage reforms.
	
	\section{Fraction-Affected and Gap Designs \label{sec:fagap}}
	
	\subsection{Definition}
	
	These designs construct time-invariant regional exposure measures $E_{r}$ using data from before the focal increase in the national minimum wage. The function $h_t(E_r)$ mapping exposure to outcomes is a linear function interacted with a post-increase indicator. The resulting estimator is a difference-in-differences design,
	\begin{equation}
		y_{r,t} = \alpha_{r} + \delta_{t} + \beta E_{r}\cdot \bm{1}\{t=1\} + \epsilon_{r,t}. \label{eq:fraction-affected-design}
	\end{equation}
	The difference between the fraction-affected and gap designs lies in the exposure measure,
	\begin{align*}
		FA_{r}=& \frac{1}{I_r}\sum_{i=1}^{I_r}\mathbf{1}\left\{w_{i,0}<mw_1\right\}
		\\
		Gap_{r}=& \frac{\sum_{i=1}^{I_r}\max\{\exp(mw_{1})-\exp(w_{i,0}),0\}}{\sum_{i=1}^{I_r} \exp(w_{i,0})},
	\end{align*}
	where $i\in\{1,\dots,I_r\}$ denotes individual workers in region $r$ at $t=0$, and $w_{i,t}$ is their log wage at $t$. The fraction-affected measure is the share of workers who, before the minimum wage increase, earn less than the new minimum. The gap measure corresponds to the relative increase in the average wage if, between periods 0 and 1, workers had their wages adjusted to match the new minimum if necessary.
	
	After estimating the regression coefficients, the predicted average treatment effect $ATE$ is obtained by multiplying the average exposure measure by $\hat{\beta}$.
	
	\citet{Card1992} first introduced the fraction-affected design in an analysis of the 1990 increase in the federal minimum wage in the US. Since then, that estimator has been applied in other contexts with no regional variation in nominal minimum wages, such as the introduction of a federal minimum wage in Germany in 2015 \citep{Ahlfeldt2018, Fedorets2021}. The gap measure was introduced by \citet{Card1994} in a firm-level econometric design, but was later applied to region-level designs as well. \citet{Dustmann2021} provide an example, again in the context of Germany. 
	
	Under correct specification---that is, if the effects of the minimum wage are well-captured by the product of the exposure measure and a constant scalar $\beta$, common across regions---identification relies on the well-known parallel trends assumption: absent the national minimum wage increase, changes in regional outcomes are orthogonal to the exposure measure.
	
	In the first part of the discussion below, I show that this design may be subject to significant functional form misspecification biases and discuss potential solutions. In the second part, I evaluate the parallel trends assumption in minimum wage studies, with the aim of guiding applied researchers on likely risks and how to screen them.	
	
	The discussion is closely related to the guidance in \citet{Dube2024}. Their handbook chapter is, to my knowledge, the clearest recent guide to exposure-based designs for national minimum wages. They formulate these designs as difference-in-differences regressions using pre-reform bite measures, discuss fraction-affected and gap variables, and emphasize that identification requires more than parallel trends: researchers must also consider spillovers, mobility across groups, and heterogeneity in treatment effects that is correlated with exposure. They also discuss grouping estimators that absorb heterogeneity along observable dimensions, while noting that those approaches require prior knowledge of the relevant dimensions of heterogeneity. My focus is on the complementary question of whether these regression designs are consistent with how minimum wages are thought to affect employment and wages in labor market models.
	
	\subsection{Functional form misspecification \label{subsec:fa-gap-meas-error}}
	
	As discussed in Section~\ref{sec:definitions}, the empirical design can be misspecified in three ways: the exposure variable may be inconsistent with the underlying theoretical model, it can be observed with error, and the function mapping exposure to outcomes may be misspecified.
	
	The use of a pre-determined measure of exposure eliminates measurement errors introduced by minimum wage effects, which I showed to be problematic in the effective minimum wage design. However, a constant exposure may partly capture temporary mean-reverting components. This issue is more easily conceptualized as a potential deviation from the well-known parallel trends assumption, which I discuss in the following subsection. Here, I focus instead on the choice of exposure variable and functional forms.
	
	Both the fraction-affected and gap measures are only correctly specified under very restrictive assumptions on the data-generating process.\footnote{The fraction-affected design is correctly specified in a structural model where the outcome is the change in employment as a share of initial employment, and the minimum wage causes a constant share of workers below the new minimum to lose their jobs. The gap design is correctly specified in a model where the outcome is the relative change in the average wage, and the minimum wage only causes wage censoring effects.} That said, they may still be good approximations. One of the objectives of my analysis is to assess whether either is a better proxy over a broad variety of DGPs. 
	
	A potentially more important point is the function mapping exposure to outcomes. In all theoretical models mentioned in this paper, the effects of minimum wages on outcomes are nonlinear. This is particularly true for employment outcomes. Marginal changes in the federal minimum wage may have negligible effects at low bindingness levels, positive effects at moderate bindingness levels, and negative effects at higher levels. If a national minimum wage change moves some regions but not others across those ranges, a linear exposure specification may be restrictive even when the parallel trends assumption holds.
	
	To isolate biases arising exclusively from functional form misspecification, I exploit simulation exercises based on the Normal-markdown model where deviations from parallel trends are ruled out. I do so in two ways. First, the only time-varying factor in the simulation model is the national minimum wage. Thus, there are no region-specific trends in structural parameters. Second, regions only differ in $\mu_{r}$, the location parameter of their latent log wage distributions. Thus, the only differences across regions are that some are permanently at lower wage levels than others.
	
	Table~\ref{tab:fa-gap-meas-error} reports results for four different DGPs. They differ in two dimensions. First, the starting minimum wage may be lower or higher, in which case the minimum wage increase has stronger effects (as the minimum bites more into the latent distribution). Second, in some DGPs, I allow for added employment mass effects (the green areas of Figure~\ref{fig:simulation-model}).
	
	\begin{table}
		\small
		\caption{Misspecification Biases in the Fraction-Affected and Gap Designs}
		\centering
		
		\label{tab:fa-gap-meas-error}
		\input{tab_new_corr_meas_error_fa_gap.tex}	
		\begin{minipage}{1.0\textwidth}
			\footnotesize
			\justify
			\textbf{Notes:}      
			See the notes below Table~\ref{tab:iss1} for an explanation of the structure of the table, and Appendix~\ref{appendix:simulation-details} for a full description of the simulation model. The national minimum wage increases by 20 log points in all panels. In Panels B and D, the initial minimum wage is higher, such that the minimum wage bites more into the latent wage distribution. In Panels C and D, added employment mass effects are allowed, as illustrated in the green region of Figure~\ref{fig:simulation-model}.
		\end{minipage}		
	\end{table}
	
	I find that functional-form misspecification biases may be statistically and economically significant. For example, the fraction-affected and gap estimators are substantially downward-biased for employment outcomes in Panel~D. In Panel~B, the fraction-affected estimator is still downward biased, while the gap estimator has a smaller---but positive---bias. Wage effects in the lower tail are also substantially biased in some specifications, and biases may appear in both directions.
	
	As \citet{Dube2024} note in their review, existing work has begun to explore nonlinearities in minimum wage effects, including differences between large and small minimum wage increases \citep[e.g.,][]{Clemens2021}. However, applications that study nonlinear regional exposure-response relationships in national minimum wage settings remain limited. Appendix~\ref{appendix:fa-other-designs} discusses several related specifications, including binary treatment definitions and the bin-level design of \citet{Giupponi2024}. The latter is especially relevant because it combines regional variation in minimum wage bindingness with flexible estimates across the wage distribution. At the same time, its flexibility is primarily across wage bins rather than across regional levels of bite. I therefore view those designs as complementary: they are useful for important purposes, but they do not directly resolve the functional-form misspecification problem studied here.
	
	One potential way to solve misspecification from nonlinearity is to estimate higher-order polynomial versions of these designs. Appendix Tables~\ref{tab:fa_gap_quad} and~\ref{tab:fa_gap_cubic} report results for quadratic and cubic models, respectively. These specifications do not always reduce biases. But for several combinations of DGP and outcomes, the higher-order polynomial versions do seem to work better.	
	
	The simulation exercises above drive home the point that functional form misspecification biases can be a serious concern, even in simulations designed to be ideal applications of these difference-in-differences designs. No particular specification performed strictly better than the others. That said, we only considered four DGPs. We will reassess our conclusions after analyzing the extended simulation exercises in Section~\ref{sec:500-dgps}.

	\subsection{Parallel trends in the national minimum wage context \label{sec:trends-dispersion}}
	
	Below, I discuss three potential deviations from the parallel trends assumption that may be particularly likely to occur in applications of the fraction-affected/gap designs at the regional level. 
	
	\paragraph{Secular trends in latent wage dispersion.} Suppose that, around the time of a national minimum wage hike, the dispersion of latent log wages is changing in the same way across all regions. That could happen, for example, because of technical change increasing the returns to unobserved ability \citep{Card1996}. Even if the latent structural shock is identical across regions, it may generate differential trends between low-wage and high-wage regions because of the minimum wage's asymmetric effects on the wage distribution.
	
	\begin{figure}
		\centering
		\textit{Panel A: Fraction-affected is larger at the top region}
		
		\includegraphics[width=0.5\linewidth]{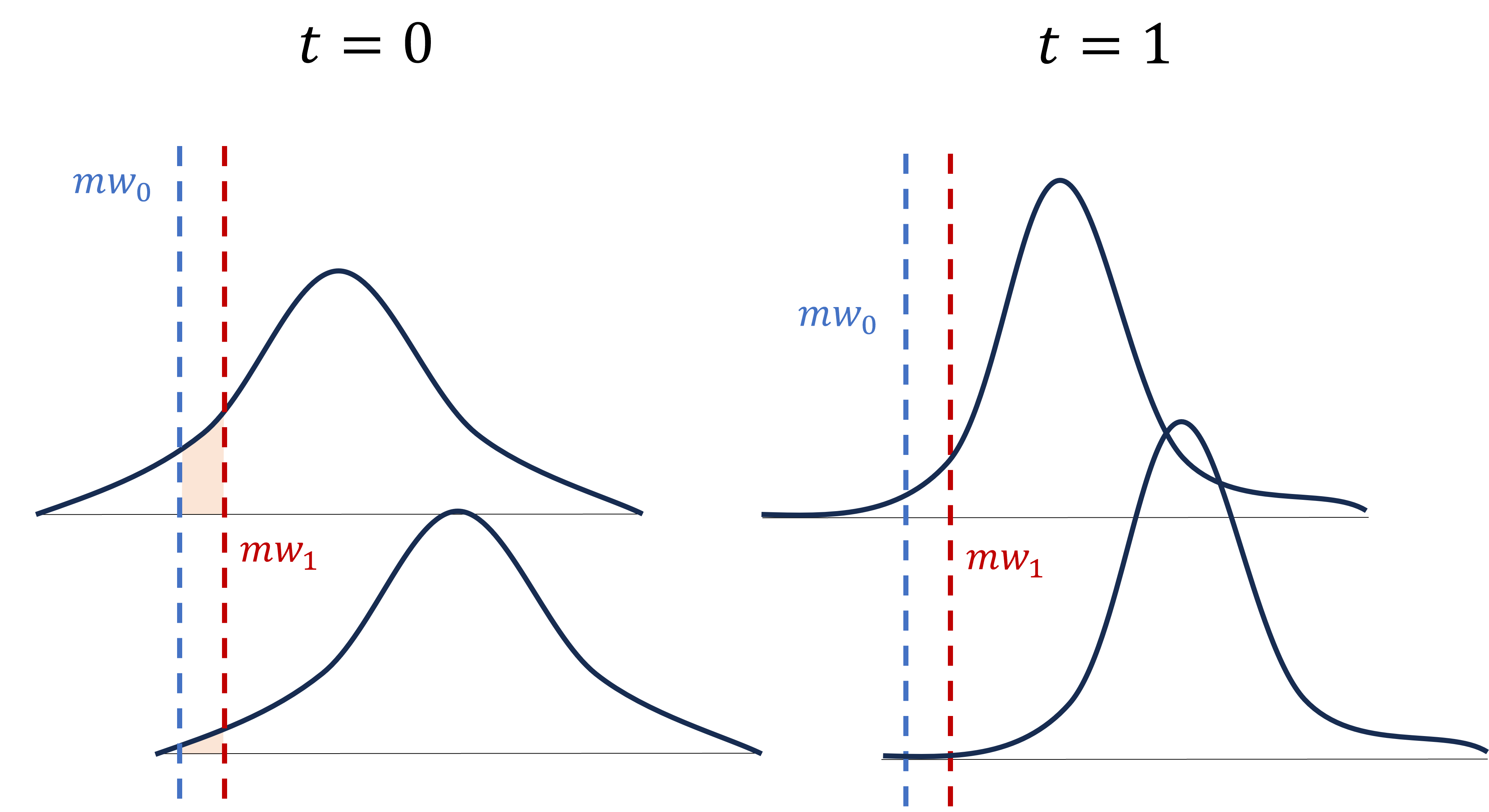}
		\vspace{0.5cm}
		
		\textit{Panel B: Observed employment rises in the top region}
		
		\includegraphics[width=0.5\linewidth]{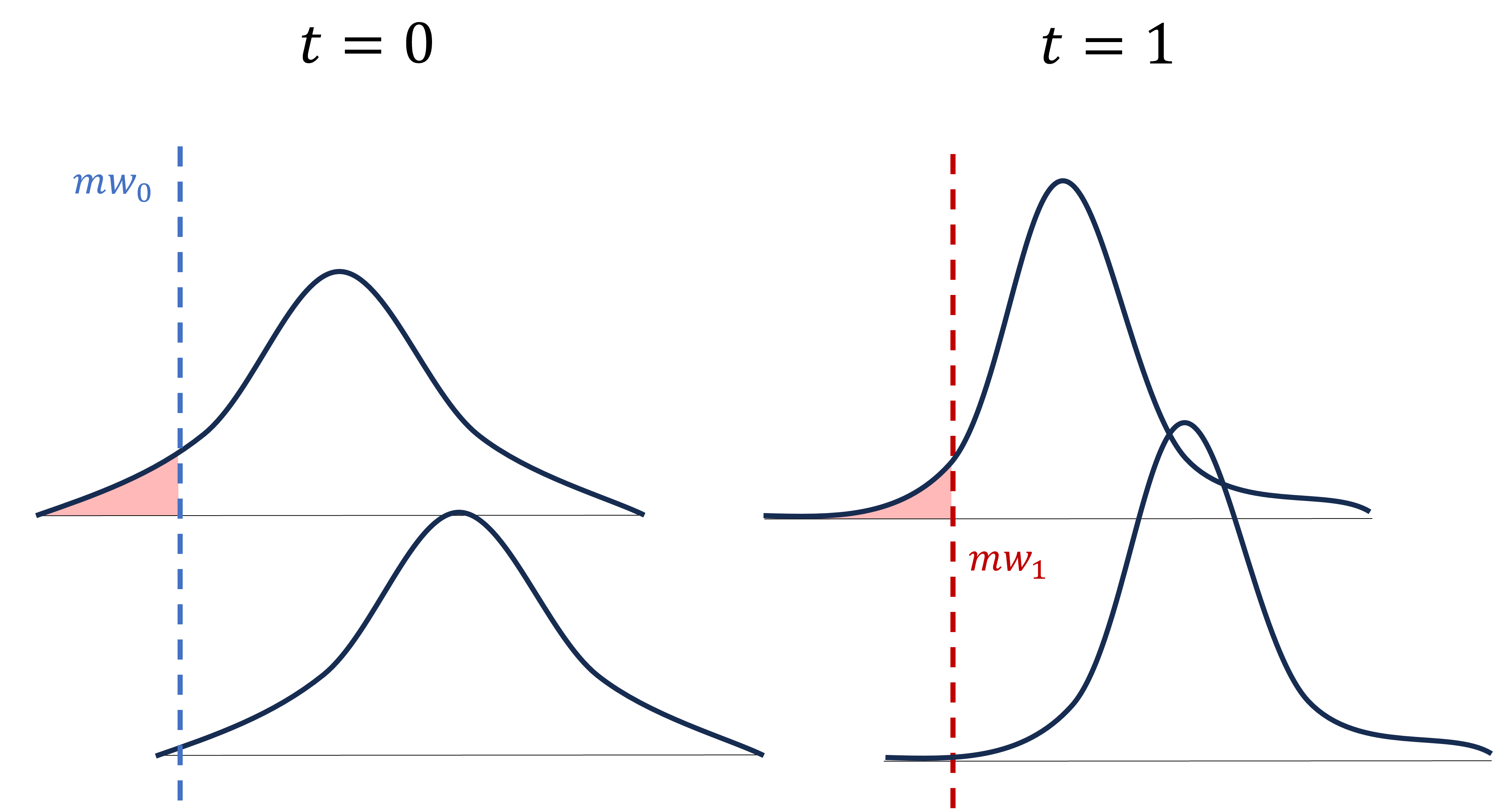}
		\caption{A truncation model with a secular decline in latent log wage dispersion}
		\label{fig:dispersion-and-fa}
	\end{figure}

	Figure~\ref{fig:dispersion-and-fa} illustrates this problem. There are two regions which differ in latent wage levels, but share the same shape. For simplicity, assume the minimum wage only causes truncation (disemployment) effects. Holding latent wages constant, the illustrated increase in the national minimum wage causes larger disemployment effects in the top (low-wage) region. Now, suppose the variance of latent wages decreases over time. That causes an \textit{increase} in employment by reducing the truncation effects. This effect is stronger where the minimum wage bites more. The overall change from $t=0$ to $t=1$ combines both effects, generating a bias that may be large enough to change the sign of the estimated minimum wage effect on employment. In Appendix~\ref{appendix:regional-convergence}, I show that the resulting biases can be quantitatively relevant using simulations calibrated to US data.
	
	\paragraph{Regression to the mean.} The fraction-affected and gap measures are constructed based on extreme wage observations: individual workers only contribute to those measures if their wages are below some threshold. Thus, these estimators may suffer regression-to-the-mean bias. This problem is well-known in minimum wage studies at the individual worker or firm levels. For example, in their worker-level analysis, \citet{Dustmann2021} use pre-treatment data to infer the extent of, and control for, regression to the mean. However, it is not always investigated in regional-level studies. In those contexts, regression to the mean can arise not only from sampling variation and individual measurement errors (which may be significant if some regions are small), but also from mean-reverting regional economic shocks. Appendix~\ref{appendix:regional-convergence} presents a detailed discussion of this issue, along with an illustration of the resulting biases using the Normal-markdown simulation model.
	
	\paragraph{Regional convergence.} Region-specific structural trends affecting wages and employment, such as productivity growth, schooling reforms, or sector-specific demand shocks, may affect low-wage regions more strongly. \citet{Caliendo2017} document the quantitative importance of regional productivity shocks in a large country (the US). \citet{Gennaioli2014} collect time-series data on regional GDP for 83 countries and document within-country regional convergence. This issue is more likely to be consequential for longer-run specifications.
	
	If this is a plausible concern in a given application, the researcher may consider adjusting the empirical model to include predetermined regional variables interacted with flexible time trends. The variables should be thoughtfully selected to reflect the structural shocks that may correlate with exposure: for example, educational composition to capture schooling-related trends, sectoral composition to capture demand-side shocks, or informality rates to capture changes in the enforcement of labor regulations. However, researchers should avoid a ``kitchen-sink'' approach of including an excessive number of such interactions. Adding too many controls makes the source of identifying variation less transparent and may increase vulnerability to misspecification biases or overfit.
	
	\paragraph{Testing for parallel pre-trends.} If the possible problems listed above are relatively smooth over time and affect regions similarly in the pre- and post-treatment periods, then they can be detected with tests for parallel pre-trends. I make this argument explicit in Appendix~\ref{appendix:regional-convergence}, particularly in the placebo regressions reported in Appendix Table~\ref{tab:fa_td_placebo}. That analysis supports a recommendation to implement those tests in every application of the fraction-affected/gap designs.
	
	When testing for parallel pre-trends, researchers should construct the exposure measure using a single pre-treatment year (typically the year before the reform). If the exposure variable is instead averaged over all pre-treatment periods, then the test will likely not detect regression-to-the-mean issues, even when those issues create a bias in the estimated treatment effects. The end of Appendix~\ref{appendix:regional-convergence} explains this issue.
	
	Sometimes, pre-trends are not parallel in the basic regression design, but can be made parallel after adjusting the empirical model to include region-specific time trends. This approach is \textit{not} recommended for two reasons. One is that, if biases are caused by regression to the mean or idiosyncratic, mean-reverting regional productivity shocks, the ``event-study'' pattern implied by them is a V shape centered on the reference period, instead of a smooth trend over time. Thus, controlling for region-specific trends may increase, rather than decrease, the bias in the estimated treatment effects. The second reason comes from \citet{Meer2016}: if minimum wages affect changes more than levels of the outcome variables, controlling for unit-specific trends causes attenuation of treatment effects.
	
	Unfortunately, testing for differential pre-trends is not possible in all potential applications of the fraction-affected/gap designs. Increases in the national minimum wage are sometimes spread smoothly over a long period, without clear pre- and post-reform periods. In those cases---or if tests show non-parallel pre-trends---researchers are discouraged from using the fraction-affected/gap designs. They may instead consider more structural methods that address the possibility of reversion to the mean, regional convergence, and secular trends in wage dispersion. 
	
	\section{Quantifying biases under alternative DGPs \label{sec:500-dgps}}
	
	I quantify biases for 500 different DGPs in the Normal-markdown class introduced in Subsection~\ref{subsec:normal-markdown}. Each DGP specifies assumptions on the distribution of region-specific parameters $[\mu_{r,0}, \sigma_{r,0},\mu_{r,1},\sigma_{r,1}]$, the markdown parameter $m$, the positive employment effect parameters $P_{base},P_{height}$, and the levels of the national minimum wage $mw_0,mw_1$ (measured relative to the center of the latent log wage distribution). They range from models with significant disemployment effects, to positive employment effects, to no employment effects but significant wage spillovers. The parameters are randomly drawn, with most distributions centered around the corresponding values observed for US states (used for previous simulations in the paper).
	
	I make conservative choices for DGP elements that have been shown to be problematic. For example, the contemporaneous correlation between location and dispersion parameters is allowed to be positive, but its value is never larger than 0.20. That number is smaller than the correlation between regional mean log wages and regional standard deviation of log wages observed for the US in 2004 (0.26) or Brazil in 1998 (0.34). See Appendix~\ref{appendix:500-gdps} for details.
	
	For each DGP, I simulate 1,000 samples, each with 200 regions and two periods. For each sample, I estimate the causal effects of the national minimum wage using the effective minimum wage and fraction-affected/gap designs. Then, I average those estimates over the 1,000 samples to obtain the expected value of those estimators under that DGP. I classify an estimator as biased for an outcome under a DGP if the expected estimated effect is significantly different from the expected causal effect of the minimum wage on that outcome under that DGP.
	
	What does ``significantly different'' mean in this context? I use an absolute-bias criterion. For employment effects, the estimator is considered biased if the absolute gap between the expected estimate and the expected causal effect exceeds 0.005 (half a percentage point) \textit{and} exceeds 25\% of the absolute true causal effect. For quantile gaps, the corresponding absolute-gap threshold is 0.05 (five log points), again combined with the requirement that the absolute bias exceed 25\% of the absolute true causal effect. For example, an estimator is considered biased for the p10-p50 quantile gap if the expected estimate is 0.03 while the expected causal effect is 0.09. It is not classified as biased if the expected estimate is 0.48 and the expected causal effect is 0.40.

	\begin{table}
		\small
		\caption{Quantifying biases under 500 alternative DGPs}
		\centering
		
		\label{tab:500-dgps-main}
		\input{Tab_rob_dgps.tex}	
		\begin{minipage}{1.0\textwidth}
			\footnotesize
			\justify
			\textbf{Notes:}      
			See the text for an explanation of the criterion used to classify an estimator as biased under a specific DGP, and Appendix~\ref{appendix:500-gdps} for a description of the DGPs used in this exercise.
		\end{minipage}		
	\end{table}

	Table~\ref{tab:500-dgps-main} reports the results. I split DGPs into three groups: positive employment effects (0.005 or more), negative employment effects (-0.005 or less), and small employment effects but significant wage effects.\footnote{To be included in the analysis, the DGP must be such that the minimum wage causes effects of magnitude at least half a percentage point on the employment-to-population ratio or 5 log points on at least one quantile gap. This choice rules out ``trivial'' cases with non-binding minimum wages. I also exclude DGPs where the absolute causal effect of the minimum wage on the employment-to-population ratio is four percentage points or more. This criterion ensures that the biases are not arising from atypical scenarios such as a large minimum wage hike in a country where latent log wages have low dispersion.} All numbers with two decimal places in the table correspond to the share of DGPs in a given group for which one of the estimators (rows) is significantly biased in either direction for a given outcome (columns). 
	
	The broad message is that, in all three categories of DGPs, each estimator has at least one outcome for which significant biases are common, and many other estimator-outcome combinations also display sizable bias rates. The average absolute values of the biases, reported below the share with significant biases, are both economically large and significantly bigger than the average standard errors in the simulated regressions. Appendix Table~\ref{tab:500-dgps-big-bias} repeats this exercise using a more stringent requirement for classifying an estimator as biased. It finds that, for every estimator and DGP group, at least one outcome still has a high incidence of large bias. 
	
	Taken together, these results suggest that the reliability of the effective minimum wage and fraction-affected/gap designs should not be taken for granted if the data do not include regional variation in minimum wage laws.
	
	\paragraph{Upper-tail effects as a diagnostic tool?} Researchers may believe that minimum wages should not affect upper-tail wage inequality. This idea suggests a two-step heuristic where the researcher first measures the upper-tail spillovers using the chosen econometric method, and only proceeds to measure other outcomes if the placebo test does not detect large upper-tail effects. Appendix~\ref{subsec:upper-tail-diagnostics} discusses this strategy in more detail in the context of the effective minimum wage design, where this procedure was first suggested by \citet{Lee1999}.
	
	Appendix Tables~\ref{tab:500-dgps-nut-kaitz} and~\ref{tab:500-dgps-nut-fa} replicate the exercise in Table~\ref{tab:500-dgps-main}, but restricting attention to scenarios where measured upper-tail effects are small. This restriction is very selective for the effective minimum wage diagnostic, leaving only 81 of the 500 DGPs, though it is less selective for the fraction-affected diagnostic, leaving 350 DGPs.\footnote{This result is consistent with the fact that papers using the effective minimum wage design commonly detect upper-tail spillovers. Using US data, both \citet{Lee1999} and \citet{Autor2016} find evidence of such spillovers in some specifications and argue that they should be used for model validation. In the Brazilian context, \citet{Hinojosa2019} and \citet{Engbom2022} also detect upper-tail spillovers but interpret them as causal effects, rather than evidence of misspecification.} 	
	Even after this restriction, biases are still common. These simulation exercises support the view that upper-tail diagnostics do not significantly improve the reliability of these regression designs, as discussed in Appendix~\ref{subsec:upper-tail-diagnostics}.
	
	\paragraph{The canonical model of labor demand.} All simulations so far used the Normal-markdown class of simulation models. That class is very flexible and can approximate a broad range of economic models. However, it imposes symmetry of the latent log wage distribution and implicitly assumes that the marginal returns to labor are invariant to the minimum wage. To address those limitations, I perform an additional simulation exercise based on an equilibrium labor market model very similar to that of \citet{Katz1992}, with parameters calibrated to reflect US states as of 1989.
	
	Appendix~\ref{appendix:simulation-details-canonical} describes that exercise and presents the results. They mirror previous results for the fraction-affected estimator regarding functional form issues, except that they now extend to the effective minimum wage design as well. Specifically, model misspecification introduces economically significant biases even in cases that, conceptually, should be ideal applications of those estimators. These results suggest that simulation results based on the Normal-markdown model may be conservative in the sense that misspecification biases can be larger in cases where minimum-wage-induced employment effects or reallocation introduce changes in marginal returns to labor.
	
	\paragraph{Assessing functional form misspecification in the fraction-affected/gap designs.} In Section~\ref{subsec:fa-gap-meas-error}, we were interested in understanding which fraction-affected/gap specifications, if any, featured the smallest functional form misspecification biases in contexts where the parallel trends assumption holds. To evaluate this question, I conduct a second simulation exercise using 500 randomly drawn Normal-markdown DGPs constructed such that the parallel trends assumption holds. I then evaluate linear, quadratic, and cubic polynomial versions of the fraction-affected and gap designs.
	
	Results are presented in Appendix Table~\ref{tab:500-dgps-func-forms}. They show that, if one restricts attention to linear versions of these estimators, the gap measure displays lower absolute biases over this broader range of DGPs. But the table also shows that higher-order polynomials can work even better. Based on those simulation results, researchers are encouraged to use a cubic fraction-affected estimator when measuring employment effects, and to favor the gap measure over the fraction-affected measure when measuring wage effects; among the nonlinear specifications, the quadratic gap estimator performs especially well. The different polynomial order choices align with the theoretical intuition that employment effects are more non-linear than wage effects. As discussed before, many models predict marginal employment effects that are initially negligible, then positive, and then turn negative as the minimum wage becomes more binding. Wage compression effects, on the other hand, are more likely to be positive, but with varying intensities.\footnote{If researchers prefer to use a single exposure measure for the sake of simplicity, the gap measure should be preferred. If equal specifications across outcomes are essential, then the researcher should use a quadratic gap estimator. But note that the cubic fraction-affected estimator performs significantly better for employment, with only half as much mean absolute bias. Additionally, the larger standard errors from the cubic fraction-affected specification line up well with the magnitude of the expected biases, meaning that researchers using that specification for employment outcomes would be less likely to report confidence intervals that exclude the true expected causal effects.}
	
	Even if those recommendations are followed, misspecification biases are not completely eliminated. Applied researchers may thus wonder whether results from multiple specifications should be reported. To assess this possible recommendation, Appendix Table~\ref{tab:500-dgps-brackets} reports how often the true causal effects lie in between expected estimates from a pair of estimators. There is no pair of estimators that consistently brackets true effects across several outcomes. In other words, misspecification biases are probably correlated across estimators. Thus, a heuristic of reporting results from two or more specifications may lead readers further away from the true effects. Instead, researchers are recommended to thoughtfully choose their primary specification.
	
	\section{The rise of the Brazilian federal minimum wage \label{sec:brazil}}
	
\begin{figure}
		\centering
		
		\includegraphics[width=0.8\linewidth]{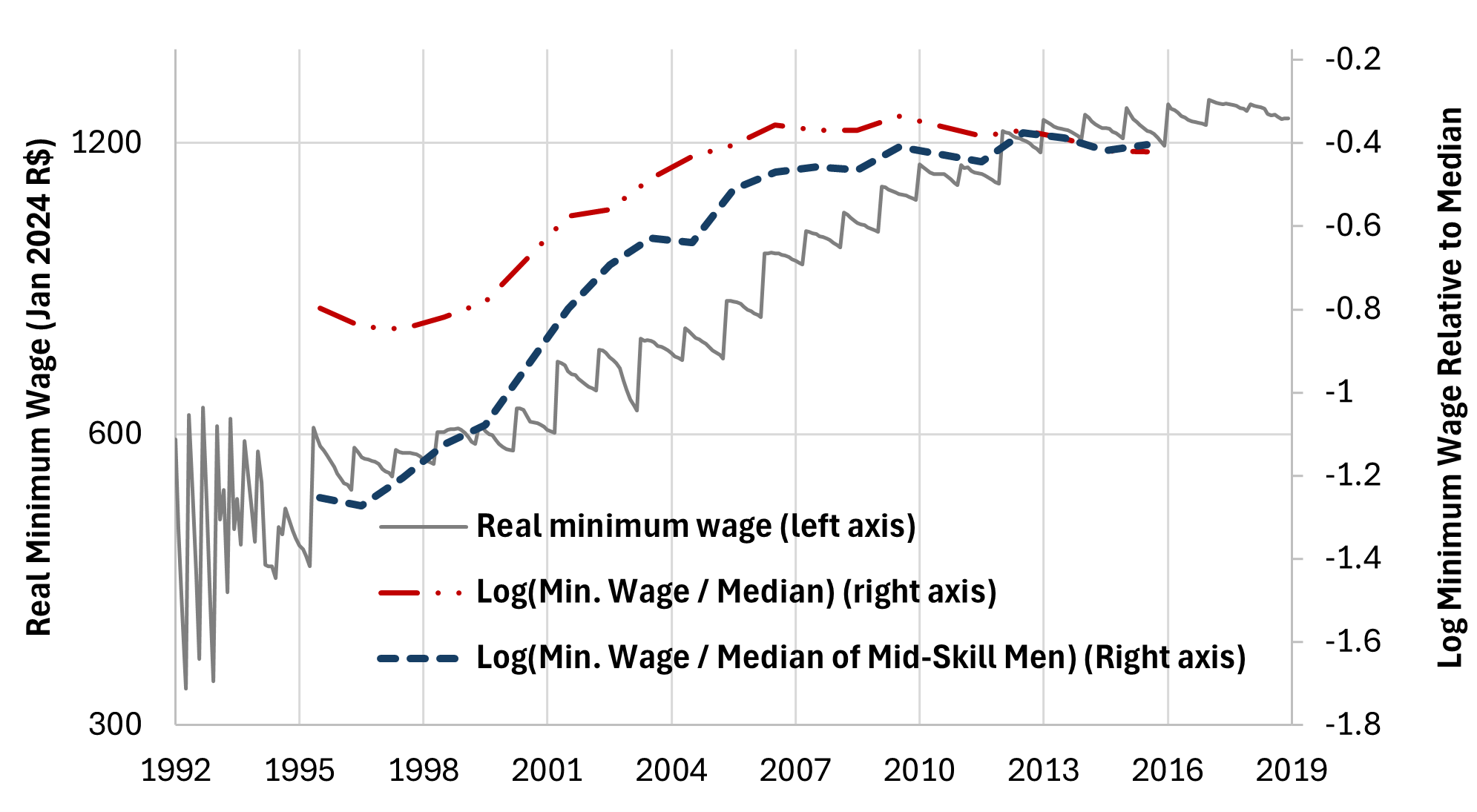}
		\caption{Evolution of the federal minimum wage in Brazil (logarithmic scale)}
		\label{fig:Brazil-mw}
		
		\bigskip
		\begin{minipage}{1.0\textwidth}
			\footnotesize
			\justify
			\textbf{Notes:}      
			The monthly real minimum wage series is constructed based on IPEA Data time series for nominal minimum wage and IPCA price index. The jagged pattern in earlier years corresponds to frequent nominal minimum wage adjustments during a hyperinflationary period. The other two series are constructed based on yearly PNAD survey data from IBGE (processed using the DataZoom tool from PUC-Rio), using the September minimum wage value for each year. They show the minimum wage---legally defined as a minimum monthly earnings level---relative to median monthly earnings in workers' main jobs. The sample includes salaried and self-employed workers between ages 18 and 54. ``Mid-Skill Men'' are those up to 30 years of age with exactly 11 years of schooling. The relative series start in 1995 because there was no PNAD survey in 1994, and before that, macroeconomic instability makes it difficult to calculate bindingness measures.
		\end{minipage}	
	\end{figure}

	This section discusses the limitations of the effective minimum wage and fraction-affected/gap estimators in the context of the rise of the Brazilian federal minimum wage beginning in 1995, illustrated in Figure~\ref{fig:Brazil-mw}. Brazil underwent a series of macroeconomic reforms in 1993 and 1994 which stabilized the economy after years of hyperinflation. Over the 22 years following stabilization, successive administrations implemented yearly adjustments that increased the minimum wage by 167 percent in real terms. Productivity also grew, but the minimum wage clearly became more binding. The red dash-dot line in Figure~\ref{fig:Brazil-mw} shows that the minimum wage grew substantially relative to median wages. Because the Brazilian workforce became more skilled over the same period, the relative change is starker if one conditions on a specific worker group, such as young men with high school degrees (the blue dashed line).\footnote{Figure~\ref{fig:Brazil-mw-share} in the appendix displays alternative bindingness measures based on shares of workers earning close to the minimum wage. They reinforce the view that the ``bite'' of the minimum wage grew substantially before 2007.}

	This setting is useful for analyzing the performance of the effective minimum wage and fraction-affected/gap estimators for three reasons. First, Brazil is a large country with substantial variation across regional labor markets. Second, there is no regional variation in minimum wage laws, making those estimators more appealing.\footnote{Brazilian law allows for state- and occupation-specific wage floors. However, they exist in few states, apply to few occupations, and are imperfectly enforced \citep{Moura2008,Corseuil2015}.} Third, there are several published papers and recent working papers using those estimators in the Brazilian context, allowing me to discuss their implementation and evaluate their results.

	\subsection{Estimates from previous literature: which one is the right one?}
	
	I focus on the six papers listed in Appendix Table~\ref{tab:lit}, which were selected based on a search on the IDEAS database.\footnote{The search was executed on 10/29/2024. I used the keywords ``Brazil minimum wage'' and included published papers from 2000 or later, or working papers from 2019 or later, that (i) used the methods studied in this paper; and (ii) used data from the post-stabilization period (i.e., after 1994). \citet{Derenoncourt2025} was not present in the IDEAS data set, but was included because it was known to me through participation in conferences.} The table summarizes the data sources for each paper, the preferred econometric specification, and the main results.
	
	\defcitealias{Derenoncourt2025}{Derenoncourt, G\'erard, Lagos, and Montialoux (2025)}
	
	Although all of those papers use regional variation in wage levels to identify the effects of the same minimum wage increase, they often reach different conclusions. \citet{Parente2025} find significant increases in informal employment relative to formal employment, \citet{Derenoncourt2025} find more modest effects, and \citet{Engbom2022} find no effects at all. \citet{Parente2025} also find that wage dispersion rises for all workers (formal and informal), while \citet{Hinojosa2019} finds the opposite. \citet{Neumark2006} find negative employment effects, while most other papers looking at that outcome do not.\footnote{Using a different design (still based on regional variation in wage levels), \citet{Hinojosa2019} also finds negative effects on employment, with smaller magnitudes than \citet{Neumark2006}. Studying the period from 1982 to 2000 with PNAD data, \citet{Lemos2009IntRev} also reports that estimates of employment effects are sensitive to the choice of estimator, with negative employment effects in some specifications and zero effects in others.} Spillover effects also differ between \citet{Hinojosa2019} and \citet{Engbom2022}.
	
	A clearer illustration comes from the sensitivity analysis in \citet{Engbom2022}. The central component of their masterful study of the Brazilian minimum wage is a structural labor market model with search frictions and two-sided heterogeneity. But their motivating descriptive analysis is also thorough and exceptionally transparent. Figure B.14 in their Online Appendix, replicated as Appendix Figure~\ref{fig:engbommoserfigb14} here, shows that their effective minimum wage estimates vary dramatically depending on specification choices.
	
	In many cases, disagreements between papers studying the same empirical context can be informative. For example, estimates of returns to education may differ depending on whether an instrument is used or not, or which specific instrument is used. Researchers can rationalize those differences based on transparent discussions of identification assumptions: the presence or absence of omitted variable bias, instrument validity, and who the complier population is for each instrument.
	
	However, for the papers in Table~\ref{tab:lit}, it is difficult to argue which specification should be preferred because the source of identification is not as transparent. In the first three papers in that table (using effective minimum wage designs), identification relies on the existence of an unobserved shock that shifts the location of the latent wage distribution, but is orthogonal to determinants of its shape and of latent employment levels. None of those papers specify what economic factor could play that role. The main shocks affecting Brazilian labor markets between 1995 and 2018---changes in schooling, trade and commodity shocks, labor-regulation enforcement, technological adoption, and social programs---are generally skill-biased rather than simple locational shifts of the log wage distribution, and they are also likely to affect employment directly.

	Similar arguments apply to the latter three papers in Table~\ref{tab:lit}, which use difference-in-differences estimators analogous to the fraction-affected design. The parallel trends assumption cannot be tested given the lack of a pre-treatment period with a stable minimum wage. In addition, the long-run estimates from the last two papers can be substantially biased if there are other structural factors causing regional convergence \citep[see][for papers documenting regional convergence in Brazil in the 2000s]{Manzi2023,Abreha2024}.
	
	\subsection{Sensitivity analysis and comparison with a structural model}
	
	Some conflicting results in existing literature may come from using different samples, different treatment of dynamic issues, or different ways to display the results (though Figure B.14 in \citealt{Engbom2022} goes a long way in addressing this limitation). To provide more direct evidence on the sources of conflicting results, this section presents original data analysis using Brazilian data. Specifically, I estimate different versions of the effective minimum wage and fraction-affected/gap designs on the same data. As an additional reference point, I also report causal effects estimated using the structural model of local labor markets from \citet{Haanwinckel2025}, which is not subject to the econometric concerns discussed in this paper (though, as with any other structural model, it may be vulnerable to other forms of misspecification). Details on that structural model are provided in Appendix~\ref{appendix:sdif}.
	
	To keep the analysis comparable to the rest of the paper, the data include exactly two periods: 1998 and 2012, the same years used in \citet{Haanwinckel2025}. This exercise aims to estimate the long-run effects of the minimum wage, starting from a period a few years after the stabilization reforms and ending in a period where the economy has had a few years to adjust to the higher minimum wage levels. The outcomes of interest are formal employment and formal wages. They are measured using an administrative matched employer-employee data set (\textit{``Rela\c{c}\~ao Anual de Informa\c{c}\~oes Sociais,''} henceforth RAIS).
	
	Each spatial unit is a ``microregion'' from the Brazilian Statistical Bureau (IBGE). These units are analogous to commuting zones in the United States in terms of size and population, and in the sense that fewer than 5\% of workers live in one microregion and work in another. They are commonly used to determine the boundaries of labor markets in studies using Brazilian data \citep[see e.g.,][]{Costa2016,DixCarneiro2017,Ponczek2021}.\footnote{Microregions are grouped to ensure constant boundaries throughout the studied period. Following the sample selection criterion in \citet{Haanwinckel2025}, I select regions with at least 15,000 workers in the RAIS data set in each year, and at least 1,000 in each educational group (less than high school, high school, and complete college or more). Those criteria select 151 out of 486 microregions that, together, account for 73\% of the adult population in Brazil.}
	
	The goal is not to choose a preferred regression specification, but to ask how much estimated effects move when the same data are analyzed using closely related exposure-based designs.
	
	\begin{table}
		\small
		\caption{Comparison of estimates using Brazilian data}
		\centering
		
		\label{tab:sdif}
		\textit{Panel A: Fraction-Affected Designs}
		
		\input{tab_sdif_fagap.tex}

		\vspace{1cm}
		\textit{Panel B: Effective Minimum Wage Designs}
		
		\input{tab_sdif_eff.tex}

		\vspace{1cm}
		\textit{Panel C: Effective Minimum Wage based on the 90th percentile}
		
		\input{tab_sdif_eff90.tex}	
		\begin{minipage}{1.0\textwidth}
			\footnotesize
			\justify
			\textbf{Notes:}      
			This table reports average causal effects from the structural model and estimated effects from exposure-based regressions applied to the same Brazilian data (from the RAIS data set, using a sample of 151 microregions and two time periods: 1998 and 2012). Panel A reports effects on log wage quantiles, while Panels B and C report effects on quantile gaps. Estimated standard errors (clustered at the microregion level) are shown in parentheses.
		\end{minipage}		
	\end{table}

	Table~\ref{tab:sdif} reports the average causal effects of the minimum wage predicted by that structural model, along with estimates of causal effects using different versions of the effective minimum wage and fraction-affected/gap estimators based on the same data. The structural model predicts sizable disemployment effects: a reduction of 3.3 percentage points in the share of the adult population (18 to 54 years old) employed in the formal sector. Both the baseline fraction-affected design and the baseline effective minimum wage design also predict reductions in formal employment that are economically and statistically significant. However, the magnitudes differ substantially. For the fraction-affected design, the decline is smaller at 1.9 percentage points, while for the effective minimum wage it is larger at 6.1 percentage points. 
	
	The structural model predicts sizable compression of wages in the lower tail of the wage distribution, but no inequality-reducing effects on the upper tail. The effective minimum wage estimate is very close for the p10-p50 quantile gap, but overestimates the compression of the p25-p50 gap relative to the structural model. It also predicts an \textit{increase} in inequality in the upper tail. The baseline fraction-affected estimate predicts large wage effects that are mostly uniform over the wage distribution. To the extent that there is wage compression, it happens in the upper tail.
	
	Table~\ref{tab:sdif} also reports estimates for alternative estimators. These include binary versions of the fraction-affected design, changes in the fixed effects included in the effective minimum wage design, an instrumental variables version of this estimator, and a version that constructs the effective minimum wage using the 90th percentile as the deflator instead of the median. Those specifications mirror the variety of implementations in existing literature, summarized in Appendix Table~\ref{tab:lit}. Each of those estimators yields different numbers for the causal effects of interest. Note that the divergence in predicted effects is large relative to estimated standard errors, even though the data only include two years. These results support the claim that, at least in the Brazilian context, most---or potentially all---of those regression-based estimators suffer from significant biases, which at least partly explain the variation in estimates reported in existing papers.
	
	\section{Recommendations for Applied Researchers \label{sec:recommendations}}
	
	This short section condenses the paper's main results into recommendations for applied researchers interested in using those regression methods. These recommendations are intended for panel studies where the unit of analysis is a labor market. This is most commonly a region, though it may also encompass cases where workers are assumed to be segregated at the region$\times$sector or region$\times$occupation levels. They focus on contexts where minimum wages are set at the national level, or where variation from market-level minimum wages is too scarce to be used for identification purposes.
	
	\begin{enumerate}
		\item Do not use the effective minimum wage design (see Section~\ref{sec:eff-min-w}).
		\item If the setting does not allow for testing for differential pre-trends---for example, because there are no pre-treatment data with a stable level of the national minimum wage---then the fraction-affected/gap designs should not be used either. The researcher should instead consider more structural methods that directly address the potential issues discussed in Section~\ref{sec:trends-dispersion}.
		\item Based on knowledge of the relevant institutional setting, consider including a small number of initial variables interacted with flexible time trends as controls. Their purpose is to capture the role of transformations that may differentially affect low- and high-wage regions (see Section~\ref{sec:trends-dispersion}). Unit-specific time trends should not be included.
		\item Test for parallel pre-trends on all outcomes of interest using the gap measure constructed using a single pre-treatment year (see Appendix~\ref{appendix:regional-convergence}). If those tests reject parallel pre-trends, researchers are discouraged from using the fraction-affected/gap designs and should consider instead a more structural approach, as mentioned above.
		\item For the main regressions, use a quadratic polynomial on the gap measure for wage outcomes and a cubic polynomial on the fraction-affected measure for employment outcomes. If a single specification must be used for all outcomes, select the quadratic gap specification (see the end of Section~\ref{sec:500-dgps}).
	\end{enumerate}
	
	Applications may require adjustments or deviations from these general prescriptions. That said, these guidelines should provide a useful starting point, and the remainder of the paper can be helpful in pointing toward potential pitfalls if alternative implementation choices are made.
	
	\section{Broader relevance}
	
	This paper studies econometric strategies for a national minimum wage when the policy does not vary across regions. I close by discussing two strands of Labor Economics literature where empirical designs used in recent work may be subject to similar concerns.
	
	\paragraph{Local exposure to recessions.}	
	Papers studying how local labor markets are affected by recessions are a close analogue to studies of a national minimum wage: there is a common aggregate shock, and researchers study its local impacts by constructing a regional exposure measure.
	
	Some of these papers use regressors analogous to the effective minimum wage design in that they are built from contemporaneous local outcomes. \citet{HershbeinStuart2023,HershbeinStuart2024} measure local recession severity using realized local employment declines, in one case to study earnings and transfers across places and in the other to study the evolution of local labor markets. These regressors are not primitive shocks, but realized local equilibrium objects that bundle the recessionary disturbance with local industry and product mix, firm-level shocks, and other demand- and supply-side adjustment margins. They may therefore be subject to some of the concerns discussed in Section~\ref{sec:eff-min-w}. The estimates in these papers are informative for the authors' stated object: the relative evolution of places that actually experienced larger employment losses. The caution is that realized local severity should not be read as a primitive recession shock without additional structure.
	
	Other recession papers use designs closer to the fraction-affected logic. \citet{GreenstoneMasNguyen2020} and \citet{MitraWei2025} construct predetermined credit-shock exposures from pre-crisis bank or lender shares and national credit-supply shifts. These designs include safeguards such as predetermined county characteristics interacted with time effects in \citet{GreenstoneMasNguyen2020} and pre-trend or placebo checks in both papers. Still, they are not clear-cut treated-control comparisons: treatment intensity is summarized by a researcher-specified exposure measure, and the empirical model imposes a functional-form mapping from that exposure to outcomes. A potential path for further research would be to investigate whether such estimators can be sensitive to functional-form misspecification in ways analogous to the fraction-affected/gap designs studied in this paper.
	
	\paragraph{Local labor market concentration.}	
	Recent labor economics research has used local employer concentration measures, often Herfindahl-Hirschman indexes (HHIs), to study wages, inequality, and job security under imperfect competition. These papers are further apart from the national minimum wage setting, as they are not interested in the local impacts of a national shock. However, they are similar to the estimators studied here in the sense that treatment intensity is not clearly observed; rather, the researcher estimates regressions where the key independent variable is constructed according to a formula that may not clearly map into an underlying theoretical model. The measured HHI is not a structural shock, but an observed equilibrium object reflecting entry and exit, mergers, demand and productivity shocks, worker mobility, and market boundaries. Researchers often interpret it as a proxy for employer market power, but the same HHI movement can reflect very different latent shocks.
	
	For example, better local transportation can make workers more responsive to wage differentials, reducing markdowns, while also allowing better firms to recruit from a wider area and raising concentration. By contrast, higher entry costs or tighter land-use constraints can also raise the HHI while strengthening employer power. The same change in measured concentration may therefore reflect latent shocks with opposite implications for wages and employment. Regressions using observed local HHIs are therefore difficult to interpret structurally without assumptions about the shocks moving concentration. The HHI may proxy for a changing mix of mechanisms, and the proxy error is unlikely to be classical (as discussed in Subsection~\ref{subsec:good-and-bad}).
	
	Instrumental variables (IV) designs such as \citet{Rinz2022} and \citet{BassaniniEtAl2024} address part of this problem by moving away from purely local contemporaneous variation. In \citet{Rinz2022}, the leave-one-out instrument uses average local concentration in other commuting zones within the same industry. In \citet{BassaniniEtAl2024}, identification comes from a quasi-leave-one-out instrument based on the number of firms hiring in other functional areas within the same occupation and year. These strategies reduce local reverse causality, which I showed to be problematic in the effective-minimum-wage design. But they still shift an observed equilibrium statistic rather than a primitive structural determinant of employer power. Interpretation therefore depends on what structural forces are moving concentration.
	
	A more directly interpretable approach is to isolate the source of variation behind measured concentration. \citet{BenmelechBergmanKim2022}, for example, use merger activity to instrument for changes in local employer concentration, focusing on merger-driven movements in HHI rather than all contemporaneous changes in concentration. A complementary approach is to give the concentration measure an explicit role in a model. \citet{Xu2023} combines the recession and concentration literatures by studying how pre-recession HHI, interpreted as a summary of firm size distribution, shaped employment declines during the Great Recession. The paper models how HHI can modulate the transmission of idiosyncratic firm shocks, illustrating one way to give economic content to a concentration-based exposure measure.
	
	These considerations suggest another direction for future work: evaluating formula-based regional concentration measures under explicit models of firm dynamics, worker mobility, and labor demand, much as this paper does for minimum wage designs.

	\singlespacing 
	
	\bibliographystyle{style_aea}
	\bibliography{refs.bib}
	
	\onehalfspacing
	
	\pagebreak
	\setcounter{page}{1}
	
	\appendix
	
	{\huge \textbf{Online Appendix}}

	\renewcommand{\thefigure}{A\arabic{figure}}
	\setcounter{figure}{0}

	\renewcommand{\thetable}{A\arabic{table}}
	\setcounter{table}{0}

	\section{What do empirical minimum wage papers identify? \label{appendix:ate-discussion}}
	
	In this appendix, I argue that the central goal in all of the recently published papers mentioned early in the introduction is to measure treatment effects in a way that is equivalent to the definition used in this paper (Equation~\ref{eq:ate}). 
	
	\paragraph{\citet{Bosch2010}:} Their goal is to explain average changes in quantile gaps, where averages are taken over local labor markets---exactly in the same way as in Equation~\ref{eq:ate}. Those effects are estimated based on coefficients obtained from the effective minimum wage, in the same way that is described in Subsection~\ref{subsec:eff-min-w-definition}. They write: \textit{``Our analysis reveals that a substantial part of the growth in inequality between
				1989 and 2001, and essentially all the growth in inequality in the bottom end of
				the distribution, is due to the steep decline in the real value of the minimum wage.''}
				
	\paragraph{\citet{Dinkelman2012}:} The paper uses an identification strategy similar to the fraction-affected/gap designs, but with the continuous treatment intensity variable being the initial effective minimum wage from \citet{Lee1999}. The authors do not explicitly specify how they infer aggregate effects from the regression coefficients, as that requires a normalization to deal with the ``missing intercept'' issue. That said, the language in the paper suggests that it attempts to measure aggregate effects similar to the definition in Equation~\ref{eq:ate}. Specifically, the paper begins with the question \textit{``What happens to wages and employment in the informal sector after
			the introduction of a minimum wage in that sector?''}, indicating that the object of interest is the aggregate impact of a ceteris paribus introduction of a minimum wage. The introduction summarizes their strategy and results as follows: \textit{``We evaluate
			the effects of South Africa's 2002 minimum wage law for domestic
			workers by exploiting time-series variation in the application of the
			law and pre-existing cross-sectional variation related to the intensity of
			the law to identify wage and employment effects. (...) We complement the before-after analysis
			with a difference-in-differences strategy that adopts the methods in
			Lee (1999) to statistically examine the effects of the law. (...)
			We find that wages increase by a statistically significant 13--15\% in the
			wake of the law.''}
	
	\paragraph{\citet{Dustmann2021}:} The authors write in the introduction: \textit{``We investigate the wage, employment, and reallocation effects of the introduction
		of a nationwide minimum wage in Germany that affected 15\% of all employees.
		Based on identification designs that exploit variation in exposure across
		individuals and local areas, we find that the minimum wage raised wages but did
		not lower employment.''} In the introduction, speaking specifically about their gap design exploiting pre-existing variation in regional wage levels, they write: \textit{``Findings from an
		analysis that exploits variation in the exposure to the minimum
		wage across 401 local areas (similar to Card 1992) corroborate our
		findings from the individual-level analysis: the minimum wage
		boosted wages but did not reduce employment in heavily affected
		areas relative to less affected. Our findings therefore do not confirm
		the fears of many economists that the minimum wage would
		cause substantial job losses. Rather, our findings support the idea
		that the minimum wage helped reduce wage inequality without
		reducing employment across individuals and across local areas.''} Given this language, it is reasonable to argue that the authors are measuring aggregate, ceteris paribus effects of the introduction of the national minimum wage in Germany.
	
	\paragraph{\citet{Bailey2021}:} Their abstract reads: \textit{``This paper examines the short- and longer-term economic effects of
		the 1966 Fair Labor Standards Act (FLSA), which increased the national
		minimum wage to its highest level of the twentieth century and
		extended coverage to an additional 9.1 million workers. Exploiting
		differences in the ``bite'' of the minimum wage owing to regional variation
		in the standard of living and industry composition, this paper
		finds that the 1966 FLSA increased wages dramatically but reduced
		aggregate employment only modestly. However, some evidence
		shows that disemployment effects were significantly larger among
		African American men, 40\% of whom earned below the new minimum
		wage.''} A passage in the introduction makes the aggregate-effects, ceteris-paribus interpretation even more explicit: \textit{``Across the United States, our estimates suggest that average wages
		increased by 6.5\% because of the 1966 FLSA.''}
	
	\paragraph{\citet{Engbom2022}:} The central results of the paper come from counterfactual simulations of a ceteris paribus minimum wage increase, based on an estimated structural model. Thus, their findings correspond to the definition in Equation~\ref{eq:ate} in a setting with a single ``national'' region in the model (such that no averaging across locations is necessary) and with a static model (such that $\mathbf{ATE}_0$ and $\mathbf{ATE}_1$ are identical).
	
	\paragraph{\citet{Giupponi2024}:} From their abstract: \textit{``We assess the impact of nationwide minimum wages on employment throughout the whole wage distribution by exploiting geographical variation in the level of wages. We find a substantial increase in wages at the bottom of the wage distribution, while we detect a small, statistically insignificant negative effect on employment.''} The authors specify in detail how they can measure aggregate effects, as opposed to only relative effects, using their methodology: \textit{``Since in practice no region is entirely unaffected by a national minimum wage policy, our approach shares the characteristic of other regional variation approaches of identifying the relative effect of the minimum wage on employment in lower-wage areas compared with higher-wage ones. Retrieving an absolute effect requires additional assumptions that we describe below.''}

	\section{Simulation details for the Normal-markdown model \label{appendix:simulation-details}}

	This appendix lists the parameters used in all simulations. Every simulation exercise is repeated 1,000 times. The increase in the log minimum wage is always 0.2, except where otherwise noted.
	
	\subsection{Microfoundation and mathematical description of the basic model}
	
	Each local labor market has a continuum of workers of different productivity levels $z$, all of whom supply their unit endowment of labor inelastically. Productivity is perfectly observable. There is a large number of identical firms with linear technologies: revenues are simply the sum of all of their workers' productivities. The indirect utility of a worker with productivity $z$ if they choose to work at firm $j$ is
	\begin{equation*}
		U(z,w_{zj})=\frac{m}{1-m} \ln w_{zj} + \eta_j,
	\end{equation*}
	where $m\in (0,1)$ is a parameter that will be discussed below, $w_{zj}$ is the wage paid to workers of productivity $z$ by firm $j$, and $\eta_j$ is a Gumbel-distributed idiosyncratic preference shock representing the ``amenity value'' of that firm for a particular worker. This preference structure is the same as in the monopsonistic model of \citet{Card2018}. Workers observe all posted wages $w_{zj}$ in the economy and choose to work for the firm that gives them the highest utility.
	
	Firms cannot discriminate wages based on a worker's preference shocks, which is why wages can be represented as only varying at the $z,j$ level. Firms cannot choose wages below the minimum: $\ln w_{zj}\ge mw$. However, firms have the option of choosing a minimum productivity threshold $\munderbar{z}$ below which workers are not offered employment opportunities.
	
	Assuming firms consider themselves small relative to the market, as in \citet{Card2018}, the equilibrium of this model has all firms posting the same wages:
	\begin{equation*}
		w_z = \begin{cases}
			m\cdot z & \text{if } z > \exp(mw) / m
			\\
			\exp(mw) & \text{if } \exp(mw) \le z \le \exp(mw) / m
			\\
			\text{Job not offered} & \text{if } z < \exp(mw).
		\end{cases}
	\end{equation*}
	The first row corresponds to workers with high productivity, for whom the minimum wage does not bind. They earn their productivity multiplied by a common markdown factor $m$. This happens because workers have idiosyncratic preferences over firms, giving firms market power over them. 	
	The second row corresponds to workers for whom the minimum wage binds, but who are still profitable for the firm if paid the minimum wage. Those are the workers who benefit from the minimum wage policy in this simple model. Workers whose productivity is strictly below the minimum wage $\exp(mw)$ will not be hired by any firm, and are thus hurt by the minimum wage.
	
	Now assume productivities are lognormal, with region-time specific means and variances:
	\begin{equation*}
		\ln z \sim \mathcal{N}\left(\mu_{r,t} - \ln m, \sigma^2_{r,t}\right).
	\end{equation*}
	Because unconstrained wages are equal to $m\cdot z$, this assumption implies that the latent log wage distribution in region $r$ at time $t$ is Normal with mean $\mu_{r,t}$ and standard deviation $\sigma_{r,t}$. The productivity cutoff $z=\exp(mw)$ corresponds to truncating latent log wages below $mw+\ln m$. In addition, workers with $z\in[\exp(mw), \exp(mw) / m]$ create a ``spike'' or mass point exactly at the minimum wage. Figure~\ref{fig:simulation-model} illustrates these effects in two regions that differ in the ``location'' parameter $\mu_{r,t}$. The truncation corresponds to the red area, while the spike corresponds to the orange area.
	
	Let $w^*$ denote latent log wages. Their cumulative distribution function (cdf) is $G^*_{r,t}(w^*)=\Phi\left(\frac{w^*-\mu_{r,t}}{\sigma_{r,t}}\right)$, where $\Phi$ is the cdf of a standard Normal distribution. The employment-to-population ratio and the distribution of observed wages depend on latent wages, the level of the national minimum wage, and a ``markdown'' parameter $m\in(0,1]$ as follows:
	\begin{align*}
		emp_{r,t} = & 1 - \Phi\left(\frac{mw_{t}+\log m-\mu_{r,t}}{\sigma_{r,t}}\right)
		\\
		G_{r,t}(w)= & \frac{\Phi\left(\frac{w-\mu_{r,t}}{\sigma_{r,t}}\right) - \Phi\left(\frac{mw_{t}+\log m-\mu_{r,t}}{\sigma_{r,t}}\right)}{1 - \Phi\left(\frac{mw_{t}+\log m-\mu_{r,t}}{\sigma_{r,t}}\right)}  \qquad \text{for } w \ge mw_{t}
	\end{align*}
	This model generates both truncation and censoring of the latent wage distribution. Workers whose latent log wages are below $mw_t+\log m$, the log of the product of the minimum wage and the markdown, become disemployed. For those with latent log wages above the log minimum wage, the observed wage is equal to the latent wage. Finally, those who remain employed but have latent log wages below the log minimum wage see a mechanical increase in their wage. The latter group corresponds to the minimum wage ``spike'' in the log wage distribution.
	
	The model can be understood as reflecting an economy with an inelastic labor supply, exogenous worker productivities, and identical monopsonistic firms paying wages that are below the marginal products of labor unless mandated to pay higher wages via the minimum wage. When the markdown $m$ is low, disemployment effects are smaller, and positive effects on wages are larger. Unless otherwise noted, I use $m=0.7$.
	
	\subsection{Calibration}
	
	The meta-parameters governing the distribution of region-specific parameters $[\mu_{r,0}, \sigma_{r,0},\mu_{r,1},\sigma_{r,1}]$ are based on data from the US Current Population Survey for 1989 (corresponding to period $t=0$) and 2004 (corresponding to $t=1$). I chose those years because the national minimum wage was small and approximately the same, in real terms, in both years and the unemployment rate was also approximately equal.
	
	The data were processed using the same procedures as in \citet{Lemieux2006}. The sample is restricted to workers between 16 and 64 years of age, with positive potential experience, and whose wages and worked hours are reported by the respondent instead of inferred. Top-coded earnings are adjusted by a factor of 1.4.
	
	Using this sample, I calculate the mean and standard deviation of log wages in each combination of state and year, weighting by the CPS sampling weights and worker hours. Then, I de-mean the $\mu_{r,t}$ elements using simple averages within the period so that the $\mu_{r,t}$ are mean zero in both periods. I treat those statistics as corresponding to the $[\mu_{r,0}, \sigma_{r,0},\mu_{r,1},\sigma_{r,1}]$ vector for each state. Thus, I calculate the corresponding covariance matrix of that vector and use it to calibrate the simulation models.
	
	Finally, I calibrate the simulations using the estimated vector of means and covariance matrix. As stated in the main text, in each simulation, the vectors $[\mu_{r,0}, \sigma_{r,0},\mu_{r,1},\sigma_{r,1}]$ for each region $r$ are drawn from a Multivariate Normal distribution. The parameters for that meta-distribution are created by either ``shutting down'' some of the correlations in the estimated covariance matrix, eliminating differences in dispersion parameters, increasing the correlation between some initial and final region parameters to one (to impose that those parameters are time-invariant), or averaging some meta-parameters between both periods so that the distributions are stable over time. Tables~\ref{tab:simparams-nm-eff} and~\ref{tab:simparams-nm-fagap} report the meta-parameters used in every simulation exercise with the Normal-markdown model.

	\begin{sidewaystable}
		\small
		\caption{Simulation meta-parameters: Normal-markdown model, effective minimum wage design}
		\centering
		
		\label{tab:simparams-nm-eff}
		\input{tab_simparams_normal_eff.tex}	
		\begin{minipage}{1.0\textwidth}
			\footnotesize
			\justify
			\textbf{Notes:}   
			This table shows meta-parameters used to draw simulation samples for the Normal-markdown model. See Appendix~\ref{appendix:simulation-details} for details.
		\end{minipage}		
	\end{sidewaystable}

	\begin{sidewaystable}
		\small
		\caption{Simulation meta-parameters: Normal-markdown model, fraction-affected/gap designs}
		\centering
		
		\label{tab:simparams-nm-fagap}
		\input{tab_simparams_normal_fagap.tex}	
		\begin{minipage}{1.0\textwidth}
			\footnotesize
			\justify
			\textbf{Notes:}   
			This table shows meta-parameters used to draw simulation samples for the Normal-markdown model. See Appendix~\ref{appendix:simulation-details} for details.
		\end{minipage}		
	\end{sidewaystable}

	\subsection{Positive employment effects \label{appendix:sim-pos-emp-eff}}
	
	For some simulation exercises, I augment the Normal-markdown model to include the possibility of positive employment effects. I add two parameters to the model: $P_{base}$ and $P_{height}$. The total employment mass added to the model is equal to $\frac{P_{base} P_{height}}{2} g^*_{r,t}(mw_t)$, where $g^*_{r,t}(w)=\frac{1}{\sigma_{r,t}}\phi\left(\frac{w-\mu_{r,t}}{\sigma_{r,t}}\right)$ is the density of the latent log wage distribution. Equivalently, the added mass is $\frac{P_{base} P_{height}}{2\sigma_{r,t}} \phi\left(\frac{mw_t-\mu_{r,t}}{\sigma_{r,t}}\right)$. The wage distribution for that extra mass is triangular, with support $[mw_t, mw_t+P_{base}]$ and peak at the left extreme of the support.
	
	Intuitively, that model corresponds to one where the minimum wage increases labor force participation of individuals with potential wages just above the minimum wage, in the interval $[mw_t, mw_t+P_{base}]$. This effect's overall intensity is assumed to be proportional to the latent log-wage density evaluated at the minimum wage level and to the $P_{height}$ parameter. In this model, a small minimum wage is likely to have positive employment effects, which are initially increasing. However, at some point, the effects of disemployment start to become more significant. Eventually, the effects of the minimum wage on employment will become negative.
	
	\subsection{The Normal-markdown model with state-level minimum wages \label{appendix:simulation-details-ams}}
	
	For the exercise shown in Table~\ref{tab:ams}, I augment the Normal-markdown model to include the possibility of state-specific minimum wages that surpass the national minimum wage. I first choose the share of regions that, in each period, are selected to have a higher local minimum wage. Those shares are 0.2 or 0.4, depending on the panel in Table~\ref{tab:ams}. For reference, the share of states in the US that had local minimum wages at least 5 log points above the federal minimum wage was 0.23 in both 1989 and 2004.
	
	When simulating the model for the initial period, I randomly draw the subset of regions with higher local minimum wages. Given the shares chosen above, those subsets have the same size in all simulations. Then, I draw a number from a Normal distribution with a mean of 0.25 and a standard deviation of 0.075. I assign a local log minimum wage that is equal to the federal minimum wage plus that number (or the federal minimum wage plus 0.05, whatever is higher). The numbers 0.25 and 0.075 above are chosen to match the mean and standard deviation of the log gap between local minimum wages and the federal minimum wage in 2004 for the subset of states for which the minimum wage is higher than the federal one (the corresponding numbers are 0.15 and 0.052 for 1989).
	
	In the second period simulation, I follow the same procedure, except that I do not allow for reductions of local minimum wages between periods. So, the local minimum wage is either calculated from the procedure above or observed in the first period, whichever is higher.
	
	The definition of the average treatment effect to be estimated is updated in the following way to account for the possibility of local minimum wages:
	\begin{align*}
		\bm{ATE}_0 & = \mathbb{E}\left[ f\left(mw_{r,1}, \theta_{r,0}\right) - f\left(mw_{r,0}, \theta_{r,0}\right) \right]
		\\
		& = \mathbb{E}\left[ f\left(mw_{r,1}, \theta_{r,0}\right) - \bm{y}_{r,0}\right]
		\\
		\bm{ATE}_1 & = \mathbb{E}\left[ f\left(mw_{r,1}, \theta_{r,1}\right) - f\left(mw_{r,0}, \theta_{r,1}\right)\right]
		\\
		& = \mathbb{E}\left[ \bm{y}_{r,1} - f\left(mw_{r,0}, \theta_{r,1}\right)\right]
		\\
		\bm{ATE} & = \frac{\bm{ATE}_0 + \bm{ATE}_1}{2}
	\end{align*}
	The only difference is that the counterfactuals being considered correspond to state-specific changes in the minimum wage, caused either by the increase in the federal minimum wage or by a random draw of a higher local minimum wage in the latter period. Calculating the estimated average treatment effects is the same, using region-specific changes in the observed effective minimum wage. That is consistent with the updated definition, as changes in the effective local minimum wage reflect both national and local minimum wage changes.
	
	\subsection{Generation of 500 DGPs for the exercise in Section~\ref{sec:500-dgps} \label{appendix:500-gdps} }
	
	Each of the 500 DGPs corresponds to the parameters determining the joint distribution of the vector of region-specific parameters $[\mu_{r,0}, \sigma_{r,0},\mu_{r,1},\sigma_{r,1}]$, coupled with a markdown parameter $m$, the positive employment parameters $P_{base}, P_{height}$, and the levels of the national minimum wage, $mw_0,mw_1$. I randomly draw these features of the DGP as specified below:
	
	\paragraph{Average location $\mu_{r,t}$:} Set to zero in both periods in all DGPs.
	
	\paragraph{Standard deviation of location $\mu_{r,t}$:} That parameter is uniformly drawn in the interval (0.056, 0.185) for each of the two periods (that is, the variability of location parameters between regions may change over time). That interval corresponds to 0.5 times the corresponding parameter for the US in 2004 and 1.5 times for the US in 1989, respectively.
	
	\paragraph{Correlation between initial and final location, $Corr(\mu_{r,0},\mu_{r,1})$:} Uniformly drawn in the interval (0.838, 0.95). The number observed for the US, 0.894, corresponds to the center of the interval. The maximum was set to 0.95 to avoid DGPs where the initial and final location parameters are too similar, which means that the variance of location \textit{shocks} would be too small. In that scenario, the effective minimum wage design performs poorly. Alternatively, if the correlation is too low, the lack of persistence across time handicaps the fraction-affected design. With this narrow range, all DGPs have persistence levels similar to what can be inferred from the US between 1989 and 2004.
	
	\paragraph{Average dispersion $\sigma_{r,t}$:} For the first period, that average is uniformly drawn in the interval (0.255, 0.813). That interval corresponds to 0.5 times the corresponding parameter for the US in 2004 and 1.5 times for the US in 1989, respectively. Thus, the simulated DGPs are substantially heterogeneous in the within-region dispersion of latent log wages, corresponding to different empirical applications (e.g., Germany vs. Brazil).
	
	The \textit{change} in average dispersion from the first to the second period is uniformly drawn in the interval (-0.047, 0.047). The extreme values correspond to 1.5 times the corresponding statistic observed in the US. Thus, trends in the dispersion of latent log wages---which I showed to be problematic for the fraction-affected design---can never be much more significant than what can be inferred from US data and are typically smaller than that.
	
	\paragraph{Standard deviation of dispersion $\sigma_{r,t}$:} Uniformly drawn in the interval (0.012, 0.074) for each of the two periods. That interval corresponds to 0.5 times the corresponding parameter for the US in 1989 and 1.5 times for the US in 2004, respectively.
	
	\paragraph{Correlation between initial and final dispersion, $Corr(\sigma_{r,0},\sigma_{r,1})$:} Uniformly drawn in the interval (0.228, 0.684). That interval is the corresponding parameter for the US multiplied by 0.5 and 1.5, respectively.
	
	\paragraph{Contemporaneous correlation between location and dispersion, $Corr(\mu_{r,t},\sigma_{r,t})$:} Uniformly drawn in the interval (0.0, 0.2). Thus, while I allow for positive correlations between location and dispersion, the magnitudes are small compared to the data from the US and Brazil. The corresponding numbers for the US are 0.076 in 1989 and 0.264 in 2004. Using Brazilian data from 1998 (see Section~\ref{sec:brazil} of the paper for details), I find a correlation of 0.339 between the mean log wage and standard deviation of log wage in the 151 Brazilian microregions included in my sample. One may be worried that the minimum wage itself generates this correlation, such that it does not reveal information about the \textit{latent} distribution of log wages. To test this hypothesis, I calculate the correlation using only 88 microregions where the share of workers earning at most the minimum wage plus 30 log points is 10\% or less. Rather than decreasing, the correlation between mean log wage and standard deviation of log wage increases to 0.528 in this subsample, suggesting that the correlation between location and dispersion of latent log wages in Brazil is significantly higher than 0.2. Thus, the simulated DGPs are conservative in that regard.
	
	\paragraph{Markdown parameter $m$:} Uniformly drawn in the interval (0.1, 0.9). The lower half of this interval may seem implausible under partial-equilibrium models of monopsonistic firms, as they would imply labor supply elasticities to the firm below one. I include such numbers to allow for the possibility that the minimum wage causes significant wage increases in the lower tail of the distribution without causing any employment effects. One could interpret the very-low-markdown model as an approximation for a richer model where the minimum wage reallocates workers from firms that would be paying less than the minimum wage to other firms paying exactly the minimum wage. 
	
	\paragraph{Positive employment effect parameters:} $P_{base}$ is uniformly drawn in the interval (0.01, 0.5), implying that the minimum wage may attract workers earning up to 50 log points above the minimum wage in the highest-draw DGP. $P_{height}$ is uniformly drawn in the interval (0.01, 0.5) as well, implying a triangular added-mass component with height up to 50\% of the latent log-wage density right above the minimum wage. See Subsection~\ref{appendix:sim-pos-emp-eff} above for details.
	
	\paragraph{Minimum wage levels:} The initial log minimum wage is uniformly drawn in the interval (-1.5, -0.5), where zero corresponds to the center of the latent log wage distribution. The increase in the log minimum wage---which should be interpreted as an increase relative to average TFP growth---is uniformly drawn in the interval (0.15, 0.5). Thus, the simulations include minimum wage increases ranging from relatively small increases of 15 log points to larger increases similar in magnitude to the increase observed in Brazil between 1998 and 2012.

	\paragraph{Simulation procedure:} Sometimes, the combination of meta-parameters may be invalid because it leads to a covariance matrix for the $[\mu_{r,0}, \sigma_{r,0},\mu_{r,1},\sigma_{r,1}]$ vector that is not positive definite. When that happens, that draw is discarded and a substitute is used.
	
	After obtaining a valid DGP candidate, I perform 1,000 simulations, with each simulation corresponding to a sample of 200 regions and two periods (as in the rest of the paper). Using this sample, I calculate average treatment effects as defined in Section~\ref{sec:definitions}. The DGP is included in the analysis if the absolute value of minimum wage effects on employment is at least 0.005 (half a percentage point), or if the absolute value of any spillover relative to the median wage is at least 0.05 (five log points). If that is not the case---typically because of a low starting minimum wage coupled with a slight minimum wage increase---then that DGP is discarded, and I randomly draw another one to substitute for it. I also discard and find a replacement if the absolute value of the minimum wage effect on employment is 0.04 or larger. The process is completed when there are 500 DGPs with non-trivial, but not excessively large, minimum wage effects.
	
	\section{Assessing variants of the effective minimum wage design \label{appendix:eff-min-w-alternatives}}
	
	In this appendix, I discuss the effectiveness of potential solutions to the problems described in Section~\ref{sec:eff-min-w} of the paper, based on alternative specifications of the effective minimum wage design.
	
	\subsection{Fixed effects, trends, controls, and confounders \label{subsec:fe-controls}}
	
	The baseline specification in \citet{Lee1999} does not include region fixed effects. Concerning the inclusion of such fixed effects, he writes: ``\textit{... the reduced
		identifying variation resulting from eliminating the "permanent"
		state effects may magnify biases due to misspecification, in the
		same way biases stemming from measurement error in the
		independent variable are magnified when true variation in the
		independent variable is reduced.}'' Using the language introduced in Subsection~\ref{subsec:good-and-bad}, the estimator without region fixed effects has another source of ``good'' variation: within-period differences in the location parameters of latent log wage distributions (instead of simply differential shocks to location). That may significantly reduce the influence of ``bad'' variation from correlated measurement error in the centrality measure, reducing the amount of bias.
	
	Table~\ref{tab:no-tfe} illustrates Lee's argument through simulations. The data-generating process for those simulations is the same as reported in Panel~B of Table~\ref{tab:iss1}. I report predicted treatment effects using the default effective minimum wage design and two alternative specifications, the first being the estimator without region fixed effects. The comparison of the no-region-fixed-effects row to the baseline effective-minimum-wage row shows that, by using more ``good'' variation from level differences in $\mu_{r,t}$, the estimator without region fixed effects can indeed perform better.
	
	Still, it is easy to contemplate omitted variable biases that could cause problems for estimators without region fixed effects. For example, a persistent shock reducing labor demand over a long period would introduce a negative correlation between the effective minimum wage and employment. Alternatively, regions may differ in the shape of their distribution of worker skills, leading to different baseline levels for quantile wage gaps. Such concerns make specifications with fixed effects overwhelmingly more popular in published literature.
	
	Indeed, the specifications in papers such as \citet{Bosch2010}, \citet{Autor2016}, and \citet{Engbom2022} go beyond region fixed effects and include region-specific trends as well. These trends may absorb region-specific supply and demand shocks that affect the median wage and the outcomes of interest. 
	
	However, region-specific trends and controls may amplify, rather than reduce, biases. The reason is analogous to Lee's argument on region fixed effects. By including the trends, the researcher may throw out the ``good variation'' with the bathwater. The fixed effects, region-specific trends, and controls may absorb much of the $\Delta \mu_{r,t}$ shocks, such that the measurement errors become a larger share of the residual variation in the effective minimum wage. Once those controls are included, it may be difficult to interpret where the variation in the effective minimum wage is coming from. This lack of intuition is problematic; ideally, the researcher should be able to defend the assumption that there exists an economic factor, separate from all trends and controls, that shifts wage levels but does not affect employment levels or the shape of the log wage distribution in any way (other than making the minimum wage more or less binding). In Section~\ref{sec:brazil}, I discuss those issues in the Brazilian context.
	
	\begin{table}
		\small
		\caption{Effective minimum wage: alternative fixed effects specifications}
		\centering
		
		\label{tab:no-tfe}
		\input{tab_new_eff_noTFE.tex}	
		\begin{minipage}{1.0\textwidth}
			\footnotesize
			\justify
			\textbf{Notes:}      
			See the notes below Table~\ref{tab:iss1} for an explanation of the table's structure. The data-generating process corresponds to Panel~B from Table~\ref{tab:iss1}.
		\end{minipage}		
	\end{table}

	One may wonder whether dropping the time effects from the design would also add more ``good'' variation. \citet{Lee1999} explains that this choice is unwise if the shape and average dispersion of latent log wages change over time. It is not warranted in the presence of inequality trends from technical change or trade shocks, for example.
	
	Table~\ref{tab:no-tfe} illustrates the sensitivity of that estimator to changes in the economic environment. The baseline data-generating process displays minor differences in the marginal distributions of $\mu_{r,t}$ and $\sigma_{r,t}$ between periods. The most salient differences are that average $\sigma_{r,t}$ falls from 0.54 to 0.51, and the standard deviation of $\sigma_{r,t}$ between regions increases from 0.026 to 0.049. Those small changes are enough to warrant the inclusion of time effects, as biases are much more prominent when the model does not include them.

	\subsection{Does using a higher quantile as the deflator help? \label{subsec:p90}}

	Sometimes, the researcher may have a prior that the minimum wage significantly impacts the median wage, making it a poor measure of centrality. In those cases, they may consider using a higher quantile of the wage distribution to construct the effective minimum wage. For example, \citet{Bosch2010} use quantile 0.7 as the deflator in a study of Mexico, and \citet{Engbom2022} use quantile 0.9 when studying Brazil.

	\begin{table}
		\small
		\caption{Effective minimum wage using the 90th percentile as the deflator}
		\centering
		
		\label{tab:kaitz90}
		\input{tab_new_eff_90.tex}	
		\begin{minipage}{1.0\textwidth}
			\footnotesize
			\justify
			\textbf{Notes:}      
			This table has the same structure as Table~\ref{tab:no-tfe}, but reports regression results where the effective minimum wage is calculated based on the 90th percentile of the observed log wage distribution.
		\end{minipage}		
	\end{table}

	\citet{Lee1999} argues that the deflator should be a good approximation for centrality $\mu_{r,t}$ instead of merely an overall measure of wages. Otherwise, the regression may yield non-zero estimates even when the observed log wage distribution is identical to the latent wage distribution.\footnote{See Lee's discussion around Equation (5) on page 996.} The discussion regarding correlated measurement error introduces another reason to be wary of choosing higher quantiles of the wage distribution. While it is true that those higher quantiles may be less affected by the minimum wage, the effects will still not be zero if the minimum wage has employment effects, positive or negative. In addition, higher quantiles are likely to be more sensitive to cross-region differences in the dispersion of latent log wages. Due to those two issues, the biases may be more significant when a quantile other than the median is used as the deflator.

	Table~\ref{tab:kaitz90} evaluates the performance of an estimator based on the 90th percentile of the log wage distribution using the baseline scenario with regional differences in location and dispersion parameters (the same as in Table~\ref{tab:no-tfe}). The biases are significantly larger than those for other estimators previously discussed.

	\subsection{Is the standard diagnostic test effective? \label{subsec:upper-tail-diagnostics}}
	
	\citet{Lee1999} proposes estimating relative effects on high log wage quantiles $q>0.5$ to validate the model. The justification for that approach is that, in many applications (such as in the US), the researcher may have a strong prior that the minimum wage should have minimal effects on the upper tail of the wage distribution. \citet{Autor2016} use the same specification test to validate their instrumental variables implementation of the effective minimum wage design.
	
	As with any test, one should consider the possibility of false positives and false negatives. False positives may arise because many plausible mechanisms could lead to minimum wage spillovers that extend beyond the median wage. \citet{Engbom2022} develop and estimate an on-the-job search model where minimum wages cause spillovers that extend far into the upper tail of the wage distribution, primarily due to worker reallocation from low- to high-wage firms. The model in \citet{Haanwinckel2025} also includes endogenous changes in within-firm returns to skill in response to reallocation flows, firm entry responses, and price effects as mechanisms that can generate spillovers in the upper parts of the wage distribution. Those channels may be quantitatively important even when net disemployment effects are minor, as in \citet{Engbom2022}. Thus, a researcher with a strict rejection rule based on effects in the upper tail may reject a valid model.
	
	False negatives may also arise for two reasons. First, true upper-tail spillovers may be offset by estimator bias of the opposite sign, so the measured upper-tail effect remains small even though the causal effect is not. This may happen if, for example, the negative upper-tail bias illustrated in Table~\ref{tab:iss1} is combined with positive bias arising from measurement error, as discussed by \citet{Autor2016}. Second, misspecification biases may affect lower-tail outcomes or employment while leaving upper-tail estimates close to zero. In that case, the diagnostic would pass even though the estimator is unreliable for the outcomes of interest.

	\section{Regional minimum wages and instrumental variables \label{appendix:ams2016}}

	If the data include region-level minimum wage changes, one may consider an instrumental variables (IV) estimator that isolates that source of good variation. One approach is to use the prevailing institutional minimum wage (and its square) as an instrument for the effective minimum wage (and its square). In their pursuit of an effective minimum wage estimator robust to measurement error, \citet[][henceforth AMS]{Autor2016} propose an IV estimator along those lines but include a third instrument: the interaction of the log minimum wage with the average median wage in each region. Because it uses observed median wages in its construction, this third instrument may be subject to some of the abovementioned concerns.
	
	\begin{table}
		\small
		\caption{State-level minimum wages and instrumental variables approaches}
		\centering
		
		\label{tab:ams}
		\input{tab_ams.tex}	
		\begin{minipage}{1.0\textwidth}
			\footnotesize
			\justify
			\textbf{Notes:}      
			Each panel displays average results for 1,000 simulations, each with 200 regions and two periods, for different assumptions on the data-generating process (see the notes below Table~\ref{tab:iss1} for an explanation of the table's structure). Models in all panels are similar to those from Panel~B in Table~\ref{tab:iss2}, where there is a small intra-temporal correlation between location ($\mu_{r,t}$) and dispersion ($\sigma_{r,t}$) parameters. Panels~B and~C introduce region-specific minimum wages. They differ in the share of regions with a local minimum wage higher than the national minimum wage. ``Two instruments'' corresponds to regressions that employ the nominal minimum wage and its square as instruments for the effective minimum wage and its square. ``Three instruments (AMS)'' adds a third instrument following \citet{Autor2016}. See Appendix~\ref{appendix:simulation-details-ams} for details.
		\end{minipage}		
	\end{table}
	
	Table~\ref{tab:ams} presents the outcomes of simulations that incorporate region-specific minimum wages and implement alternative instrumental variables estimators. As with the previous simulations, the parameters of the data-generating process are tailored to mirror the US context; for more details, refer to Appendix~\ref{appendix:simulation-details-ams}. Panel A shows the baseline model, where there is a slight correlation between location and dispersion parameters (as in Table~\ref{tab:iss2}). Panels B and C introduce region-specific minimum wages that surpass the national minimum wage. The distinction between the panels is the proportion of regions with local minimum wages exceeding the national minimum wage. In Panels B and C, I present results not only for the regular effective minimum wage design but also for instrumental variables specifications, with either two or three instruments.
	
	From the table, three key findings emerge. First, the more variation derived from state-level minimum wages, the smaller the biases, even when using the ordinary least squares estimator. This is evident when comparing the ``Effective min. wage'' rows across panels, which gradually align with the corresponding ``True average causal effect'' rows. However, some bias persists. Second, the use of instrumental variables approaches significantly mitigates this bias. Third, the two-instrument estimator performs especially well, with biases comparable to or smaller than the AMS three-instrument version, albeit at the expense of precision.
	
	Therefore, the issues discussed in this section are an additional reason to adopt instrumental variables regressions in the style of AMS when the data include regional-level variation in minimum wage laws. Such estimators circumvent previously discussed biases by avoiding the potentially endogenous variation from median wages. This section also provides a rationale for avoiding the ``interaction'' instrument in AMS if the minimum wage instruments alone offer sufficient identifying variation.
	
	\section{Alternative Fraction-Affected/Gap Specifications \label{appendix:fa-other-designs}}
	
	Below, I discuss whether alternative specifications of the fraction-affected and gap designs can effectively solve the functional form misspecification biases discussed in Subsection~\ref{subsec:fa-gap-meas-error} of the paper.
	
	\paragraph{Treatment effect heterogeneity by observable subgroups.} Section 3.2.4 in \citet{Dube2024} provides a useful benchmark for thinking about the specifications considered here. They discuss fraction-affected, gap, and related exposure-based estimators for national minimum wages, and clearly explain why treatment-effect heterogeneity correlated with bite can bias those designs. They also discuss grouping estimators that allow the researcher to absorb heterogeneity along observable dimensions, such as age, industry, or location.
	
	Those approaches are valuable because many empirical threats in minimum wage studies do reflect observable differences across groups. The issue emphasized in this paper is different. Economic models of the minimum wage generally do not imply that treatment effects are constant within a sufficiently fine group. Instead, they imply that the marginal effect of a minimum wage increase depends on where the minimum wage lies relative to the latent wage distribution and to the relevant labor supply and labor demand schedules. Thus, even after controlling flexibly for observable group-level heterogeneity, the researcher still has to specify how bite maps into outcomes within those groups.
	
	The simulations in this paper make that point concrete. Even within a fine observable group, regions may differ in latent wage levels, wage dispersion, or the extent to which the national minimum wage lies near the margin where employment responses change sign. One way to see this is to reinterpret the simulations above as featuring region-industry combinations as the unit of analysis, rather than regions---under the assumption that workers are completely immobile across regions.\footnote{If workers do move across groups and the design uses group-specific exposure measures, then there may be additional misspecification arising from not capturing ``spillovers'' across groups. Evaluating this potential issue would require a more involved simulation model, and is thus left to future work.} Thus, the analysis here should be read as complementary to the guidance in \citet{Dube2024}: it studies the economic content of the exposure variable itself, quantifies the importance of nonlinear bite-to-outcome mappings, and evaluates which simple functional forms perform best in simulations calibrated to minimum wage applications.
	
	\paragraph{Binary treatment definitions.} The functional form misspecification issues I highlight are related to more general problems arising from treatment effect heterogeneity in difference-in-differences models with continuous treatment variables, recently discussed by \citet{Callaway2024}. These authors argue that, if there is a control group composed of entirely untreated units, then a binary difference-in-differences specification can recover an average treatment effect on treated units. In the minimum wage context, it may be hard to argue that any single region is entirely untreated by a national minimum wage without imposing strong structural assumptions on the data-generating process. That said, one plausible approach is to group regions that are less treated according to the chosen exposure measure and consider them to be the control group.
	
	In Appendix Table~\ref{tab:binary}, I show that rather than solving the problem, a binary treatment specification can suffer from more substantial biases than the baseline models. In this exercise, I split regions into treatment or control groups based on whether the initial median wages are below a given threshold. I choose thresholds such that either half or 90\% of the regions are in the treatment group. Consistent with the logic of \citet{Callaway2024}, biases are often smaller when 90\% of the sample is in the treatment group since it makes the ``zero'' group closer to being entirely untreated. However, there is a loss of precision, and significant biases remain.
	
	\paragraph{Bin-level estimators.} Another alternative specification discussed by \citet{Dube2024} is the bin-level estimator of \citet{Giupponi2024}. This estimator is an important recent contribution because it adapts frequency-distribution methods to a setting with a nationwide minimum wage. It differs from the baseline fraction-affected design in three main ways. First, it uses a binary regional comparison: regions in the top decile of estimated regional wage premia serve as the control group, while regions in the lower nine deciles are treated as more exposed to the national minimum wage. Second, this split is based not on raw regional wage levels but on location wage premia estimated from pre-reform wages, netting out observable worker and job characteristics and, in some specifications, using person effects. Third, it estimates impacts separately across wage or ``skill'' bins, where skill is defined using wages purged of the location wage premium. Thus, rather than imposing a single linear relationship between an exposure measure and an aggregate outcome, the estimator flexibly traces where effects appear in the wage distribution.
	
	This flexibility is valuable, but it addresses a somewhat different problem from the one studied here. The bin-level design is especially useful for measuring missing and excess mass around the minimum wage, detecting spillovers above the minimum, and checking whether estimated effects appear far from the policy-relevant part of the wage distribution. However, its baseline regional comparison is still binary: lower-wage-premium regions are compared to high-wage-premium regions. It therefore does not directly estimate how aggregate effects vary across multiple levels of regional bite. Moreover, because comparable skill bins are constructed by shifting wage distributions according to regional wage premia, the design relies on restrictions on how regional wage distributions differ beyond their location. If regions differ systematically in dispersion or other shape parameters, the mapping from treated wage bins to control skill bins may itself be misspecified. For these reasons, I view the estimator of \citet{Giupponi2024} as complementary to the analysis here: it provides a powerful way to study the distributional incidence of nationwide minimum wages, while the simulations above focus on functional-form restrictions in the mapping from regional exposure measures to aggregate outcomes.
	
	\paragraph{Instrumental variable approaches.} A final potential solution would be to use the regressor in the fraction-affected design as an instrument for the regressor based on the gap measure, or vice versa. This strategy could be justified if each of those regressors were equal to an unobserved metric of propensity to be affected by the minimum wage plus some random noise and if those noise terms are uncorrelated with one another. Appendix Table~\ref{tab:fa_gap_iv} reports the results of using such strategies. They have no impact on the estimates compared to the basic OLS estimator.

	\section{Simulations of the canonical model of labor demand \label{appendix:simulation-details-canonical}}
	
	\subsection{Model description}

	\paragraph{Overview.} Consider a competitive economy with two types of labor: low- or high-skilled. Each worker is characterized by their type and by an idiosyncratic productivity component. The log wage gap between the two labor types is pinned down by the ratio of marginal products of labor. Marginal products of labor are, in turn, determined by a representative production function with constant returns and constant elasticity of substitution.
	
	Now, consider the implications of a binding minimum wage in that model. The representative firm will not employ workers whose productivity is below the minimum wage. Thus, one can think of this model as one where each worker has a latent log wage given by the sum of a type-specific equilibrium location parameter and an idiosyncratic productivity component. The observed wage distribution is the truncated version of that latent mixture distribution.
	
	The critical difference between this model and the Normal-markdown model used in other simulations in the paper is that the shape of the latent log wage distribution responds to the minimum wage. As the minimum wage causes more disemployment for low-skill workers, returns to skill are expected to fall. These equilibrium responses generate wage spillovers for other low-skill workers and attenuate the minimum wage's disemployment effects.
	
	\paragraph{Mathematical description.} The only production factors are skilled ($i=1$) and unskilled ($i=2$) labor, both of which have inelastic supply. A representative firm produces the numeraire good in the economy using a constant elasticity of substitution (CES) production function:
	\begin{equation*}
		F(L_1,L_2)=\left[\alpha {L_1}^{\frac{E-1}{E}}+(1-\alpha) {L_2}^{\frac{E-1}{E}}\right]^{\frac{E}{E-1}}
	\end{equation*}
	The measure of workers is normalized to one, and the region-specific share of skilled workers is $s_r$. Each worker has an idiosyncratic productivity component $e$, which has a Normal distribution with mean zero and standard deviation $D$. Conditional on type $i$, a worker's latent log wage is $p_{i,r,t}+e$, where $p_{i,r,t}$ is a type-specific equilibrium log wage location parameter. Workers whose latent log wage falls below the log minimum wage $mw_t$ are not employed by the representative firm. The equilibrium location parameters are given by the solution to a system of two equations:
	\begin{align*}
		p_{i,r,t} &= \log F_i\left(s_r \mathcal{E}(p_{1,r,t},mw_t),(1-s_r)\mathcal{E}(p_{2,r,t},mw_t)\right) \qquad i\in\{1,2\}
		\\
		\text{where } \mathcal{E}(p, mw) &= \int_{mw - p}^{\infty} \exp(p+e) \frac{1}{D}\phi\left(\frac{e}{D}\right)de
		\\
		\text{and } F_i(L_1, L_2) &= \frac{d F(L_1, L_2)}{d L_i}
	\end{align*}
	In the expressions above, $\phi$ denotes the density of a standard Normal distribution. The function $\mathcal{E}(\cdot)$ calculates the aggregate input contributed by workers of a given type, taking into account the disemployment effects of the minimum wage.
	
	The resulting employment-to-population ratio in a given region and period is given by:
	\begin{equation*}
		emp_{r,t} = s_r \left[1-\Phi\left(\frac{mw_t - p_{1,r,t}}{D}\right)\right] + (1-s_r) \left[1-\Phi\left(\frac{mw_t - p_{2,r,t}}{D}\right)\right],
	\end{equation*}
	Define $q_{i,r,t}=1-\Phi\left((mw_t-p_{i,r,t})/D\right)$ as the employment rate for type $i$, and define the employment shares
	\[
		\omega_{1,r,t}=\frac{s_r q_{1,r,t}}{emp_{r,t}}, \qquad 
		\omega_{2,r,t}=\frac{(1-s_r) q_{2,r,t}}{emp_{r,t}}.
	\]
	The corresponding cumulative distribution function for log wages among employed workers is:
	\begin{equation*}
		G_{r,t}\left(w\right) = \omega_{1,r,t} \frac{\Phi\left(\frac{w - p_{1,r,t}}{D}\right)-\Phi\left(\frac{mw_t - p_{1,r,t}}{D}\right)}{q_{1,r,t}}
		+ \omega_{2,r,t} \frac{\Phi\left(\frac{w - p_{2,r,t}}{D}\right)-\Phi\left(\frac{mw_t - p_{2,r,t}}{D}\right)}{q_{2,r,t}} \qquad \text{for } w \ge mw_{t}
	\end{equation*}
	where $\Phi$ is the cumulative distribution function of a standard Normal. The mixture weights $\omega_{i,r,t}$ are the shares of each type among employed workers, not population shares.

	\subsection{Calibration}
	
	I also use US Current Population Survey data to calibrate the simulations. Using the same sample restrictions described in the previous subsection and data for 1989, I define a worker as belonging to the skilled group $i=1$ if they have at least four years of college education. Then, I calculate the mean and standard deviation of log wages by skill group for each state and the share of workers in each group.
	
	On the labor supply side, the (unweighted) average of the share of skilled workers across states is 0.224, and the standard deviation is 0.047. Then, in the simulations, I draw the share of skilled workers in each region from a Normal distribution with the corresponding mean and standard deviation, trimming the draws so that the share of each worker type can never be below 0.01. The average standard deviation of log wages within states is close to 0.5 for both educational groups. Thus, I set $D=0.5$.
	
	On the demand side, the mean log wage gap between skilled and unskilled workers is also close to 0.5, corresponding to a wage ratio of approximately $\exp(0.5)$. Thus, I choose the $\alpha$ parameter such that $p_{1,r,t}-p_{2,r,t}$ is approximately 0.5 in an equilibrium of the model with the share of skilled workers equal to the cross-state average, and at the lowest initial value of the minimum wage used (see below). That corresponds to $\alpha=0.563$ when the elasticity of substitution used in the simulation is $E=3$, and $\alpha=0.493$ for $E=1.4$.
	
	The simulations are run for six scenarios. They combine the two values for the elasticity of substitution in production and three initial values of the minimum wage: -2.2, -1.8, and -1.5. The corresponding initial employment-to-population ratios given the average share of skilled workers are around 0.995, 0.966, and 0.896, respectively.
	
	\subsection{Results}

	\begin{table}
		\small
		\caption{Canonical Model of Labor Demand, Fraction-Affected and Gap Designs}
		\centering
		
		\label{tab:canonical-fa-gap}
		\input{tab_canonical_fa_gap.tex}	
		\begin{minipage}{1.0\textwidth}
			\footnotesize
			\justify
			\textbf{Notes:}      
			This table is similar in structure to Table~\ref{tab:fa-gap-meas-error}, but the simulation is based on the Canonical model instead of the Normal-markdown model. Panels differ in the initial level of the minimum wage and the elasticity of substitution between skill levels in the Canonical model. See Appendix~\ref{appendix:simulation-details-canonical} for details.
		\end{minipage}		
	\end{table}

	\begin{table}
		\small
		\caption{Canonical model of labor demand, Effective Minimum Wage design}
		\centering
		
		\label{tab:canonical-eff}
		\input{tab_canonical_kaitz.tex}	
		\begin{minipage}{1.0\textwidth}
			\footnotesize
			\justify
			\textbf{Notes:}      
			This table is similar in structure to Table~\ref{tab:iss1}, but the simulation is based on the Canonical model instead of the Normal-markdown model. Panels differ in the initial level of the minimum wage and the elasticity of substitution between skill levels in the Canonical model. See Appendix~\ref{appendix:simulation-details-canonical} for details.
		\end{minipage}		
	\end{table}

	Table~\ref{tab:canonical-fa-gap} reports simulations that assess the effectiveness of the fraction-affected and gap designs using the Canonical model as the data-generating process. In those simulations, regions only differ in the initial share of workers in the high-skill group, and the only time-varying factor is the minimum wage. In that sense, they parallel Table~\ref{tab:fa-gap-meas-error} above, illustrating ideal scenarios for the fraction-affected and gap designs. Also, as specified above, I report results for different model specifications, varying the bindingness of the initial minimum wage and the elasticity of substitution between low- and high-skill workers.
	
	The takeaways from that exercise are the same as those from Table~\ref{tab:fa-gap-meas-error}. The predicted average causal effects estimated using those approaches do not always equal the true ones, and those misspecification biases are more significant when the minimum wage is more binding.
	
	Table~\ref{tab:canonical-eff} evaluates the effective minimum wage design using the same simulations. The baseline effective minimum wage specification displays severe biases when both region and time fixed effects are included. That is because the model does not include region-specific wage shocks, the source of ``good'' variation in that design. For that reason, I also report estimates for the model without region fixed effects \citep[the baseline specification in][]{Lee1999}, which exploit differences in wage levels instead of changes. The model is almost ideal for that design in that there are no sources of ``endogeneity'' (like systematically lower latent employment levels in low-wage regions) and because the dispersion of efficiency units for both low- and high-skill workers is the same (which reduces the correlation between ``location'' and ``dispersion'' parameters of latent log wages). Even so, that model displays significant biases in some specifications.

	\section{Additional discussion on parallel trends \label{appendix:regional-convergence}}
			
			The discussion below complements Section~\ref{sec:trends-dispersion} in the paper, which discusses potential deviations of the parallel trends assumption in the national minimum wage context.			
			
			\begin{table}
					\small
					\caption{Sensitivity of the Fraction-Affected Design}
					\centering
					
					\label{tab:fa_td}
					\input{tab_fa_td.tex}	
					\begin{minipage}{1.0\textwidth}
							\footnotesize
							\justify
							\textbf{Notes:}    
							All panels illustrate scenarios where the national minimum wage increases by 20 log points. Each panel displays average results for 1,000 simulations with 200 regions and two periods. For each outcome, the numbers correspond to the mean true ATE across simulations, the mean estimates of causal effects based on the regressions listed on the left, and the average standard error associated with the estimates (in parentheses, clustered at the region level). Panel~A includes only permanent differences in location parameters. Panel~B adds mean-reverting location shocks. Panel~C adds time-varying dispersion heterogeneity. Panel~D allows average dispersion to fall over time. Table~\ref{tab:simparams-nm-fagap} reports the corresponding meta-parameters.  
						\end{minipage}		
				\end{table}

			The comparison between Panels A and B in Table~\ref{tab:fa_td} illustrates biases arising from mean-reverting regional shocks in the Normal-markdown simulation model. It reports results for the fraction-affected design; Table~\ref{tab:gap_td} shows similar results for the gap design. In Panel~A, regions only differ in a time-invariant location parameter $\mu_{r}$. Panel~B introduces time-specific location parameters $\mu_{r,t}$, with each region's initial and final parameters being jointly Normal with a correlation of 0.894 (equal to the correlation between state-level mean log wages in the US for 1989 and 2004). As expected, biases on wage effects become substantially more positive, because regions with particularly unlucky draws of $\mu_{r,t}$ in the initial period are more treated but also more likely to have higher growth in that parameter.
			
			Panel~C further explores this issue by including time-varying heterogeneity in dispersion parameters $\sigma_{r,t}$ between regions. Those parameters are assumed to be independent of the location parameters but have an autocorrelation of 0.456 between periods. That magnifies the positive bias in the lower tail, though it reduces it in the upper tail.
			
			Fortunately, biases arising from regression to the mean can be detected with tests for differential pre-trends if the context allows such tests. Table~\ref{tab:fa_td_placebo} illustrates this concept using a placebo exercise that parallels Table~\ref{tab:fa_td}. The fraction-affected measure is calculated as if the national minimum wage would increase by 0.2, but in reality, the minimum wage remains stable. That placebo exercise shows positive employment effects that have about the same size as the biases discussed above.
			
			In the main text, we stated that tests for parallel pre-trends will only detect biases from regression to the mean if the exposure measure is calculated using only a single pre-treatment year, as opposed to being an average of exposure in all pre-treatment periods. Below, we formalize this argument using two statistical models. The first has an arbitrary number of pre-treatment periods and the observed exposure measure is subject to idiosyncratic, i.i.d. shocks. In the second model, the idiosyncratic shocks to the exposure measure can be persistent, but we only consider three pre-treatment periods for simplicity.
			
			\paragraph{Inefficacy of pre-trend checks for regression to the mean when the treatment variable is averaged over all pre-treatment periods.} 			
			Suppose there are $m \geq 2$ pre-treatment periods, indexed by $t \in \{-m,\ldots,-1\}$, and one post-treatment period $t=0$. Note that this is a change of time notation relative to the remainder of the paper. We use this alternative notation in this model to make it more consistent with event-study specifications in applied microeconometrics.
			
			Let the region-level treatment intensity measure that would be observed absent any minimum wage change be
			\begin{equation}
				E_{r,t}=E_r^*+u_{r,t},
			\end{equation}
			where $E_r^*$ is a time-invariant component and $u_{r,t}$ is a temporary region-level shock. Assume
			\begin{equation}
				\mathbb{E}[u_{r,t}\mid E_r^*]=0,\qquad \operatorname{Var}(u_{r,t})=\sigma_u^2>0,\qquad \operatorname{Cov}(u_{r,t},u_{r,s})=0\text{ for }t\neq s,
			\end{equation}
			and $\operatorname{Var}(E_r^*)>0$. Suppose untreated outcomes satisfy
			\begin{equation}
				y_{r,t}(0)=\alpha_r+ \delta_t -\psi u_{r,t},\qquad \psi>0. \label{eq:outcome-model}
			\end{equation}
			Thus, a temporary shock that raises the exposure measure also depresses the untreated outcome. For example, a high $u_{r,t}$ could represent positive measurement error in regional wage statistics, which would imply a positive bias on measured exposure and a negative bias on wage outcomes. Alternatively, $u_{r,t}$ could represent a real, but idiosyncratic, negative productivity shock at the regional level, which would also depress wages (mechanically increasing exposure measures) and reduce employment (the outcome $y_{r,t}$).
			
			For any time-invariant regressor $T_r$, consider the population event-study coefficients $\beta_k(T)$ from
			\begin{equation}
				y_{r,t}(0)=\alpha_r+\delta_t+\sum_{k\in\{-m,\ldots,-2,0\}}\beta_k(T)\,T_r\,\mathbf{1}\{t=k\}+\varepsilon_{r,t},
				\label{eq:event-study}
			\end{equation}
			where $t=-1$ is the omitted period.
			
			\textbf{Proposition.} Under the assumptions above,
			\begin{equation}
				\beta_k(T)=\psi\,\frac{\operatorname{Cov}(T_r,u_{r,-1})-\operatorname{Cov}(T_r,u_{r,k})}{\operatorname{Var}(T_r)}.
				\label{eq:r2m-betak}
			\end{equation}
			Consequently:
			\begin{enumerate}
				\item If $T_r=E_{r,-1}$, then for every $k\in\{-m,\ldots,-2,0\}$,
				\begin{equation}
					\beta_k(T)=\psi\,\frac{\sigma_u^2}{\operatorname{Var}(E_r^*)+\sigma_u^2}>0.
					\label{eq:part1}
				\end{equation}
				Hence a placebo event-study displays a mechanical dip at $t=-1$: all other coefficients are positive relative to the omitted year.
				
				\item If $T_r=\bar E_r\equiv m^{-1}\sum_{j=1}^{m} E_{r,-j}$, then
				\begin{equation}
					\beta_k(T)=0\ \text{for every } k\in\{-m,\ldots,-2\},\qquad
					\beta_0(T)=\psi\,\frac{\sigma_u^2/m}{\operatorname{Var}(E_r^*)+\sigma_u^2/m}>0.
					\label{eq:part2}
				\end{equation}
				Thus averaging the treatment intensity over all pre-treatment years attenuates the post-treatment bias but eliminates it from the placebo pre-trends. In an actual post-treatment regression with a genuine causal effect, this bias term would simply be added to the causal coefficient.
			\end{enumerate}
			
			\medskip
			
			\textit{Proof.} The proof has four steps: (i) reduce the event-study coefficient $\beta_k(T)$ to a cross-sectional regression by a Frisch--Waugh--Lovell (FWL)/differencing argument; (ii) substitute the model for $Y_{r,t}(0)$ to obtain \eqref{eq:r2m-betak}; (iii) plug in the two specific choices of $T_r$; and (iv) show that averaging attenuates the post-treatment bias.
			
			\medskip
			
			\textit{Step 1: Reduction to a cross-sectional difference-in-means.}
			We show that for every $k\in\{-m,\ldots,-2,0\}$,
			\begin{equation}
				\beta_k(T)=\frac{\operatorname{Cov}\bigl(T_r,\,Y_{r,k}(0)-Y_{r,-1}(0)\bigr)}{\operatorname{Var}(T_r)}.
				\label{eq:fwl-claim}
			\end{equation}
			This is the difference-in-differences interpretation of the event-study coefficient, but because the regression \eqref{eq:event-study} contains both region and time fixed effects together with several event-study interactions, the identity warrants a brief verification.
			
			Fix $k\ne -1$ and consider the problem of minimizing the sum of squared errors in \eqref{eq:event-study}. Because $T_r\,\mathbf{1}\{t=k'\}$ vanishes whenever $t\ne k'$, the population first-order condition (FOC) for $\beta_k(T)$ in \eqref{eq:event-study} contributes only at $t=k$, and reads
			\begin{equation}
				\mathbb{E}\bigl[\bigl(Y_{r,k}(0)-\alpha_r-\delta_k-\beta_k(T)\,T_r\bigr)\,T_r\bigr]=0.
				\label{eq:foc-betak}
			\end{equation}
			The FOC for $\delta_t$ at any $t$ gives $\delta_t=\mathbb{E}[Y_{r,t}(0)]-\mathbb{E}[\alpha_r]-\beta_t(T)\,\mathbb{E}[T_r]\,\mathbf{1}\{t\ne -1\}$, with the convention $\beta_{-1}(T):=0$. Substituting $\delta_k$ into \eqref{eq:foc-betak} and rearranging,
			\begin{equation}
				\beta_k(T)\operatorname{Var}(T_r)=\operatorname{Cov}\bigl(Y_{r,k}(0)-\alpha_r,\,T_r\bigr).
				\label{eq:beta-form-1}
			\end{equation}
			
			Next, the FOC for $\alpha_r$ yields $\alpha_r=\bar Y_{r,\cdot}(0)-\bar\delta-\tfrac{T_r}{m+1}\sum_{t\ne -1}\beta_t(T)$, where $\bar Y_{r,\cdot}(0)$ averages over the $m+1$ periods. Hence
			\begin{equation}
				\operatorname{Cov}(\alpha_r,T_r)
				=\operatorname{Cov}\bigl(\bar Y_{r,\cdot}(0),\,T_r\bigr)-\frac{S}{m+1}\operatorname{Var}(T_r),\qquad S:=\sum_{t\ne -1}\beta_t(T).
				\label{eq:alpha-cov}
			\end{equation}
			Summing \eqref{eq:beta-form-1} over $k\ne -1$ (and noting $\sum_{k\ne -1}Y_{r,k}(0)=(m+1)\bar Y_{r,\cdot}(0)-Y_{r,-1}(0)$) gives
			\begin{equation}
				S\operatorname{Var}(T_r)
				=\operatorname{Cov}\bigl((m+1)\bar Y_{r,\cdot}(0)-Y_{r,-1}(0)-m\,\alpha_r,\,T_r\bigr).
			\end{equation}
			Combining with \eqref{eq:alpha-cov} and solving for $S$ yields
			\begin{equation}
				\frac{S}{m+1}\operatorname{Var}(T_r)=\operatorname{Cov}\bigl(\bar Y_{r,\cdot}(0)-Y_{r,-1}(0),\,T_r\bigr).
				\label{eq:S-form}
			\end{equation}
			Substituting \eqref{eq:alpha-cov} and \eqref{eq:S-form} back into \eqref{eq:beta-form-1},
			\begin{equation}
				\beta_k(T)\operatorname{Var}(T_r)
				=\operatorname{Cov}\bigl(Y_{r,k}(0)-\bar Y_{r,\cdot}(0),\,T_r\bigr)+\operatorname{Cov}\bigl(\bar Y_{r,\cdot}(0)-Y_{r,-1}(0),\,T_r\bigr)
				=\operatorname{Cov}\bigl(Y_{r,k}(0)-Y_{r,-1}(0),\,T_r\bigr),
			\end{equation}
			which is \eqref{eq:fwl-claim}.
			
			\medskip
			
			\textit{Step 2: General formula for $\beta_k(T)$.}
			From the model (Equation~\ref{eq:outcome-model}),
			\begin{equation}
				Y_{r,k}(0)-Y_{r,-1}(0)=(\delta_k-\delta_{-1})-\psi\bigl(u_{r,k}-u_{r,-1}\bigr).
			\end{equation}
			The first term is a constant in $r$ and contributes nothing to $\operatorname{Cov}(\cdot,T_r)$. Therefore \eqref{eq:fwl-claim} becomes
			\begin{equation}
				\beta_k(T)=\frac{-\psi\,\operatorname{Cov}(T_r,u_{r,k})+\psi\,\operatorname{Cov}(T_r,u_{r,-1})}{\operatorname{Var}(T_r)}
				=\psi\,\frac{\operatorname{Cov}(T_r,u_{r,-1})-\operatorname{Cov}(T_r,u_{r,k})}{\operatorname{Var}(T_r)},
			\end{equation}
			which is \eqref{eq:r2m-betak}.
			
			\medskip
			
			\textit{Step 3: The two choices of $T_r$.}
			For both choices, $T_r=E_r^*+\xi_r$ where $\xi_r$ is a linear combination of pre-treatment shocks. Since $\operatorname{Cov}(E_r^*,u_{r,t})=0$ for every $t$,
			\begin{equation}
				\operatorname{Cov}(T_r,u_{r,t})=\operatorname{Cov}(\xi_r,u_{r,t})\quad\text{and}\quad\operatorname{Var}(T_r)=\operatorname{Var}(E_r^*)+\operatorname{Var}(\xi_r).
				\label{eq:Tr-decomp}
			\end{equation}
			
			\emph{Case 1: $T_r=E_{r,-1}$, so $\xi_r=u_{r,-1}$.}
			Using the orthogonality $\operatorname{Cov}(u_{r,s},u_{r,t})=\sigma_u^2\,1\{s=t\}$,
			\begin{equation}
				\operatorname{Cov}(T_r,u_{r,-1})=\sigma_u^2,\qquad \operatorname{Cov}(T_r,u_{r,k})=0\quad\text{for every }k\in\{-m,\ldots,-2,0\}.
			\end{equation}
			Also $\operatorname{Var}(\xi_r)=\sigma_u^2$. Substituting into \eqref{eq:r2m-betak} via \eqref{eq:Tr-decomp} gives \eqref{eq:part1}.
			
			\emph{Case 2: $T_r=\bar E_r$, so $\xi_r=m^{-1}\sum_{j=1}^{m}u_{r,-j}$.}
			For any $t$,
			\begin{equation}
				\operatorname{Cov}(\xi_r,u_{r,t})=\frac{1}{m}\sum_{j=1}^{m}\operatorname{Cov}(u_{r,-j},u_{r,t})
				=\frac{\sigma_u^2}{m}\,1\{t\in\{-m,\ldots,-1\}\}.
				\label{eq:cov-bar}
			\end{equation}
			For $\operatorname{Var}(\xi_r)$, the cross-covariance terms vanish, so
			\begin{equation}
				\operatorname{Var}(\xi_r)=\frac{1}{m^2}\sum_{j=1}^{m}\operatorname{Var}(u_{r,-j})=\frac{\sigma_u^2}{m}.
			\end{equation}
			For every pre-period $k\in\{-m,\ldots,-2\}$, \eqref{eq:cov-bar} gives $\operatorname{Cov}(T_r,u_{r,k})=\sigma_u^2/m=\operatorname{Cov}(T_r,u_{r,-1})$, so $\beta_k(T)=0$ by \eqref{eq:r2m-betak}. For the post-treatment period, \eqref{eq:cov-bar} gives $\operatorname{Cov}(T_r,u_{r,0})=0$, while $\operatorname{Cov}(T_r,u_{r,-1})=\sigma_u^2/m$. Therefore
			\begin{equation}
				\beta_0(T)=\psi\,\frac{\sigma_u^2/m}{\operatorname{Var}(E_r^*)+\sigma_u^2/m},
			\end{equation}
			which is \eqref{eq:part2}.
			
			\medskip
			
			\textit{Step 4: Attenuation.}
			The function $f(x)=x/(\operatorname{Var}(E_r^*)+x)$ is strictly increasing on $(0,\infty)$, since $f'(x)=\operatorname{Var}(E_r^*)/(\operatorname{Var}(E_r^*)+x)^2>0$. Because $m\geq 2$ implies $\sigma_u^2/m<\sigma_u^2$,
			\begin{equation}
				\frac{\sigma_u^2/m}{\operatorname{Var}(E_r^*)+\sigma_u^2/m}<\frac{\sigma_u^2}{\operatorname{Var}(E_r^*)+\sigma_u^2},
			\end{equation}
			so averaging attenuates, but does not remove, the post-treatment bias. \hfill$\square$
			
			\paragraph{Allowing for persistent shocks to the exposure measure.}
			The exact disappearance of placebo pre-trends in part 2 uses the assumption that the shocks $u_{r,t}$ are transitory. If the shocks are mean-reverting but serially correlated, averaging no longer makes the pre-treatment coefficients exactly zero; however, it still weakens them relative to the post-treatment bias, so the diagnostic is still not very effective.
			
			To illustrate, suppose there are three pre-treatment years and that $u_{r,t}$ follows a stationary AR(1) with autocorrelation $\rho\in(0,1)$, so that $\operatorname{Cov}(u_{r,t},u_{r,s})=\rho^{|t-s|}\sigma_u^2$. Take $T_r=(E_{r,-3}+E_{r,-2}+E_{r,-1})/3$. Then for any $t$,
			\begin{equation}
				\operatorname{Cov}(T_r,u_{r,t})=\frac{\sigma_u^2}{3}\bigl(\rho^{|t+1|}+\rho^{|t+2|}+\rho^{|t+3|}\bigr).
			\end{equation}
			Evaluating at the four periods of interest gives
			\begin{equation}
				\operatorname{Cov}(T_r,u_{r,-1})=\operatorname{Cov}(T_r,u_{r,-3})=\frac{\sigma_u^2}{3}(1+\rho+\rho^2),
			\end{equation}
			\begin{equation}
				\operatorname{Cov}(T_r,u_{r,-2})=\frac{\sigma_u^2}{3}(1+2\rho),\qquad
				\operatorname{Cov}(T_r,u_{r,0})=\frac{\sigma_u^2}{3}\rho(1+\rho+\rho^2).
			\end{equation}
			Substituting into \eqref{eq:r2m-betak} and using $(1+\rho+\rho^2)(1-\rho)=1-\rho^3$,
			\begin{equation}
				\beta_{-3}(T)=0,\qquad
				\beta_{-2}(T)=-\psi\,\frac{\rho(1-\rho)\sigma_u^2}{3\operatorname{Var}(T_r)},\qquad
				\beta_0(T)=\psi\,\frac{(1-\rho^3)\sigma_u^2}{3\operatorname{Var}(T_r)}>0.
			\end{equation}
			Pre-trends will look mostly parallel, given the zero coefficient for $t=-3$. One may be concerned that this zero is an artifact of the model's restriction to three pre-periods, and that applied researchers would in fact detect a positive pre-trend based on the growth of estimated coefficients from $t=-2$ to $t=-1$. However, the implied magnitudes matter. Specifically, the growth from $t=-2$ to $t=-1$ is significantly smaller than that for the post-treatment bias (from $t=-1$ to $t=0$). To see this, note that the ratio of the post-treatment coefficient for $t=0$ to the magnitude of the pre-trend coefficient for $t=-2$ is equal to $\frac{1-\rho^3}{\rho(1-\rho)}$. That ratio is strictly decreasing in $(0,1)$ but always above 3. So even with a very persistent shock, the event-study graph would still show a post-treatment effect of much larger magnitude than the differential pre-trend. As $\rho\to 0$, the ratio grows to infinity, as we return to the benchmark case with idiosyncratic shocks.
			
	\section{Details on the \citet{Haanwinckel2025} model \label{appendix:sdif}}

	In \citet{Haanwinckel2025}, I develop an imperfectly competitive labor market model with worker and firm heterogeneity to study the impact of rising schooling achievement, labor demand shocks, and the federal minimum wage on Brazilian labor markets. The predicted impacts of the federal minimum wage incorporate several channels discussed in previous literature: disemployment of very low productivity workers (i.e., truncation of the latent log wage distribution); mechanical wage increases for some workers (i.e., censoring); increases in employment for some worker types due to the presence of monopsony power; reallocation of workers to firms with higher wage premiums (as in \citealt{Dustmann2021}); changes in returns to skill, both in the aggregate economy and differential changes within firms; effects on the composition of firms operating in the economy (as in \citealt{Aaronson2018}); and pass-through effects to consumer prices (as in \citealt{Harasztosi2018}).

	The magnitudes of those mechanisms are governed by parameters estimated with a simultaneous-equation nonlinear least squares procedure. The estimator targets several endogenous outcomes: employment rates, measures of wage inequality between and within educational groups, the contribution of firm wage premiums to inequality, and a measure of sorting of high-wage workers to high-wage firms \citep[the latter two being measured using the methodology developed by][]{Kline2018}. To account for potential confounders on the labor supply side, the model directly incorporates the role of education in affecting labor market outcomes such as the distribution of, and returns to, skill. The model also includes three types of unobserved labor demand shifters at the regional level, which are partly determined by initial sectoral shares in agriculture and manufacturing. These labor demand shocks absorb the influence of important transformations mentioned above, such as trade liberalization and technical change.
	
	Table~\ref{tab:sdif-fit} reports the quality of fit. The first panel shows that the model can replicate levels and changes in several wage inequality measures, both between and within educational groups. Note that the model not only matches average levels, but also has predictive power over outcomes at the region level. This is shown in the last column, which reports the $R^2$ metric for the sample of 151 microregions and two time periods.

	\begin{table}
		\small
		\caption{Quality of fit of the task-based, monopsonistic model}
		\centering
		
		\begin{tabular}{lccccc}
			\toprule 
			& \multicolumn{2}{c}{Data} & \multicolumn{2}{c}{Model} & $R^2$  \\
			& 1998 & 2012 & 1998 & 2012 & Model  \\
			Moments & (1) & (2) & (3) & (4) & (5)  \\
			\midrule 
			\multicolumn{6}{l}{\textit{Wage inequality measures}} \\
			\hspace{0.5cm} Secondary / less than secondary & 0.498 & 0.168 & 0.478 & 0.168 & 0.755  \\
			\hspace{0.5cm} Tertiary / secondary & 0.965 & 1.038 & 0.981 & 0.954 & 0.127 \\
			\hspace{0.5cm} Within less than secondary & 0.41 & 0.241 & 0.401 & 0.233 & 0.607  \\
			\hspace{0.5cm} Within secondary & 0.684 & 0.355 & 0.67 & 0.331 & 0.848 \\
			\hspace{0.5cm} Within tertiary & 0.702 & 0.624 & 0.701 & 0.637 & 0.139  \\
			\multicolumn{6}{l}{\textit{Two-way fixed effects decomposition}} \\
			\hspace{0.5cm} Variance of establishment effects & 0.126 & 0.054 & 0.126 & 0.04 & 0.586  \\
			\hspace{0.5cm} Covariance of worker and estab. effects & 0.052 & 0.046 & 0.056 & 0.059 & 0.374  \\ 
			\multicolumn{6}{l}{\textit{Formal employment rates}} \\
			\hspace{0.5cm} Less than secondary & 0.266 & 0.337 & 0.266 & 0.335 & 0.953  \\
			\hspace{0.5cm} Secondary & 0.435 & 0.508 & 0.435 & 0.508 & 1.0  \\
			\hspace{0.5cm} Tertiary & 0.539 & 0.629 & 0.539 & 0.63 & 0.89 \\
			\multicolumn{6}{l}{\textit{Minimum wage bindingness}} \\
			\hspace{0.5cm} Log min. wage - mean log wage & -1.418 & -0.922 & -1.418 & -0.922 & 1.0  \\
			\hspace{0.5cm} Share < log min. wage + 0.05 & 0.031 & 0.053 & 0.046 & 0.084 & 0.528  \\
			\hspace{0.5cm} Share < log min. wage + 0.30 & 0.086 & 0.212 & 0.107 & 0.201 & 0.873  \\
			\bottomrule 
		\end{tabular}
		
		\label{tab:sdif-fit} 		
		\begin{minipage}{1.0\textwidth}
			\footnotesize
			\justify
			\textbf{Notes:} This table is adapted from \citet{Haanwinckel2025}. The columns labeled ``Data'' report statistics calculated at the \textit{microregion} level using Brazilian data. ``Model'' corresponds to the model fit using those data as input. Columns (1) through (4) report averages for all regions for each of the two years, using region weights based on total formal employment. Column (5) reports the usual $R^2$ metric \linebreak $r^2_{e}=1-\left[\sum_{t=1}^2 \sum_{r=1}^{151}s_r\left(Y_{e,r,t}-\hat{Y}_{e,r,t}\right)^2\right]/\left[\sum_{t=1}^2 \sum_{r=1}^{151}s_r\left(Y_{e,r,t}-\bar{Y}_{e}\right)^2\right]$, where $e$ indexes the specific target moment, $\hat{Y}_{e,r,t}$ is the model prediction, and $\bar{Y}_{e}$ is the sample average using the region weights employed in the estimation of the model. See \citet{Haanwinckel2025} for details.
		\end{minipage}
	\end{table}

	The structural model emphasizes cross-firm wage differentials for different workers as a key component of wage dispersion, and a possible margin through which minimum wages can affect labor market outcomes. But those channels are only captured effectively if the model can replicate empirical measures of ``firm wage premiums.'' Specifically, I estimate statistical models of firm-level wage premiums at the region-time level using the \citet{Kline2018} approach. From those models, I obtain the bias-corrected variance of firm wage effects and the covariance between worker and establishment effects. I then compare those empirical estimates to the simulated versions of those statistics from the structural model. The second panel in Table~\ref{tab:sdif-fit} shows that, again, the model goes beyond successfully matching averages; it also has predictive power in the cross-section.
	
	The other two panels report high quality of fit along two other margins. The first is formal employment rates by educational group. The second corresponds to different measures of minimum wage bindingness at the regional level. Matching bindingness metrics is particularly important for the exercises in this paper, as the predicted causal effects of a change in the minimum wage depend crucially on observed bindingness in both periods.
	
	\clearpage

	\section{Additional Tables and Figures \label{appendix:additional-tables}}
	
%
%
%
%
%
%
%
%
	
	\begin{table}[ht!]
		\small
		\caption{Difference-in-differences with binary treatment}
		\centering
		
		\label{tab:binary}
		\input{tab_new_corr_meas_error_binary.tex}	
		\begin{minipage}{1.0\textwidth}
			\footnotesize
			\justify
			\textbf{Notes:}  This table is analogous to Table~\ref{tab:fa-gap-meas-error}, except that it reports results for a difference-in-differences estimator based on a binary version of treatment. Treated status is based on initial median wages being below some simulation-specific threshold, chosen such that the share of treated units corresponds to the desired level.
		\end{minipage}		
	\end{table}

	\begin{table}
		\small
		\caption{Difference-in-differences with instrumental variables}
		\centering
		
		\label{tab:fa_gap_iv}
		\input{tab_new_corr_meas_error_fagap_inst.tex}	
		\begin{minipage}{1.0\textwidth}
			\footnotesize
			\justify
			\textbf{Notes:}  This table is analogous to Table~\ref{tab:fa-gap-meas-error}, except that it reports results for a difference-in-differences estimator where the main regressor (the interaction between one treatment intensity variable and an indicator for the post period) is instrumented with an alternative treatment intensity variable interacted with the dummy for the post period.
		\end{minipage}		
	\end{table}

	\begin{table}
		\small
		\caption{Difference-in-differences with quadratic treatment intensity}
		\centering
		
		\label{tab:fa_gap_quad}
		\input{tab_new_corr_meas_error_fagap_quad.tex}	
		\begin{minipage}{1.0\textwidth}
			\footnotesize
			\justify
			\textbf{Notes:}  This table is analogous to Table~\ref{tab:fa-gap-meas-error}, except that it reports results for a difference-in-differences estimator that allows treatment effects to vary with the treatment intensity through a quadratic polynomial functional form.
		\end{minipage}		
	\end{table}

	\begin{table}
		\small
		\caption{Difference-in-differences with cubic treatment intensity}
		\centering
		
		\label{tab:fa_gap_cubic}
		\input{tab_new_corr_meas_error_fagap_cubic.tex}	
		\begin{minipage}{1.0\textwidth}
			\footnotesize
			\justify
			\textbf{Notes:}  This table is analogous to Table~\ref{tab:fa-gap-meas-error}, except that it reports results for a difference-in-differences estimator that allows treatment effects to vary with the treatment intensity through a cubic polynomial functional form.
		\end{minipage}		
	\end{table}

	\begin{table}
		\small
		\caption{Sensitivity of the Gap Design}
		\centering
		
		\label{tab:gap_td}
		\input{tab_gap_td.tex}	
		\begin{minipage}{1.0\textwidth}
			\footnotesize
			\justify
			\textbf{Notes:}      
			This table is analogous to Table~\ref{tab:fa_td}, except that it shows results for the gap design instead of the fraction-affected design.
		\end{minipage}		
	\end{table}

	\begin{table}
		\small
		\caption{Sensitivity of the Fraction-Affected Design: Placebo}
		\centering
		
		\label{tab:fa_td_placebo}
		\input{tab_fa_td_placebo.tex}	
		\begin{minipage}{1.0\textwidth}
			\footnotesize
			\justify
			\textbf{Notes:} This table is analogous to Table~\ref{tab:fa_td}, but it reports a placebo scenario with no increase in the national minimum wage. The fraction-affected measure, however, is calculated as if the national log minimum wage were going to increase by 0.2 between periods (as is the case in Table~\ref{tab:fa_td}).
		\end{minipage}		
	\end{table}

	\begin{table}
		\small
		\caption{Quantifying \textit{large} biases under 500 alternative DGPs}
		\centering
		
		\label{tab:500-dgps-big-bias}
		\input{Tab_rob_dgps_big_bias.tex}	
		\begin{minipage}{1.0\textwidth}
			\footnotesize
			\justify
			\textbf{Notes:}      
			This table is similar to Table~\ref{tab:500-dgps-main}, except the classification of estimators into the ``biased'' category is more stringent: the minimum absolute-bias thresholds are doubled, and the required bias share rises from 25\% to 50\% of the absolute true causal effect.
		\end{minipage}		
	\end{table}

	\begin{table}
		\small
		\caption{Quantifying biases: small measured upper-tail spillovers}
		\centering
		
		\label{tab:500-dgps-nut-kaitz}
		\input{Tab_rob_dgps_nut_kaitz.tex}	
		\begin{minipage}{1.0\textwidth}
			\footnotesize
			\justify
			\textbf{Notes:}      
			This table is similar to Table~\ref{tab:500-dgps-main}, except that it only considers the subset of DGPs where the effective minimum wage estimate of p90-p50 effects is less than 0.05 in magnitude.
		\end{minipage}		
	\end{table}

	\begin{table}
		\small
		\caption{Quantifying biases: small measured upper-tail wage effects}
		\centering
		
		\label{tab:500-dgps-nut-fa}
		\input{Tab_rob_dgps_nut_fa.tex}	
		\begin{minipage}{1.0\textwidth}
			\footnotesize
			\justify
			\textbf{Notes:}      
			This table is similar to Table~\ref{tab:500-dgps-main}, except that it only considers the subset of DGPs where the fraction-affected estimate of wage effects on the 90th quantile is less than 0.05 in magnitude.
		\end{minipage}		
	\end{table}

	\begin{table}
		\small
		\caption{Functional Form Misspecification Biases in Fraction-Affected and Gap Designs}
		\centering
		
		\label{tab:500-dgps-func-forms}
		\input{Tab_rob_dgps_pt_fa_bias.tex}	
		\begin{minipage}{1.0\textwidth}
			\footnotesize
			\justify
			\textbf{Notes:}      
			This table reports results from the simulation of 500 DGPs constructed such that the parallel-trends assumption holds by construction. Specifically, the distributions of location and dispersion parameters are stable over time; initial and final location/dispersion parameters are nearly constant over time (correlations between 0.995 and 0.999); and all cross-correlations between location and dispersion are set to zero.
		\end{minipage}		
	\end{table}

	\begin{table}
		\small
		\caption{Are Causal Effects Bracketed by Pairs of Fraction-Affected/Gap Estimators?}
		\centering
		
		\label{tab:500-dgps-brackets}
		\input{Tab_rob_dgps_pt_fa_brackets.tex}	
		\begin{minipage}{1.0\textwidth}
			\footnotesize
			\justify
			\textbf{Notes:}      
			This table reports results from the simulation of 500 DGPs constructed such that the parallel-trends assumption holds by construction, as in Table~\ref{tab:500-dgps-func-forms}. It reports the share of DGPs where the true average causal effects are bracketed by the expected estimates for a pair of estimators.
		\end{minipage}		
	\end{table}

	\begin{figure}[h]
		\centering
		
		\includegraphics[width=0.8\linewidth]{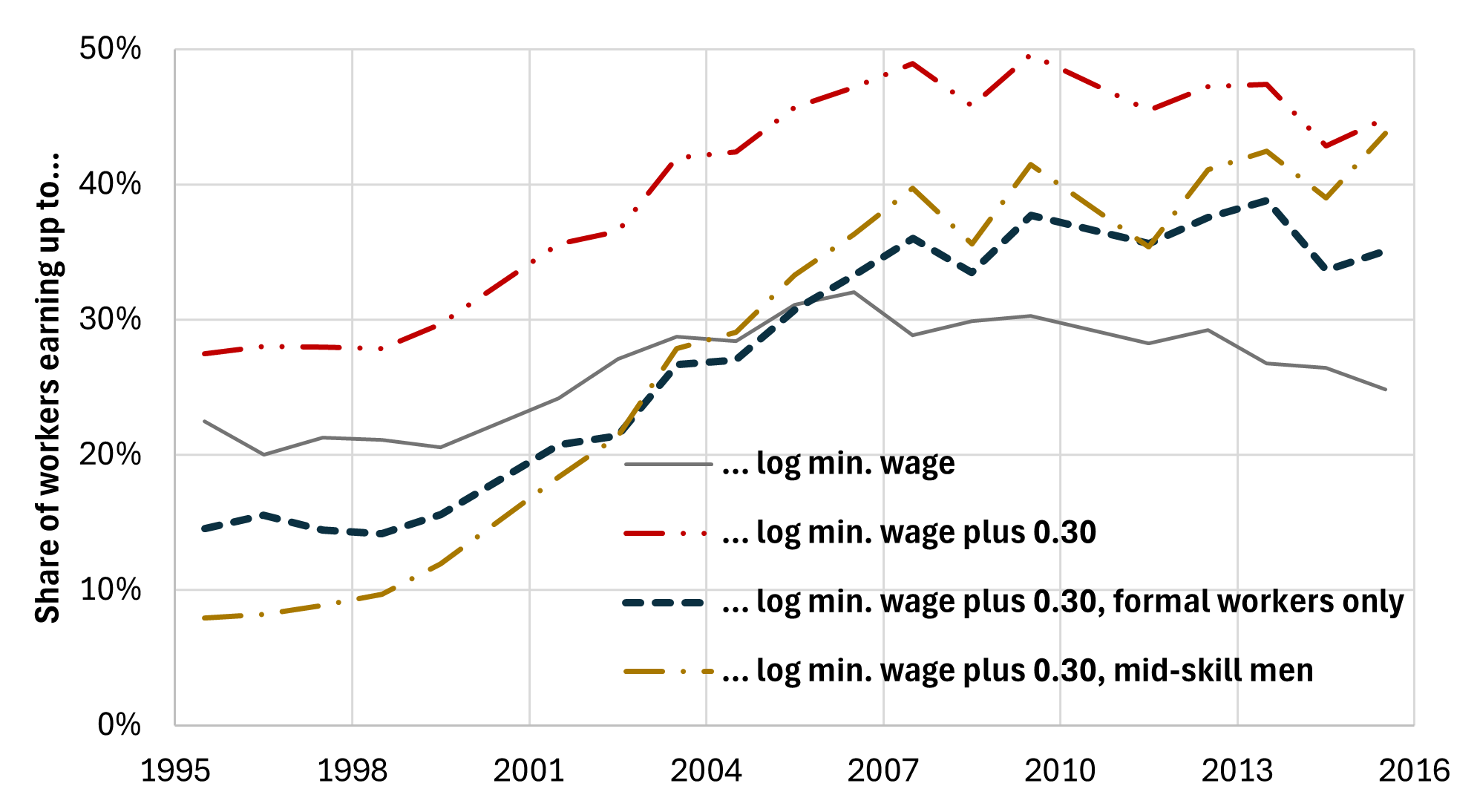}
		\caption{Alternative bindingness measures for the federal minimum wage in Brazil}
		\label{fig:Brazil-mw-share}
		
		\bigskip
		\begin{minipage}{1.0\textwidth}
			\footnotesize
			\justify
			\textbf{Notes:}      
			All series are constructed based on yearly PNAD survey data from IBGE (processed using the DataZoom tool from PUC-Rio), using the September minimum wage value for each year. They show the share of workers with monthly earnings in their main jobs up to a given threshold based on the minimum wage. The base sample includes all workers and self-employed individuals between ages 18 and 54. The third series conditions on workers being employed in the formal sector. The fourth series conditions on workers being male, up to 30 years old, and with exactly 11 years of schooling (i.e., complete high school).
		\end{minipage}	
	\end{figure}

	\begin{sidewaystable}
		\small
		\caption{Summary of existing literature studying the Brazilian minimum wage}\label{tab:lit}
		\begin{tabular*}{20cm}{@{\extracolsep{\fill}}p{3cm}p{1.8cm}p{7.2cm}p{6.8cm}}
			\toprule
			Paper & Data  & Preferred Specification & Results \\ \midrule 
			\addlinespace
			\citet{Hinojosa2019} & PME, 2002-2016; PNAD, 2002-2016 & Eff. min. wage based on {p60}; region and time FEs; region-specific trends; some specifications instrument with lagged features of the wage distribution. & Significant {wage compression} both for formal workers and all workers; spillovers up to the median.  \\ \addlinespace
			\citet{Saltiel2022} & RAIS, 2003-2012 & Eff. min. wage based on {p50}; region and time FEs; region-specific trends; control for regional GDP and adult population. & No effects on employment. \\ \addlinespace
			\citet{Engbom2022} & RAIS, 1996-2018 & Eff. min. wage based on {p90}; region FEs; region-specific trends. & Significant {wage compression} for formal workers; spillovers up to p80 or more. \\
			& PNAD, 1996-2012 &  As above, but based on {p50}. & No employment effects. \\
			& PME, 2002-2012 &  As above, but based on {p50}. & No effects on transitions between formal and informal sectors. \\ \addlinespace
			\citet{Neumark2006} & PME, 1996-2001 & ``Rolling'' fraction-affected measure based on wage distribution from previous quarter; region and time FEs; includes lags. & Large, significant reduction in employment of head of household; reduction in household income for families in the bottom 30\%.  \\ \addlinespace
			\citetalias{Derenoncourt2025} & PNAD, 1995-2015 & Fraction-affected at state$\times$industry$\times$formality status level; state-industry and industry-year FEs; controls for regional GDP and demographic composition. & Wage increases for both formal and informal workers; modest reallocation toward the informal sector. \\ \addlinespace
			\citet{Parente2025} & PNAD, 1996-2012 & Binary version of fraction-affected with multiple treatments, grouping states according to ``share under'' as of 1999; region and year FEs; controls for demographic composition. & Large, significant increases in informality; increase in wage dispersion for all workers. \\
			\bottomrule
		\end{tabular*}
	\end{sidewaystable}

	\begin{figure}[h]
		\centering
		\includegraphics[width=1.0\linewidth]{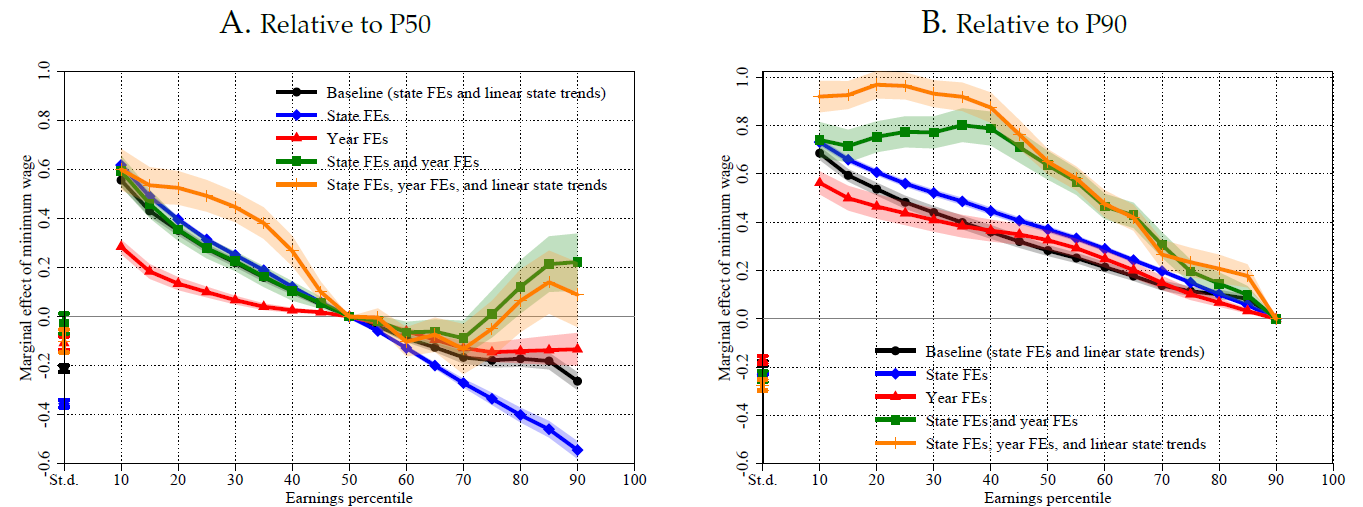}
		\caption{Copy of Figure B.14 in \citet{Engbom2022}---sensitivity of the effective minimum wage design to specification choices}
		\label{fig:engbommoserfigb14}
	\end{figure}

\end{document}

%% file: preamble.tex
    \usepackage[american]{babel}
    \usepackage[utf8]{inputenc}
    \usepackage[T1]{fontenc}

    \usepackage{amsfonts}
    \usepackage{amssymb}
    \usepackage{eufrak}
    \usepackage{amsmath}
    \usepackage{amsthm}
    \usepackage{bbm}
    \usepackage{bm}
    \usepackage{dsfont}
    \usepackage{enumerate}
    \usepackage{mathrsfs}

    \usepackage{graphicx}
    \usepackage{booktabs}
    \usepackage{multirow}
    \usepackage{array}
    \usepackage[labelfont=bf,textfont=md]{caption}
    \usepackage{subcaption}
    \usepackage[flushleft,online,para]{threeparttable}
    \usepackage{multirow}
    \usepackage{ragged2e}

    \usepackage{parskip}                       

    \usepackage{tabularx}
        \newcolumntype{Z}{>{\centering\arraybackslash}X}
        \newcolumntype{L}{>{\raggedright\arraybackslash}X}

    \usepackage{dcolumn}
        \newcolumntype{d}[1]{D{.}{.}{#1}}

    \usepackage{rotating}                      
    \usepackage{lscape}
    \usepackage{pdflscape}

    \usepackage{colortbl}

    \usepackage[round]{natbib}

    \usepackage[dvipsnames]{xcolor}
        \definecolor{darkblue}{rgb}{0,0,0.5}
    \usepackage{hyperref}
        \hypersetup{
            colorlinks = true,
            linkcolor = darkblue,
            citecolor = darkblue,
            pdfborder = 0 0 0,
            pdfdisplaydoctitle = true,
            pdfhighlight = /N,
            pdfpagelayout = OneColumn,
            pdfpagemode = UseNone,
            pdfstartview = {FitH},
            pdfauthor = {{}},
            pdfsubject = {{}},
            breaklinks=true
        }

    \usepackage[textsize=footnotesize, colorinlistoftodos, textwidth=4cm, obeyDraft]{todonotes}

    \usepackage{geometry}
    \usepackage{setspace}
    \usepackage[bottom, multiple]{footmisc}    

    \usepackage{verbatim}
    \usepackage[normalem]{ulem}     
    \usepackage{mathptmx}

    \usepackage[toc,page]{appendix}

\usepackage{accents}
\newcommand\munderbar[1]{%
	\underaccent{\bar}{#1}}

\usepackage{mathtools}

\newcommand{\distras}[1]{%
	\savebox{\mybox}{\hbox{\kern3pt$\scriptstyle#1$\kern3pt}}%
	\savebox{\mysim}{\hbox{$\sim$}}%
	\mathbin{\overset{#1}{\kern\z@\resizebox{\wd\mybox}{\ht\mysim}{$\sim$}}}%
}

%% file: tab_new_eff_iss1.tex
\begin{tabular}{lcccc}
\toprule 
 & \multicolumn{4}{c}{Outcome} \\
 & Emp. & p10 - p50 & p25 - p50 & p90 - p50 \\
\midrule 
\multicolumn{5}{l}{\textit{Panel A: Regions differ only in location}} \\
\hspace{0.5cm} True average causal effect & -0.010 & 0.019 & 0.006 & -0.004 \\
\hspace{0.5cm} Effective min. wage & -0.010 & 0.020 & 0.006 & -0.004 \\
\hspace{0.5cm}  & (0.000) & (0.001) & (0.000) & (0.000) \\
\multicolumn{5}{l}{\textit{Panel B: Regions differ in location and dispersion}} \\
\hspace{0.5cm} True average causal effect & -0.010 & 0.020 & 0.006 & -0.004 \\
\hspace{0.5cm} Effective min. wage & -0.007 & 0.034 & 0.015 & -0.023 \\
\hspace{0.5cm}  & (0.002) & (0.011) & (0.007) & (0.014) \\
\multicolumn{5}{l}{\textit{Panel C: As above, but larger increase in min. wage}} \\
\hspace{0.5cm} True average causal effect & -0.032 & 0.078 & 0.017 & -0.012 \\
\hspace{0.5cm} Effective min. wage & -0.014 & 0.117 & 0.046 & -0.080 \\
\hspace{0.5cm}  & (0.007) & (0.020) & (0.012) & (0.026) \\
\multicolumn{5}{l}{\textit{Panel D: St. dev. of dispersion is 50\% larger}} \\
\hspace{0.5cm} True average causal effect & -0.010 & 0.020 & 0.006 & -0.004 \\
\hspace{0.5cm} Effective min. wage & -0.003 & 0.050 & 0.025 & -0.047 \\
\hspace{0.5cm}  & (0.003) & (0.017) & (0.010) & (0.021) \\
\bottomrule 
\end{tabular}

%% file: tab_new_eff_iss2.tex
\begin{tabular}{lcccc}
\toprule 
 & \multicolumn{4}{c}{Outcome} \\
 & Emp. & p10 - p50 & p25 - p50 & p90 - p50 \\
\midrule 
\multicolumn{5}{l}{\textit{Panel A: No correlation between location and dispersion}} \\
\hspace{0.5cm} True average causal effect & -0.010 & 0.020 & 0.006 & -0.004 \\
\hspace{0.5cm} Effective min. wage & -0.007 & 0.033 & 0.014 & -0.022 \\
\hspace{0.5cm}  & (0.002) & (0.011) & (0.007) & (0.014) \\
\multicolumn{5}{l}{\textit{Panel B: Contemporaneous correlation of 0.076}} \\
\hspace{0.5cm} True average causal effect & -0.010 & 0.020 & 0.006 & -0.004 \\
\hspace{0.5cm} Effective min. wage & -0.002 & 0.077 & 0.040 & -0.076 \\
\hspace{0.5cm}  & (0.002) & (0.010) & (0.006) & (0.013) \\
\bottomrule 
\end{tabular}

%% file: tab_new_corr_meas_error_fa_gap.tex
\begin{tabular}{lccccc}
\toprule 
 & \multicolumn{5}{c}{Outcome} \\
 & Emp. & p10 & p25 & p50 & p90 \\
\midrule 
\multicolumn{6}{l}{\textit{Panel A: Small initial min. wage, truncation/censoring only}} \\
\hspace{0.5cm} True average causal effect & -0.006 & 0.016 & 0.008 & 0.004 & 0.002 \\
\hspace{0.5cm} Fraction affected & -0.008 & 0.020 & 0.010 & 0.006 & 0.003 \\
\hspace{0.5cm}  & (0.000) & (0.001) & (0.001) & (0.001) & (0.001) \\
\hspace{0.5cm} Gap measure & -0.006 & 0.015 & 0.007 & 0.004 & 0.002 \\
\hspace{0.5cm}  & (0.000) & (0.001) & (0.001) & (0.001) & (0.001) \\
\multicolumn{6}{l}{\textit{Panel B: Large initial min. wage, truncation/censoring only}} \\
\hspace{0.5cm} True average causal effect & -0.031 & 0.117 & 0.036 & 0.020 & 0.010 \\
\hspace{0.5cm} Fraction affected & -0.039 & 0.185 & 0.044 & 0.026 & 0.013 \\
\hspace{0.5cm}  & (0.000) & (0.009) & (0.001) & (0.001) & (0.001) \\
\hspace{0.5cm} Gap measure & -0.028 & 0.127 & 0.031 & 0.019 & 0.009 \\
\hspace{0.5cm}  & (0.000) & (0.009) & (0.001) & (0.001) & (0.001) \\
\multicolumn{6}{l}{\textit{Panel C: Small initial min. wage, added emp. mass allowed}} \\
\hspace{0.5cm} True average causal effect & 0.010 & -0.003 & -0.012 & -0.006 & -0.003 \\
\hspace{0.5cm} Fraction affected & 0.002 & 0.067 & -0.002 & -0.001 & -0.000 \\
\hspace{0.5cm}  & (0.000) & (0.003) & (0.001) & (0.001) & (0.001) \\
\hspace{0.5cm} Gap measure & 0.001 & 0.052 & -0.001 & -0.001 & -0.000 \\
\hspace{0.5cm}  & (0.000) & (0.002) & (0.001) & (0.001) & (0.001) \\
\multicolumn{6}{l}{\textit{Panel D: Large initial min. wage, added emp. mass allowed}} \\
\hspace{0.5cm} True average causal effect & -0.002 & 0.148 & 0.038 & 0.002 & 0.001 \\
\hspace{0.5cm} Fraction affected & -0.038 & 0.132 & 0.143 & 0.026 & 0.012 \\
\hspace{0.5cm}  & (0.001) & (0.007) & (0.004) & (0.002) & (0.001) \\
\hspace{0.5cm} Gap measure & -0.027 & 0.091 & 0.101 & 0.019 & 0.008 \\
\hspace{0.5cm}  & (0.001) & (0.006) & (0.003) & (0.001) & (0.001) \\
\bottomrule 
\end{tabular}

%% file: Tab_rob_dgps.tex
\begin{tabular}{lcccc}
\toprule 
Outcome & Emp & p10-p50 & p25-p50 & p90-p50 \\
\midrule 
\multicolumn{5}{l}{\textit{Panel A: 125 DGPs with positive employment effects}} \\
\hspace*{0.2cm}Average causal effects & 0.011 & 0.140 & 0.030 & 0.001 \\
\hspace*{0.2cm}Eff. min. wage: share with significant bias & 0.17 & 0.52 & 0.52 & 0.79 \\
\hspace*{0.2cm}\phantom{Eff. min. wage: }average magnitude of bias, cond. on sig. bias & 0.014 & 0.142 & 0.110 & 0.184 \\
\hspace*{0.2cm}\phantom{Eff. min. wage: }average standard error & 0.001 & 0.024 & 0.016 & 0.027 \\
\hspace*{0.2cm}Fraction affected: share with significant bias & 0.79 & 0.62 & 0.43 & 0.27 \\
\hspace*{0.2cm}\phantom{Fraction affected: }average magnitude of bias, cond. on sig. bias & 0.015 & 0.197 & 0.132 & 0.092 \\
\hspace*{0.2cm}\phantom{Fraction affected: }average standard error & 0.001 & 0.012 & 0.011 & 0.020 \\
\multicolumn{5}{l}{\textit{Panel B: 251 DGPs with negative employment effects}} \\
\hspace*{0.2cm}Average causal effects & -0.018 & 0.142 & 0.038 & -0.009 \\
\hspace*{0.2cm}Eff. min. wage: share with significant bias & 0.88 & 0.47 & 0.49 & 0.87 \\
\hspace*{0.2cm}\phantom{Eff. min. wage: }average magnitude of bias, cond. on sig. bias & 0.026 & 0.126 & 0.099 & 0.165 \\
\hspace*{0.2cm}\phantom{Eff. min. wage: }average standard error & 0.004 & 0.016 & 0.011 & 0.021 \\
\hspace*{0.2cm}Fraction affected: share with significant bias & 0.77 & 0.59 & 0.37 & 0.28 \\
\hspace*{0.2cm}\phantom{Fraction affected: }average magnitude of bias, cond. on sig. bias & 0.028 & 0.209 & 0.169 & 0.097 \\
\hspace*{0.2cm}\phantom{Fraction affected: }average standard error & 0.003 & 0.012 & 0.013 & 0.026 \\
\multicolumn{5}{l}{\textit{Panel C: 124 DGPs with small emp. effects but large wage effects}} \\
\hspace*{0.2cm}Average causal effects & 0.001 & 0.191 & 0.041 & -0.001 \\
\hspace*{0.2cm}Eff. min. wage: share with significant bias & 0.30 & 0.44 & 0.49 & 0.77 \\
\hspace*{0.2cm}\phantom{Eff. min. wage: }average magnitude of bias, cond. on sig. bias & 0.012 & 0.132 & 0.124 & 0.198 \\
\hspace*{0.2cm}\phantom{Eff. min. wage: }average standard error & 0.001 & 0.021 & 0.016 & 0.027 \\
\hspace*{0.2cm}Fraction affected: share with significant bias & 0.50 & 0.77 & 0.48 & 0.29 \\
\hspace*{0.2cm}\phantom{Fraction affected: }average magnitude of bias, cond. on sig. bias & 0.012 & 0.232 & 0.198 & 0.106 \\
\hspace*{0.2cm}\phantom{Fraction affected: }average standard error & 0.001 & 0.014 & 0.014 & 0.027 \\
\bottomrule 
\end{tabular}

%% file: tab_sdif_fagap.tex
\begin{tabular}{lccccc}
\toprule 
 & \multicolumn{5}{c}{Outcome} \\
 & Emp. & p10 & p25 & p50 & p90 \\
\midrule 
Structural model & -0.033 & 0.297 & 0.159 & 0.090 & 0.095 \\
Fraction affected & -0.019 & 0.248 & 0.235 & 0.239 & 0.183 \\
 & (0.007) & (0.013) & (0.014) & (0.015) & (0.020) \\
Quadratic on FA & -0.012 & 0.350 & 0.304 & 0.267 & 0.184 \\
 & (0.013) & (0.016) & (0.021) & (0.028) & (0.037) \\
Cubic on FA & -0.024 & 0.372 & 0.294 & 0.277 & 0.213 \\
 & (0.020) & (0.024) & (0.031) & (0.039) & (0.053) \\
Binary measure, 50\% treated & -0.004 & 0.111 & 0.107 & 0.115 & 0.078 \\
 & (0.004) & (0.008) & (0.009) & (0.010) & (0.013) \\
\bottomrule 
\end{tabular}

%% file: tab_sdif_eff.tex
\begin{tabular}{lcccc}
\toprule 
 & \multicolumn{4}{c}{Outcome} \\
 & Emp. & p10 - p50 & p25 - p50 & p90 - p50 \\
\midrule 
Structural model & -0.033 & 0.208 & 0.070 & 0.005 \\
Effective min. wage & -0.061 & 0.188 & 0.127 & 0.076 \\
 & (0.022) & (0.020) & (0.015) & (0.033) \\
Effective min. wage, no region FE & -0.090 & 0.158 & 0.091 & -0.059 \\
 & (0.009) & (0.014) & (0.009) & (0.022) \\
Effective min. wage, no time FE & 0.101 & 0.200 & 0.103 & -0.127 \\
 & (0.004) & (0.006) & (0.005) & (0.013) \\
AMS, no time FE & 0.118 & 0.201 & 0.101 & -0.149 \\
 & (0.006) & (0.006) & (0.005) & (0.013) \\
\bottomrule 
\end{tabular}

%% file: tab_sdif_eff90.tex
\begin{tabular}{lcccc}
\toprule 
 & \multicolumn{4}{c}{Outcome} \\
 & Emp. & p10 - p90 & p25 - p90 & p50 - p90 \\
\midrule 
Structural model & -0.033 & 0.202 & 0.064 & -0.005 \\
Effective min. wage, p90 & 0.005 & 0.357 & 0.341 & 0.296 \\
 & (0.022) & (0.033) & (0.037) & (0.039) \\
\bottomrule 
\end{tabular}

%% file: tab_simparams_normal_eff.tex
\begin{tabular}{lcccccccccccccccc}
\toprule 
 &  & \multicolumn{2}{c}{Min. Wage} & \multicolumn{2}{c}{Means} & \multicolumn{4}{c}{Std. Dev.} & \multicolumn{6}{c}{Correlations} & Corr. \\
Model & $m$ & $mw_0$ & $mw_1$ & $\sigma_{r,0}$ & $\sigma_{r,1}$ & $\mu_0$ & $\sigma_0$ & $\mu_1$ & $\sigma_1$ & $\mu_0,\sigma_0$ & $\mu_0,\mu_1$ & $\mu_0,\sigma_1$ & $\sigma_0,\mu_1$ & $\sigma_0,\sigma_1$ & $\mu_1,\sigma_1$ & $\mu_{r,t},w_{0.5,r,t}$ \\
\midrule 
\multicolumn{17}{l}{Table 1} \\
\hspace*{0.2cm}\textit{Panel A} & 0.7 & -1.0 & -0.8 & 0.542 & 0.510 & 0.123 & 0.000 & 0.112 & 0.000 & 0.000 & 0.894 & 0.000 & 0.000 & 0.000 & 0.000 & 1.000 \\
\hspace*{0.2cm}\textit{Panel B} & 0.7 & -1.0 & -0.8 & 0.542 & 0.510 & 0.123 & 0.026 & 0.112 & 0.049 & 0.000 & 0.894 & 0.000 & 0.000 & 0.456 & 0.000 & 0.999 \\
\hspace*{0.2cm}\textit{Panel C} & 0.7 & -1.0 & -0.6 & 0.542 & 0.510 & 0.123 & 0.026 & 0.112 & 0.049 & 0.000 & 0.894 & 0.000 & 0.000 & 0.456 & 0.000 & 0.994 \\
\hspace*{0.2cm}\textit{Panel D} & 0.7 & -1.0 & -0.8 & 0.542 & 0.510 & 0.123 & 0.039 & 0.112 & 0.074 & 0.000 & 0.894 & 0.000 & 0.000 & 0.456 & 0.000 & 0.998 \\
\multicolumn{17}{l}{Table 2} \\
\hspace*{0.2cm}\textit{Panel A} & 0.7 & -1.0 & -0.8 & 0.542 & 0.510 & 0.123 & 0.026 & 0.112 & 0.049 & 0.000 & 0.894 & 0.000 & 0.000 & 0.456 & 0.000 & 0.999 \\
\hspace*{0.2cm}\textit{Panel B} & 0.7 & -1.0 & -0.8 & 0.542 & 0.510 & 0.123 & 0.026 & 0.112 & 0.049 & 0.076 & 0.894 & 0.000 & 0.000 & 0.456 & 0.076 & 0.999 \\
\multicolumn{17}{l}{Tables A3 and A4} \\
\hspace*{0.2cm}\textit{All panels} & 0.7 & -1.0 & -0.8 & 0.542 & 0.510 & 0.123 & 0.026 & 0.112 & 0.049 & 0.000 & 0.894 & 0.000 & 0.000 & 0.456 & 0.000 & 0.999 \\
\multicolumn{17}{l}{Table A5} \\
\hspace*{0.2cm}\textit{All panels} & 0.7 & -1.0 & -0.8 & 0.542 & 0.510 & 0.123 & 0.026 & 0.112 & 0.049 & 0.076 & 0.894 & 0.000 & 0.000 & 0.456 & 0.076 & 0.999 \\
\bottomrule 
\end{tabular}

%% file: tab_simparams_normal_fagap.tex
\begin{tabular}{lcccccccccccccccc}
\toprule 
 &  & \multicolumn{2}{c}{Min. Wage} & \multicolumn{2}{c}{Means} & \multicolumn{4}{c}{Std. Dev.} & \multicolumn{6}{c}{Correlations} & Corr. \\
Model & $m$ & $mw_0$ & $mw_1$ & $\sigma_{r,0}$ & $\sigma_{r,1}$ & $\mu_0$ & $\sigma_0$ & $\mu_1$ & $\sigma_1$ & $\mu_0,\sigma_0$ & $\mu_0,\mu_1$ & $\mu_0,\sigma_1$ & $\sigma_0,\mu_1$ & $\sigma_0,\sigma_1$ & $\mu_1,\sigma_1$ & $\mu_{r,t},w_{0.5,r,t}$ \\
\midrule 
\multicolumn{17}{l}{Table 3} \\
\hspace*{0.2cm}\textit{Panel A} & 0.7 & -1.1 & -0.9 & 0.526 & 0.526 & 0.118 & 0.000 & 0.118 & 0.000 & 0.000 & 0.999 & 0.000 & 0.000 & 0.000 & 0.000 & 1.000 \\
\hspace*{0.2cm}\textit{Panel B} & 0.7 & -0.7 & -0.5 & 0.526 & 0.526 & 0.118 & 0.000 & 0.118 & 0.000 & 0.000 & 0.999 & 0.000 & 0.000 & 0.000 & 0.000 & 0.994 \\
\hspace*{0.2cm}\textit{Panel C} & 0.7 & -1.1 & -0.9 & 0.526 & 0.526 & 0.118 & 0.000 & 0.118 & 0.000 & 0.000 & 0.999 & 0.000 & 0.000 & 0.000 & 0.000 & 1.000 \\
\hspace*{0.2cm}\textit{Panel D} & 0.7 & -0.7 & -0.5 & 0.526 & 0.526 & 0.118 & 0.000 & 0.118 & 0.000 & 0.000 & 0.999 & 0.000 & 0.000 & 0.000 & 0.000 & 0.999 \\
\multicolumn{17}{l}{Tables A8 and A14} \\
\hspace*{0.2cm}\textit{Panel A} & 0.7 & -1.0 & -0.8 & 0.526 & 0.526 & 0.118 & 0.000 & 0.118 & 0.000 & 0.000 & 0.999 & 0.000 & 0.000 & 0.000 & 0.000 & 0.999 \\
\hspace*{0.2cm}\textit{Panel B} & 0.7 & -1.0 & -0.8 & 0.526 & 0.526 & 0.118 & 0.000 & 0.118 & 0.000 & 0.000 & 0.894 & 0.000 & 0.000 & 0.000 & 0.000 & 0.999 \\
\hspace*{0.2cm}\textit{Panel C} & 0.7 & -1.0 & -0.8 & 0.526 & 0.526 & 0.118 & 0.038 & 0.118 & 0.038 & 0.000 & 0.894 & 0.000 & 0.000 & 0.456 & 0.000 & 0.999 \\
\hspace*{0.2cm}\textit{Panel D} & 0.7 & -1.0 & -0.8 & 0.542 & 0.510 & 0.118 & 0.038 & 0.118 & 0.038 & 0.000 & 0.894 & 0.000 & 0.000 & 0.456 & 0.000 & 0.999 \\
\multicolumn{17}{l}{Table A15} \\
\hspace*{0.2cm}\textit{Panel A} & 0.7 & -1.0 & -1.0 & 0.526 & 0.526 & 0.118 & 0.000 & 0.118 & 0.000 & 0.000 & 0.999 & 0.000 & 0.000 & 0.000 & 0.000 & 1.000 \\
\hspace*{0.2cm}\textit{Panel B} & 0.7 & -1.0 & -1.0 & 0.526 & 0.526 & 0.118 & 0.000 & 0.118 & 0.000 & 0.000 & 0.894 & 0.000 & 0.000 & 0.000 & 0.000 & 1.000 \\
\hspace*{0.2cm}\textit{Panel C} & 0.7 & -1.0 & -1.0 & 0.526 & 0.526 & 0.118 & 0.038 & 0.118 & 0.038 & 0.000 & 0.894 & 0.000 & 0.000 & 0.456 & 0.000 & 1.000 \\
\hspace*{0.2cm}\textit{Panel D} & 0.7 & -1.0 & -1.0 & 0.542 & 0.510 & 0.118 & 0.038 & 0.118 & 0.038 & 0.000 & 0.894 & 0.000 & 0.000 & 0.456 & 0.000 & 1.000 \\
\bottomrule 
\end{tabular}

%% file: tab_new_eff_noTFE.tex
\begin{tabular}{lcccc}
\toprule 
 & \multicolumn{4}{c}{Outcome} \\
 & Emp. & p10 - p50 & p25 - p50 & p90 - p50 \\
\midrule 
True average causal effect & -0.010 & 0.020 & 0.006 & -0.004 \\
Effective min. wage & -0.007 & 0.034 & 0.015 & -0.023 \\
 & (0.002) & (0.011) & (0.007) & (0.014) \\
Effective min. wage, no region FE & -0.010 & 0.022 & 0.007 & -0.007 \\
 & (0.001) & (0.004) & (0.002) & (0.005) \\
Effective min. wage, no time FE & -0.007 & 0.052 & 0.024 & -0.041 \\
 & (0.001) & (0.003) & (0.002) & (0.004) \\
\bottomrule 
\end{tabular}

%% file: tab_new_eff_90.tex
\begin{tabular}{lcccc}
\toprule 
 & \multicolumn{4}{c}{Outcome} \\
 & Emp. & p10 - p90 & p25 - p90 & p50 - p90 \\
\midrule 
True average causal effect & -0.010 & 0.024 & 0.009 & 0.004 \\
Effective min. wage, p90 & 0.009 & 0.220 & 0.177 & 0.119 \\
 & (0.002) & (0.013) & (0.011) & (0.007) \\
\bottomrule 
\end{tabular}

%% file: tab_ams.tex
\begin{tabular}{lcccc}
\toprule 
 & \multicolumn{4}{c}{Outcome} \\
 & Emp. & p10 - p50 & p25 - p50 & p90 - p50 \\
\midrule 
\multicolumn{5}{l}{\textit{Panel A: No regional variation in minimum wage.}} \\
\hspace{0.5cm} True average causal effect & -0.010 & 0.020 & 0.006 & -0.004 \\
\hspace{0.5cm} Effective min. wage & -0.002 & 0.076 & 0.040 & -0.075 \\
\hspace{0.5cm}  & (0.002) & (0.010) & (0.006) & (0.013) \\
\multicolumn{5}{l}{\textit{Panel B: 20\% of regions with local min. wage}} \\
\hspace{0.5cm} True average causal effect & -0.015 & 0.035 & 0.008 & -0.005 \\
\hspace{0.5cm} Effective min. wage & -0.015 & 0.050 & 0.015 & -0.018 \\
\hspace{0.5cm}  & (0.001) & (0.005) & (0.003) & (0.006) \\
\hspace{0.5cm} Two instruments & -0.015 & 0.036 & 0.009 & -0.006 \\
\hspace{0.5cm}  & (0.002) & (0.006) & (0.004) & (0.008) \\
\hspace{0.5cm} Three instruments (AMS) & -0.017 & 0.041 & 0.008 & -0.005 \\
\hspace{0.5cm}  & (0.002) & (0.005) & (0.003) & (0.006) \\
\multicolumn{5}{l}{\textit{Panel C: 40\% of regions with local min. wage}} \\
\hspace{0.5cm} True average causal effect & -0.020 & 0.053 & 0.011 & -0.007 \\
\hspace{0.5cm} Effective min. wage & -0.019 & 0.059 & 0.015 & -0.016 \\
\hspace{0.5cm}  & (0.001) & (0.004) & (0.002) & (0.005) \\
\hspace{0.5cm} Two instruments & -0.020 & 0.051 & 0.011 & -0.007 \\
\hspace{0.5cm}  & (0.002) & (0.004) & (0.002) & (0.006) \\
\hspace{0.5cm} Three instruments (AMS) & -0.021 & 0.053 & 0.010 & -0.007 \\
\hspace{0.5cm}  & (0.001) & (0.004) & (0.002) & (0.005) \\
\bottomrule 
\end{tabular}

%% file: tab_canonical_fa_gap.tex
\begin{tabular}{lccccc}
\toprule 
 & \multicolumn{5}{c}{Outcome} \\
 & Emp. & p10 & p25 & p50 & p90 \\
\midrule 
\multicolumn{6}{l}{\textit{Panel A: Initial minimum wage is low, elast. subs. is 3.0}} \\
\hspace{0.5cm} True average causal effect & -0.009 & 0.024 & 0.012 & 0.007 & 0.003 \\
\hspace{0.5cm} Fraction affected & -0.009 & 0.022 & 0.011 & 0.007 & 0.004 \\
\hspace{0.5cm}  & (0.000) & (0.000) & (0.000) & (0.000) & (0.000) \\
\hspace{0.5cm} Gap measure & -0.008 & 0.018 & 0.009 & 0.005 & 0.003 \\
\hspace{0.5cm}  & (0.000) & (0.000) & (0.000) & (0.000) & (0.000) \\
\multicolumn{6}{l}{\textit{Panel B: Initial minimum wage is low, elast. subs. is 1.4}} \\
\hspace{0.5cm} True average causal effect & -0.009 & 0.022 & 0.011 & 0.006 & 0.003 \\
\hspace{0.5cm} Fraction affected & -0.009 & 0.020 & 0.011 & 0.007 & 0.004 \\
\hspace{0.5cm}  & (0.000) & (0.000) & (0.000) & (0.000) & (0.000) \\
\hspace{0.5cm} Gap measure & -0.007 & 0.017 & 0.009 & 0.005 & 0.003 \\
\hspace{0.5cm}  & (0.000) & (0.000) & (0.000) & (0.000) & (0.000) \\
\multicolumn{6}{l}{\textit{Panel C: Initial minimum wage is high, elast. subs. is 3.0}} \\
\hspace{0.5cm} True average causal effect & -0.042 & 0.087 & 0.050 & 0.030 & 0.014 \\
\hspace{0.5cm} Fraction affected & -0.040 & 0.058 & 0.044 & 0.030 & 0.017 \\
\hspace{0.5cm}  & (0.000) & (0.001) & (0.000) & (0.000) & (0.000) \\
\hspace{0.5cm} Gap measure & -0.031 & 0.044 & 0.034 & 0.023 & 0.013 \\
\hspace{0.5cm}  & (0.000) & (0.001) & (0.001) & (0.000) & (0.000) \\
\multicolumn{6}{l}{\textit{Panel D: Initial minimum wage is high, elast. subs. is 1.4}} \\
\hspace{0.5cm} True average causal effect & -0.039 & 0.083 & 0.048 & 0.029 & 0.013 \\
\hspace{0.5cm} Fraction affected & -0.036 & 0.057 & 0.044 & 0.031 & 0.017 \\
\hspace{0.5cm}  & (0.000) & (0.001) & (0.001) & (0.000) & (0.000) \\
\hspace{0.5cm} Gap measure & -0.028 & 0.045 & 0.035 & 0.025 & 0.014 \\
\hspace{0.5cm}  & (0.001) & (0.002) & (0.001) & (0.001) & (0.000) \\
\multicolumn{6}{l}{\textit{Panel E: Initial minimum wage is very high, elast. subs. is 3.0}} \\
\hspace{0.5cm} True average causal effect & -0.086 & 0.138 & 0.096 & 0.063 & 0.031 \\
\hspace{0.5cm} Fraction affected & -0.068 & 0.060 & 0.069 & 0.061 & 0.039 \\
\hspace{0.5cm}  & (0.000) & (0.001) & (0.001) & (0.001) & (0.000) \\
\hspace{0.5cm} Gap measure & -0.051 & 0.045 & 0.052 & 0.046 & 0.029 \\
\hspace{0.5cm}  & (0.001) & (0.001) & (0.001) & (0.001) & (0.000) \\
\multicolumn{6}{l}{\textit{Panel F: Initial minimum wage is very high, elast. subs. is 1.4}} \\
\hspace{0.5cm} True average causal effect & -0.081 & 0.134 & 0.093 & 0.061 & 0.029 \\
\hspace{0.5cm} Fraction affected & -0.058 & 0.064 & 0.073 & 0.065 & 0.038 \\
\hspace{0.5cm}  & (0.001) & (0.002) & (0.002) & (0.001) & (0.001) \\
\hspace{0.5cm} Gap measure & -0.045 & 0.049 & 0.057 & 0.050 & 0.029 \\
\hspace{0.5cm}  & (0.001) & (0.002) & (0.002) & (0.001) & (0.000) \\
\bottomrule 
\end{tabular}

%% file: tab_canonical_kaitz.tex
\begin{tabular}{lcccc}
\toprule 
 & \multicolumn{4}{c}{Outcome} \\
 & Emp. & p10 - p50 & p25 - p50 & p90 - p50 \\
\midrule 
\multicolumn{5}{l}{\textit{Panel A: Initial minimum wage is low, elast. subs. is 3.0}} \\
\hspace{0.5cm} True average causal effect & -0.009 & 0.017 & 0.005 & -0.004 \\
\hspace{0.5cm} Effective min. wage & 0.415 & -0.237 & -0.102 & -0.040 \\
\hspace{0.5cm}  & (0.011) & (0.020) & (0.009) & (0.003) \\
\hspace{0.5cm} Effective min. wage, no region FE & -0.009 & 0.018 & 0.003 & 0.044 \\
\hspace{0.5cm}  & (0.000) & (0.001) & (0.001) & (0.002) \\
\multicolumn{5}{l}{\textit{Panel B: Initial minimum wage is low, elast. subs. is 1.4}} \\
\hspace{0.5cm} True average causal effect & -0.009 & 0.016 & 0.005 & -0.003 \\
\hspace{0.5cm} Effective min. wage & 0.284 & -0.183 & -0.076 & -0.014 \\
\hspace{0.5cm}  & (0.006) & (0.011) & (0.006) & (0.003) \\
\hspace{0.5cm} Effective min. wage, no region FE & -0.010 & 0.002 & -0.007 & 0.078 \\
\hspace{0.5cm}  & (0.000) & (0.002) & (0.001) & (0.002) \\
\multicolumn{5}{l}{\textit{Panel C: Initial minimum wage is high, elast. subs. is 3.0}} \\
\hspace{0.5cm} True average causal effect & -0.042 & 0.057 & 0.020 & -0.016 \\
\hspace{0.5cm} Effective min. wage & 0.450 & 0.483 & 0.109 & -0.198 \\
\hspace{0.5cm}  & (0.023) & (0.013) & (0.008) & (0.012) \\
\hspace{0.5cm} Effective min. wage, no region FE & -0.043 & 0.054 & 0.017 & 0.032 \\
\hspace{0.5cm}  & (0.000) & (0.001) & (0.001) & (0.002) \\
\multicolumn{5}{l}{\textit{Panel D: Initial minimum wage is high, elast. subs. is 1.4}} \\
\hspace{0.5cm} True average causal effect & -0.039 & 0.054 & 0.019 & -0.016 \\
\hspace{0.5cm} Effective min. wage & 0.233 & 0.275 & 0.066 & -0.098 \\
\hspace{0.5cm}  & (0.007) & (0.019) & (0.008) & (0.007) \\
\hspace{0.5cm} Effective min. wage, no region FE & -0.045 & 0.042 & 0.008 & 0.066 \\
\hspace{0.5cm}  & (0.001) & (0.001) & (0.001) & (0.002) \\
\multicolumn{5}{l}{\textit{Panel E: Initial minimum wage is very high, elast. subs. is 3.0}} \\
\hspace{0.5cm} True average causal effect & -0.086 & 0.075 & 0.033 & -0.032 \\
\hspace{0.5cm} Effective min. wage & 0.255 & 0.313 & 0.153 & -0.199 \\
\hspace{0.5cm}  & (0.015) & (0.001) & (0.000) & (0.018) \\
\hspace{0.5cm} Effective min. wage, no region FE & -0.091 & 0.071 & 0.029 & 0.018 \\
\hspace{0.5cm}  & (0.001) & (0.000) & (0.000) & (0.002) \\
\multicolumn{5}{l}{\textit{Panel F: Initial minimum wage is very high, elast. subs. is 1.4}} \\
\hspace{0.5cm} True average causal effect & -0.081 & 0.074 & 0.032 & -0.032 \\
\hspace{0.5cm} Effective min. wage & 0.114 & 0.215 & 0.105 & -0.100 \\
\hspace{0.5cm}  & (0.003) & (0.009) & (0.004) & (0.010) \\
\hspace{0.5cm} Effective min. wage, no region FE & -0.097 & 0.064 & 0.022 & 0.051 \\
\hspace{0.5cm}  & (0.001) & (0.001) & (0.001) & (0.002) \\
\bottomrule 
\end{tabular}

%% file: tab_fa_td.tex
\begin{tabular}{lccccc}
\toprule 
 & \multicolumn{5}{c}{Outcome} \\
 & Emp. & p10 & p25 & p50 & p90 \\
\midrule 
\multicolumn{6}{l}{\textit{Panel A: Only permanent differences in location}} \\
\hspace{0.5cm} True average causal effect & -0.010 & 0.026 & 0.012 & 0.007 & 0.003 \\
\hspace{0.5cm} Fraction affected & -0.013 & 0.036 & 0.015 & 0.009 & 0.004 \\
\hspace{0.5cm}  & (0.000) & (0.003) & (0.001) & (0.001) & (0.001) \\
\multicolumn{6}{l}{\textit{Panel B: Adding location shocks, stable distributions}} \\
\hspace{0.5cm} True average causal effect & -0.010 & 0.026 & 0.012 & 0.007 & 0.003 \\
\hspace{0.5cm} Fraction affected & -0.010 & 0.059 & 0.042 & 0.036 & 0.033 \\
\hspace{0.5cm}  & (0.001) & (0.007) & (0.008) & (0.008) & (0.009) \\
\multicolumn{6}{l}{\textit{Panel C: Adding dispersion differences and shocks, stable distributions}} \\
\hspace{0.5cm} True average causal effect & -0.010 & 0.026 & 0.012 & 0.007 & 0.003 \\
\hspace{0.5cm} Fraction affected & -0.008 & 0.072 & 0.047 & 0.031 & 0.005 \\
\hspace{0.5cm}  & (0.001) & (0.008) & (0.008) & (0.008) & (0.012) \\
\multicolumn{6}{l}{\textit{Panel D: Average dispersion falls over time}} \\
\hspace{0.5cm} True average causal effect & -0.010 & 0.027 & 0.013 & 0.007 & 0.003 \\
\hspace{0.5cm} Fraction affected & -0.005 & 0.065 & 0.044 & 0.030 & 0.006 \\
\hspace{0.5cm}  & (0.001) & (0.008) & (0.008) & (0.008) & (0.013) \\
\bottomrule 
\end{tabular}

%% file: tab_new_corr_meas_error_binary.tex
\begin{tabular}{lccccc}
\toprule 
 & \multicolumn{5}{c}{Outcome} \\
 & Emp. & p10 & p25 & p50 & p90 \\
\midrule 
\multicolumn{6}{l}{\textit{Panel A: Small initial min. wage, truncation/censoring only}} \\
\hspace{0.5cm} True average causal effect & -0.006 & 0.016 & 0.008 & 0.004 & 0.002 \\
\hspace{0.5cm} Binary measure, 50\% treated & -0.003 & 0.007 & 0.003 & 0.002 & 0.001 \\
\hspace{0.5cm}  & (0.000) & (0.001) & (0.000) & (0.000) & (0.000) \\
\hspace{0.5cm} Binary measure, 90\% treated & -0.004 & 0.011 & 0.006 & 0.003 & 0.002 \\
\hspace{0.5cm}  & (0.000) & (0.001) & (0.001) & (0.001) & (0.001) \\
\multicolumn{6}{l}{\textit{Panel B: Large initial min. wage, truncation/censoring only}} \\
\hspace{0.5cm} True average causal effect & -0.031 & 0.118 & 0.036 & 0.020 & 0.010 \\
\hspace{0.5cm} Binary measure, 50\% treated & -0.009 & 0.053 & 0.010 & 0.006 & 0.003 \\
\hspace{0.5cm}  & (0.001) & (0.002) & (0.001) & (0.000) & (0.000) \\
\hspace{0.5cm} Binary measure, 90\% treated & -0.017 & 0.084 & 0.020 & 0.012 & 0.006 \\
\hspace{0.5cm}  & (0.001) & (0.004) & (0.001) & (0.001) & (0.001) \\
\multicolumn{6}{l}{\textit{Panel C: Small initial min. wage, added emp. mass allowed}} \\
\hspace{0.5cm} True average causal effect & 0.010 & -0.003 & -0.012 & -0.006 & -0.003 \\
\hspace{0.5cm} Binary measure, 50\% treated & 0.001 & 0.021 & -0.001 & -0.001 & -0.000 \\
\hspace{0.5cm}  & (0.000) & (0.002) & (0.000) & (0.000) & (0.000) \\
\hspace{0.5cm} Binary measure, 90\% treated & 0.003 & 0.017 & -0.004 & -0.002 & -0.001 \\
\hspace{0.5cm}  & (0.000) & (0.003) & (0.001) & (0.001) & (0.001) \\
\multicolumn{6}{l}{\textit{Panel D: Large initial min. wage, added emp. mass allowed}} \\
\hspace{0.5cm} True average causal effect & -0.003 & 0.148 & 0.038 & 0.002 & 0.001 \\
\hspace{0.5cm} Binary measure, 50\% treated & -0.008 & 0.033 & 0.035 & 0.006 & 0.003 \\
\hspace{0.5cm}  & (0.001) & (0.002) & (0.002) & (0.001) & (0.000) \\
\hspace{0.5cm} Binary measure, 90\% treated & -0.013 & 0.078 & 0.051 & 0.009 & 0.004 \\
\hspace{0.5cm}  & (0.001) & (0.005) & (0.003) & (0.001) & (0.001) \\
\bottomrule 
\end{tabular}

%% file: tab_new_corr_meas_error_fagap_inst.tex
\begin{tabular}{lccccc}
\toprule 
 & \multicolumn{5}{c}{Outcome} \\
 & Emp. & p10 & p25 & p50 & p90 \\
\midrule 
\multicolumn{6}{l}{\textit{Panel A: Small initial min. wage, truncation/censoring only}} \\
\hspace{0.5cm} True average causal effect & -0.006 & 0.016 & 0.008 & 0.004 & 0.002 \\
\hspace{0.5cm} FA instrumented by GAP & -0.008 & 0.021 & 0.010 & 0.006 & 0.003 \\
\hspace{0.5cm}  & (0.000) & (0.001) & (0.001) & (0.001) & (0.001) \\
\hspace{0.5cm} GAP instrumented by FA & -0.006 & 0.015 & 0.007 & 0.004 & 0.002 \\
\hspace{0.5cm}  & (0.000) & (0.001) & (0.001) & (0.001) & (0.001) \\
\multicolumn{6}{l}{\textit{Panel B: Large initial min. wage, truncation/censoring only}} \\
\hspace{0.5cm} True average causal effect & -0.031 & 0.118 & 0.036 & 0.021 & 0.010 \\
\hspace{0.5cm} FA instrumented by GAP & -0.040 & 0.182 & 0.044 & 0.027 & 0.013 \\
\hspace{0.5cm}  & (0.000) & (0.010) & (0.001) & (0.001) & (0.001) \\
\hspace{0.5cm} GAP instrumented by FA & -0.028 & 0.132 & 0.031 & 0.019 & 0.009 \\
\hspace{0.5cm}  & (0.000) & (0.008) & (0.001) & (0.001) & (0.001) \\
\multicolumn{6}{l}{\textit{Panel C: Small initial min. wage, added emp. mass allowed}} \\
\hspace{0.5cm} True average causal effect & 0.010 & -0.003 & -0.012 & -0.006 & -0.003 \\
\hspace{0.5cm} FA instrumented by GAP & 0.002 & 0.069 & -0.001 & -0.001 & -0.000 \\
\hspace{0.5cm}  & (0.000) & (0.003) & (0.002) & (0.001) & (0.001) \\
\hspace{0.5cm} GAP instrumented by FA & 0.001 & 0.051 & -0.001 & -0.001 & -0.000 \\
\hspace{0.5cm}  & (0.000) & (0.002) & (0.001) & (0.001) & (0.001) \\
\multicolumn{6}{l}{\textit{Panel D: Large initial min. wage, added emp. mass allowed}} \\
\hspace{0.5cm} True average causal effect & -0.002 & 0.148 & 0.038 & 0.002 & 0.001 \\
\hspace{0.5cm} FA instrumented by GAP & -0.039 & 0.129 & 0.144 & 0.027 & 0.012 \\
\hspace{0.5cm}  & (0.002) & (0.007) & (0.004) & (0.002) & (0.001) \\
\hspace{0.5cm} GAP instrumented by FA & -0.027 & 0.094 & 0.102 & 0.019 & 0.008 \\
\hspace{0.5cm}  & (0.001) & (0.006) & (0.003) & (0.001) & (0.001) \\
\bottomrule 
\end{tabular}

%% file: tab_new_corr_meas_error_fagap_quad.tex
\begin{tabular}{lccccc}
\toprule 
 & \multicolumn{5}{c}{Outcome} \\
 & Emp. & p10 & p25 & p50 & p90 \\
\midrule 
\multicolumn{6}{l}{\textit{Panel A: Small initial min. wage, truncation/censoring only}} \\
\hspace{0.5cm} True average causal effect & -0.006 & 0.016 & 0.008 & 0.004 & 0.002 \\
\hspace{0.5cm} Quadratic on FA & -0.007 & 0.017 & 0.009 & 0.005 & 0.002 \\
\hspace{0.5cm}  & (0.000) & (0.002) & (0.002) & (0.002) & (0.002) \\
\hspace{0.5cm} Quadratic on GAP & -0.006 & 0.015 & 0.008 & 0.004 & 0.002 \\
\hspace{0.5cm}  & (0.000) & (0.001) & (0.001) & (0.001) & (0.001) \\
\multicolumn{6}{l}{\textit{Panel B: Large initial min. wage, truncation/censoring only}} \\
\hspace{0.5cm} True average causal effect & -0.031 & 0.118 & 0.036 & 0.021 & 0.010 \\
\hspace{0.5cm} Quadratic on FA & -0.034 & 0.281 & 0.038 & 0.023 & 0.011 \\
\hspace{0.5cm}  & (0.001) & (0.019) & (0.003) & (0.003) & (0.003) \\
\hspace{0.5cm} Quadratic on GAP & -0.030 & 0.211 & 0.034 & 0.020 & 0.010 \\
\hspace{0.5cm}  & (0.000) & (0.008) & (0.002) & (0.002) & (0.002) \\
\multicolumn{6}{l}{\textit{Panel C: Small initial min. wage, added emp. mass allowed}} \\
\hspace{0.5cm} True average causal effect & 0.010 & -0.002 & -0.012 & -0.006 & -0.003 \\
\hspace{0.5cm} Quadratic on FA & 0.007 & 0.029 & -0.010 & -0.004 & -0.001 \\
\hspace{0.5cm}  & (0.000) & (0.006) & (0.002) & (0.002) & (0.002) \\
\hspace{0.5cm} Quadratic on GAP & 0.004 & 0.040 & -0.006 & -0.002 & -0.001 \\
\hspace{0.5cm}  & (0.000) & (0.004) & (0.001) & (0.001) & (0.001) \\
\multicolumn{6}{l}{\textit{Panel D: Large initial min. wage, added emp. mass allowed}} \\
\hspace{0.5cm} True average causal effect & -0.002 & 0.148 & 0.038 & 0.002 & 0.001 \\
\hspace{0.5cm} Quadratic on FA & -0.013 & 0.246 & 0.126 & 0.007 & 0.004 \\
\hspace{0.5cm}  & (0.001) & (0.004) & (0.014) & (0.004) & (0.004) \\
\hspace{0.5cm} Quadratic on GAP & -0.019 & 0.167 & 0.117 & 0.012 & 0.006 \\
\hspace{0.5cm}  & (0.001) & (0.005) & (0.006) & (0.002) & (0.002) \\
\bottomrule 
\end{tabular}

%% file: tab_new_corr_meas_error_fagap_cubic.tex
\begin{tabular}{lccccc}
\toprule 
 & \multicolumn{5}{c}{Outcome} \\
 & Emp. & p10 & p25 & p50 & p90 \\
\midrule 
\multicolumn{6}{l}{\textit{Panel A: Small initial min. wage, truncation/censoring only}} \\
\hspace{0.5cm} True average causal effect & -0.006 & 0.016 & 0.008 & 0.004 & 0.002 \\
\hspace{0.5cm} Cubic on FA & -0.007 & 0.019 & 0.009 & 0.005 & 0.002 \\
\hspace{0.5cm}  & (0.000) & (0.003) & (0.003) & (0.003) & (0.003) \\
\hspace{0.5cm} Cubic on GAP & -0.006 & 0.017 & 0.008 & 0.005 & 0.002 \\
\hspace{0.5cm}  & (0.000) & (0.002) & (0.002) & (0.002) & (0.002) \\
\multicolumn{6}{l}{\textit{Panel B: Large initial min. wage, truncation/censoring only}} \\
\hspace{0.5cm} True average causal effect & -0.031 & 0.118 & 0.036 & 0.020 & 0.010 \\
\hspace{0.5cm} Cubic on FA & -0.032 & 0.043 & 0.043 & 0.022 & 0.010 \\
\hspace{0.5cm}  & (0.001) & (0.028) & (0.006) & (0.007) & (0.007) \\
\hspace{0.5cm} Cubic on GAP & -0.030 & 0.160 & 0.037 & 0.020 & 0.010 \\
\hspace{0.5cm}  & (0.001) & (0.014) & (0.003) & (0.003) & (0.003) \\
\multicolumn{6}{l}{\textit{Panel C: Small initial min. wage, added emp. mass allowed}} \\
\hspace{0.5cm} True average causal effect & 0.010 & -0.003 & -0.012 & -0.006 & -0.003 \\
\hspace{0.5cm} Cubic on FA & 0.008 & -0.033 & -0.008 & -0.005 & -0.002 \\
\hspace{0.5cm}  & (0.000) & (0.005) & (0.003) & (0.003) & (0.003) \\
\hspace{0.5cm} Cubic on GAP & 0.006 & 0.005 & -0.006 & -0.003 & -0.001 \\
\hspace{0.5cm}  & (0.000) & (0.003) & (0.002) & (0.002) & (0.002) \\
\multicolumn{6}{l}{\textit{Panel D: Large initial min. wage, added emp. mass allowed}} \\
\hspace{0.5cm} True average causal effect & -0.002 & 0.148 & 0.038 & 0.002 & 0.001 \\
\hspace{0.5cm} Cubic on FA & -0.009 & 0.234 & -0.043 & 0.011 & 0.003 \\
\hspace{0.5cm}  & (0.001) & (0.014) & (0.013) & (0.009) & (0.009) \\
\hspace{0.5cm} Cubic on GAP & -0.015 & 0.190 & 0.059 & 0.012 & 0.005 \\
\hspace{0.5cm}  & (0.001) & (0.007) & (0.007) & (0.004) & (0.004) \\
\bottomrule 
\end{tabular}

%% file: tab_gap_td.tex
\begin{tabular}{lccccc}
\toprule 
 & \multicolumn{5}{c}{Outcome} \\
 & Emp. & p10 & p25 & p50 & p90 \\
\midrule 
\multicolumn{6}{l}{\textit{Panel A: Only permanent differences in location}} \\
\hspace{0.5cm} True average causal effect & -0.010 & 0.026 & 0.012 & 0.007 & 0.003 \\
\hspace{0.5cm} Gap measure & -0.009 & 0.027 & 0.011 & 0.006 & 0.003 \\
\hspace{0.5cm}  & (0.000) & (0.002) & (0.001) & (0.001) & (0.001) \\
\multicolumn{6}{l}{\textit{Panel B: Adding location shocks, stable distributions}} \\
\hspace{0.5cm} True average causal effect & -0.010 & 0.026 & 0.012 & 0.007 & 0.003 \\
\hspace{0.5cm} Gap measure & -0.007 & 0.043 & 0.030 & 0.026 & 0.023 \\
\hspace{0.5cm}  & (0.001) & (0.005) & (0.006) & (0.006) & (0.006) \\
\multicolumn{6}{l}{\textit{Panel C: Adding dispersion differences and shocks, stable distributions}} \\
\hspace{0.5cm} True average causal effect & -0.010 & 0.026 & 0.013 & 0.007 & 0.003 \\
\hspace{0.5cm} Gap measure & -0.007 & 0.052 & 0.034 & 0.024 & 0.009 \\
\hspace{0.5cm}  & (0.001) & (0.006) & (0.006) & (0.006) & (0.009) \\
\multicolumn{6}{l}{\textit{Panel D: Average dispersion falls over time}} \\
\hspace{0.5cm} True average causal effect & -0.010 & 0.027 & 0.013 & 0.007 & 0.003 \\
\hspace{0.5cm} Gap measure & -0.004 & 0.045 & 0.031 & 0.023 & 0.009 \\
\hspace{0.5cm}  & (0.001) & (0.006) & (0.006) & (0.006) & (0.009) \\
\bottomrule 
\end{tabular}

%% file: tab_fa_td_placebo.tex
\begin{tabular}{lccccc}
\toprule 
 & \multicolumn{5}{c}{Outcome} \\
 & Emp. & p10 & p25 & p50 & p90 \\
\midrule 
\multicolumn{6}{l}{\textit{Panel A: Only permanent differences in location}} \\
\hspace{0.5cm} True average causal effect & -0.000 & 0.000 & 0.000 & 0.000 & 0.000 \\
\hspace{0.5cm} Fraction affected & 0.000 & 0.000 & 0.000 & 0.000 & 0.000 \\
\hspace{0.5cm}  & (0.000) & (0.001) & (0.001) & (0.001) & (0.001) \\
\multicolumn{6}{l}{\textit{Panel B: Adding location shocks, stable distributions}} \\
\hspace{0.5cm} True average causal effect & -0.000 & 0.000 & 0.000 & 0.000 & 0.000 \\
\hspace{0.5cm} Fraction affected & 0.001 & 0.027 & 0.029 & 0.029 & 0.030 \\
\hspace{0.5cm}  & (0.000) & (0.008) & (0.009) & (0.009) & (0.009) \\
\multicolumn{6}{l}{\textit{Panel C: Adding dispersion differences and shocks, stable distributions}} \\
\hspace{0.5cm} True average causal effect & -0.000 & 0.000 & 0.000 & 0.000 & 0.000 \\
\hspace{0.5cm} Fraction affected & 0.002 & 0.041 & 0.034 & 0.024 & 0.003 \\
\hspace{0.5cm}  & (0.001) & (0.010) & (0.009) & (0.008) & (0.012) \\
\multicolumn{6}{l}{\textit{Panel D: Average dispersion falls over time}} \\
\hspace{0.5cm} True average causal effect & -0.000 & 0.000 & 0.000 & 0.000 & 0.000 \\
\hspace{0.5cm} Fraction affected & 0.005 & 0.035 & 0.032 & 0.023 & 0.002 \\
\hspace{0.5cm}  & (0.001) & (0.010) & (0.009) & (0.009) & (0.012) \\
\bottomrule 
\end{tabular}

%% file: Tab_rob_dgps_big_bias.tex
\begin{tabular}{lcccc}
\toprule 
Outcome & Emp & p10-p50 & p25-p50 & p90-p50 \\
\midrule 
\multicolumn{5}{l}{\textit{Panel A: 125 DGPs with positive employment effects}} \\
\hspace*{0.2cm}Average causal effects & 0.011 & 0.140 & 0.030 & 0.001 \\
\hspace*{0.2cm}Eff. min. wage: share with significant bias & 0.10 & 0.28 & 0.23 & 0.56 \\
\hspace*{0.2cm}\phantom{Eff. min. wage: }average magnitude of bias, cond. on sig. bias & 0.018 & 0.201 & 0.159 & 0.230 \\
\hspace*{0.2cm}\phantom{Eff. min. wage: }average standard error & 0.001 & 0.024 & 0.016 & 0.027 \\
\hspace*{0.2cm}Fraction affected: share with significant bias & 0.50 & 0.39 & 0.19 & 0.07 \\
\hspace*{0.2cm}\phantom{Fraction affected: }average magnitude of bias, cond. on sig. bias & 0.019 & 0.262 & 0.206 & 0.157 \\
\hspace*{0.2cm}\phantom{Fraction affected: }average standard error & 0.001 & 0.012 & 0.011 & 0.020 \\
\multicolumn{5}{l}{\textit{Panel B: 251 DGPs with negative employment effects}} \\
\hspace*{0.2cm}Average causal effects & -0.018 & 0.142 & 0.038 & -0.009 \\
\hspace*{0.2cm}Eff. min. wage: share with significant bias & 0.70 & 0.22 & 0.18 & 0.58 \\
\hspace*{0.2cm}\phantom{Eff. min. wage: }average magnitude of bias, cond. on sig. bias & 0.031 & 0.187 & 0.150 & 0.210 \\
\hspace*{0.2cm}\phantom{Eff. min. wage: }average standard error & 0.004 & 0.016 & 0.011 & 0.021 \\
\hspace*{0.2cm}Fraction affected: share with significant bias & 0.59 & 0.39 & 0.19 & 0.10 \\
\hspace*{0.2cm}\phantom{Fraction affected: }average magnitude of bias, cond. on sig. bias & 0.034 & 0.271 & 0.258 & 0.147 \\
\hspace*{0.2cm}\phantom{Fraction affected: }average standard error & 0.003 & 0.012 & 0.013 & 0.026 \\
\multicolumn{5}{l}{\textit{Panel C: 124 DGPs with small emp. effects but large wage effects}} \\
\hspace*{0.2cm}Average causal effects & 0.001 & 0.191 & 0.041 & -0.001 \\
\hspace*{0.2cm}Eff. min. wage: share with significant bias & 0.16 & 0.22 & 0.31 & 0.58 \\
\hspace*{0.2cm}\phantom{Eff. min. wage: }average magnitude of bias, cond. on sig. bias & 0.015 & 0.185 & 0.156 & 0.241 \\
\hspace*{0.2cm}\phantom{Eff. min. wage: }average standard error & 0.001 & 0.021 & 0.016 & 0.027 \\
\hspace*{0.2cm}Fraction affected: share with significant bias & 0.23 & 0.63 & 0.34 & 0.13 \\
\hspace*{0.2cm}\phantom{Fraction affected: }average magnitude of bias, cond. on sig. bias & 0.018 & 0.266 & 0.252 & 0.151 \\
\hspace*{0.2cm}\phantom{Fraction affected: }average standard error & 0.001 & 0.014 & 0.014 & 0.027 \\
\bottomrule 
\end{tabular}

%% file: Tab_rob_dgps_nut_kaitz.tex
\begin{tabular}{lcccc}
\toprule 
Outcome & Emp & p10-p50 & p25-p50 & p90-p50 \\
\midrule 
\multicolumn{5}{l}{\textit{Panel A: 25 DGPs with positive employment effects}} \\
\hspace*{0.2cm}Average causal effects & 0.013 & 0.108 & 0.019 & 0.002 \\
\hspace*{0.2cm}Eff. min. wage: share with significant bias & 0.00 & 0.12 & 0.00 & 0.00 \\
\hspace*{0.2cm}\phantom{Eff. min. wage: }average magnitude of bias, cond. on sig. bias & -- & 0.066 & -- & -- \\
\hspace*{0.2cm}\phantom{Eff. min. wage: }average standard error & 0.001 & 0.017 & 0.011 & 0.019 \\
\multicolumn{5}{l}{\textit{Panel B: 28 DGPs with negative employment effects}} \\
\hspace*{0.2cm}Average causal effects & -0.018 & 0.102 & 0.029 & -0.009 \\
\hspace*{0.2cm}Eff. min. wage: share with significant bias & 0.57 & 0.00 & 0.04 & 0.00 \\
\hspace*{0.2cm}\phantom{Eff. min. wage: }average magnitude of bias, cond. on sig. bias & 0.017 & -- & 0.056 & -- \\
\hspace*{0.2cm}\phantom{Eff. min. wage: }average standard error & 0.003 & 0.007 & 0.005 & 0.010 \\
\multicolumn{5}{l}{\textit{Panel C: 28 DGPs with small emp. effects but large wage effects}} \\
\hspace*{0.2cm}Average causal effects & 0.002 & 0.150 & 0.040 & -0.001 \\
\hspace*{0.2cm}Eff. min. wage: share with significant bias & 0.07 & 0.18 & 0.00 & 0.00 \\
\hspace*{0.2cm}\phantom{Eff. min. wage: }average magnitude of bias, cond. on sig. bias & 0.011 & 0.059 & -- & -- \\
\hspace*{0.2cm}\phantom{Eff. min. wage: }average standard error & 0.001 & 0.010 & 0.009 & 0.015 \\
\bottomrule 
\end{tabular}

%% file: Tab_rob_dgps_nut_fa.tex
\begin{tabular}{lcccc}
\toprule 
Outcome & Emp & p10-p50 & p25-p50 & p90-p50 \\
\midrule 
\multicolumn{5}{l}{\textit{Panel A: 92 DGPs with positive employment effects}} \\
\hspace*{0.2cm}Average causal effects & 0.011 & 0.134 & 0.025 & 0.002 \\
\hspace*{0.2cm}Fraction affected: share with significant bias & 0.77 & 0.55 & 0.33 & 0.01 \\
\hspace*{0.2cm}\phantom{Fraction affected: }average magnitude of bias, cond. on sig. bias & 0.014 & 0.177 & 0.132 & 0.051 \\
\hspace*{0.2cm}\phantom{Fraction affected: }average standard error & 0.001 & 0.012 & 0.010 & 0.018 \\
\multicolumn{5}{l}{\textit{Panel B: 170 DGPs with negative employment effects}} \\
\hspace*{0.2cm}Average causal effects & -0.019 & 0.134 & 0.030 & -0.009 \\
\hspace*{0.2cm}Fraction affected: share with significant bias & 0.75 & 0.44 & 0.23 & 0.01 \\
\hspace*{0.2cm}\phantom{Fraction affected: }average magnitude of bias, cond. on sig. bias & 0.023 & 0.208 & 0.151 & 0.066 \\
\hspace*{0.2cm}\phantom{Fraction affected: }average standard error & 0.003 & 0.012 & 0.011 & 0.023 \\
\multicolumn{5}{l}{\textit{Panel C: 88 DGPs with small emp. effects but large wage effects}} \\
\hspace*{0.2cm}Average causal effects & 0.001 & 0.183 & 0.031 & -0.000 \\
\hspace*{0.2cm}Fraction affected: share with significant bias & 0.50 & 0.70 & 0.36 & 0.01 \\
\hspace*{0.2cm}\phantom{Fraction affected: }average magnitude of bias, cond. on sig. bias & 0.012 & 0.208 & 0.170 & 0.051 \\
\hspace*{0.2cm}\phantom{Fraction affected: }average standard error & 0.001 & 0.014 & 0.013 & 0.024 \\
\bottomrule 
\end{tabular}

%% file: Tab_rob_dgps_pt_fa_bias.tex
\begin{tabular}{lcccc}
\toprule 
Estimator & Emp & p10-p50 & p25-p50 & p90-p50 \\
\midrule 
\multicolumn{5}{l}{\textit{Panel A: Mean absolute bias}} \\
FA & 0.013 & 0.104 & 0.055 & 0.012 \\
GAP & 0.008 & 0.087 & 0.027 & 0.007 \\
FA quadratic & 0.005 & 0.139 & 0.078 & 0.012 \\
GAP quadratic & 0.006 & 0.081 & 0.033 & 0.004 \\
FA cubic & 0.003 & 0.183 & 0.104 & 0.017 \\
GAP cubic & 0.005 & 0.089 & 0.029 & 0.005 \\
\multicolumn{5}{l}{\textit{Panel B: 90th percentile of absolute bias}} \\
FA & 0.029 & 0.229 & 0.162 & 0.025 \\
GAP & 0.019 & 0.238 & 0.068 & 0.016 \\
FA quadratic & 0.011 & 0.351 & 0.188 & 0.020 \\
GAP quadratic & 0.015 & 0.185 & 0.094 & 0.008 \\
FA cubic & 0.007 & 0.472 & 0.282 & 0.023 \\
GAP cubic & 0.012 & 0.230 & 0.073 & 0.008 \\
\multicolumn{5}{l}{\textit{Panel C: Average standard error}} \\
FA & 0.001 & 0.008 & 0.005 & 0.002 \\
GAP & 0.001 & 0.007 & 0.004 & 0.001 \\
FA quadratic & 0.002 & 0.018 & 0.013 & 0.007 \\
GAP quadratic & 0.001 & 0.009 & 0.006 & 0.003 \\
FA cubic & 0.006 & 0.037 & 0.031 & 0.021 \\
GAP cubic & 0.002 & 0.015 & 0.011 & 0.007 \\
\bottomrule 
\end{tabular}

%% file: Tab_rob_dgps_pt_fa_brackets.tex
\begin{tabular}{lcccc}
\toprule 
Estimator pair & Emp & p10-p50 & p25-p50 & p90-p50 \\
\midrule 
FA and GAP & 0.14 & 0.25 & 0.14 & 0.12 \\
FA and FA quadratic & 0.04 & 0.28 & 0.52 & 0.33 \\
FA and GAP quadratic & 0.06 & 0.13 & 0.29 & 0.16 \\
FA and FA cubic & 0.03 & 0.53 & 0.43 & 0.09 \\
FA and GAP cubic & 0.03 & 0.29 & 0.30 & 0.06 \\
GAP and FA quadratic & 0.17 & 0.52 & 0.65 & 0.45 \\
GAP and GAP quadratic & 0.07 & 0.33 & 0.33 & 0.21 \\
GAP and FA cubic & 0.17 & 0.77 & 0.54 & 0.21 \\
GAP and GAP cubic & 0.10 & 0.50 & 0.40 & 0.15 \\
FA quadratic and GAP quadratic & 0.10 & 0.19 & 0.32 & 0.24 \\
FA quadratic and FA cubic & 0.06 & 0.31 & 0.49 & 0.30 \\
FA quadratic and GAP cubic & 0.07 & 0.13 & 0.35 & 0.30 \\
GAP quadratic and FA cubic & 0.09 & 0.48 & 0.62 & 0.24 \\
GAP quadratic and GAP cubic & 0.03 & 0.20 & 0.34 & 0.16 \\
FA cubic and GAP cubic & 0.06 & 0.28 & 0.28 & 0.08 \\
\bottomrule 
\end{tabular}